\PassOptionsToPackage{usenames,dvipsnames}{xcolour}
\documentclass[
twocolumn,
twocolappendix,nofootinbib,iop]{openjournal}

\pdfoutput=1 
\usepackage{amsmath,amssymb,amstext}
\usepackage[T1]{fontenc}
\usepackage{apjfonts}
\usepackage{ae,aecompl}
\usepackage[utf8]{inputenc}
\usepackage{xcolor}
\usepackage[colorlinks,allcolors=blue]{hyperref}
\usepackage[figure,figure*]{hypcap}
\usepackage{natbib}
\usepackage{url}
\usepackage{mdwlist}
\usepackage{multirow}
\usepackage{lineno}
\usepackage{fontawesome}

\usepackage[caption=false]{subfig}
\usepackage{mathrsfs}
\usepackage{bm}
\usepackage{braket}
\usepackage{listings}
\usepackage{cases}
\usepackage{here}
\usepackage{comment}
\usepackage{soul}
\usepackage{cancel}
\usepackage{cases}
\usepackage{longtable}
\usepackage[normalem]{ulem}
\usepackage{xspace}
\usepackage{lineno}

\usepackage{adjustbox}

\definecolor{MONZA}{HTML}{CF000F}
\definecolor{DARKBLUE}{HTML}{00008b}
\definecolor{DARKMAGENTA}{HTML}{8b008b}

\newcommand{\eq}[1]{\begin{equation}\begin{split} #1 \end{split}\end{equation}}

\newcommand{\lr}[1]{\left( #1 \right)}

\newcommand{\Ms}{M_{\odot}}

\newcommand{\mr}{\mathrm}

\defcitealias{2010MNRAS.405.2579O}{OM10}
\defcitealias{2024ApJ...960...53V}{vdW24}
\defcitealias{2016ApJ...821..101Z}{Z16}
\defcitealias{lensed_qso_database_clemon}{Gravitationally Lensed Quasar Database}
\defcitealias{Abe_SL-Hammocks_2024}{SL-Hammocks}
\newcommand{\cta}{\citetalias}
\newcommand{\cpa}{\citepalias}

\begin{document}
\title{A halo model approach for mock catalogs of time-variable strong gravitational lenses}
\shorttitle{Halo model approach for strong lens mock catalogs}
\shortauthors{Abe et al.}

\author{\vspace{-1.0cm}Katsuya T. Abe$^{1}$}
\author{Masamune Oguri$^{1,2}$}
\author{Simon Birrer$^{3}$}
\author{Narayan Khadka$^{3}$}
\author{Philip J. Marshall$^{4,5}$}
\author{Cameron Lemon$^{6}$}
\author{Anupreeta More$^{7,8}$}
\author{the LSST Dark Energy Science Collaboration}

\affiliation{$^{1}$Center for Frontier Science, Chiba University, 1-33 Yayoi-cho, Inage-ku, Chiba 263-8522, Japan}
\affiliation{$^{2}$Department of Physics, Graduate School of Science,
Chiba University, 1-33 Yayoi-Cho, Inage-Ku, Chiba 263-8522, Japan}
\affiliation{$^{3}$Department of Physics and Astronomy, Stony Brook University, Stony Brook, NY 11794, USA}
\affiliation{$^{4}$Kavli Institute for Particle Astrophysics and Cosmology, Department of Physics, Stanford University, Stanford, CA, USA}
\affiliation{$^{5}$SLAC National Accelerator Laboratory, Menlo Park, CA, USA}
\affiliation{$^{6}$The Oskar Klein Centre, Department of Physics, Stockholm University, SE-10691 Stockholm, Sweden}
\affiliation{$^{7}$The Inter-University Centre for Astronomy and Astrophysics, Post Bag 4, Ganeshkhind, Pune 411007, India}
\affiliation{$^{8}$Kavli Institute for the Physics and Mathematics of the Universe (WPI), 5-1-5 Kashiwanoha, Kashiwa-shi, Chiba 277-8583, Japan}

\email{ktabecosmology@gmail.com}
\email{masamune.oguri@chiba-u.jp}

\begin{abstract}
Time delays in both galaxy- and cluster-scale strong gravitational lenses have recently attracted a lot of attention in the context of the Hubble tension. 
Future wide-field cadenced surveys, such as the Legacy Survey of Space and Time (LSST), are anticipated to discover strong lenses across various scales.
We generate mock catalogs of strongly lensed quasars (QSOs) and supernovae (SNe) on galaxy-, group-, and cluster-scales based on a halo model that incorporates dark matter halos, galaxies, and subhalos. 
For the upcoming LSST survey, we predict that approximately 4000 lensed QSOs and 200 lensed SNe with resolved multiple images will be discovered. Among these, about 80 lensed QSOs and 10 lensed SNe will have maximum image separations larger than $10~\mathrm{arcsec}$, which roughly correspond to cluster-scale strong lensing.
We find that adopting the Chabrier stellar initial mass function (IMF) instead of the fiducial Salpeter IMF reduces the predicted number of strong lenses approximately by half,  while the distributions of lens and source redshifts and image separations are not significantly changed. 
In addition to mock catalogs of multiple-image lens systems, we create mock catalogs of highly magnified systems, including both multiple-image and single-image systems. We find that such highly magnified systems are typically produced by massive galaxies, but non-negligible fraction of them are located in the outskirt of galaxy groups and clusters. Furthermore, we compare subsamples of our mock catalogs with lensed QSO samples constructed from the Sloan Digital Sky Survey and Gaia to find that our mock catalogs with the fiducial Salpeter IMF reproduce the observation quite well. In contrast, our mock catalogs with the Chabrier IMF predict a significantly smaller number of lensed QSOs compared with observations, which adds evidence that the stellar IMF of massive galaxies is Salpeter-like.  Our python code to generate the mock catalogs, \texttt{Strong Lensing Halo model-based mock catalogs~(SL-Hammocks}), as well as mock catalogs of lensed QSOs and SNe are made available online.
\end{abstract}

\maketitle

\section{Introduction}
In recent years, the field of cosmology has been confronted with a significant challenge known as the \textit{Hubble tension} problem \citep[see, e.g.,][for a review]{2021CQGra..38o3001D}. The Hubble tension is an empirical problem in which two types of observations have conflicting measurements of the Hubble constant $H_0$. In the first approach, $H_0$ is estimated from the early-universe measurements, in particular cosmic microwave background~(CMB), assuming the so-called standard cosmology, $\Lambda$-dominated cold dark matter~($\Lambda$CDM) model.
For example, \citet{2020A&A...641A...6P} provides a tight constraint of $H_0=67.4 \pm 0.5 \mathrm{~km} \mathrm{~s}^{-1} \mathrm{Mpc}^{-1}$ assuming the flat $\Lambda$CDM model. The other approach measures $H_0$ more directly from observations of the local Universe. Recently, Supernovae and $H_0$ for the Equation of State of dark energy~(SH0ES) collaboration has obtained $H_0=73.04 \pm 1.04 \mathrm{~km} \mathrm{~s}^{-1} \mathrm{Mpc}^{-1}$ by type Ia supernovae (SNe Ia) calibrated via the distance ladder in the local Universe \citep{2022ApJ...934L...7R}. The two types of observations exhibit inconsistency with significant $5\sigma$ tension of the value of $H_0$, which may pose a problem for the validity of the $\Lambda$CDM model.

One of the important avenues for resolving this problem is the estimation of the Hubble constant with other independent methodologies. The Tip of the Red Giant Branch~(TRGB), a waypoint in the evolutionary state for giant stars, offers another way to determine $H_0$ as a local standard candle. Some work \citep[see e.g.,][]{2020ApJ...891...57F} shows a consistent result with both to $1.5\sigma$ or less of $H_0\sim 70~\mathrm{~km} \mathrm{~s}^{-1} \mathrm{Mpc}^{-1}$, while others \citep[see e.g.,][]{2020AAS...23633601A, 2021ApJ...911...65B, 2022ApJ...933..172J} yield somewhat higher value at $H_0\sim 71.5 - 73~\mathrm{~km} \mathrm{~s}^{-1} \mathrm{Mpc}^{-1}$. Recently, \citet{2023ApJ...954L..31S} measure $H_0=73.22 \pm 2.06 \mathrm{~km} \mathrm{~s}^{-1} \mathrm{Mpc}^{-1}$ using TRGB, which further clarifies the tension, by analyzing in conjunction with the Pantheon$+$ SN Ia sample while considering a correction for differences in SN surveys and local flows.
Various other methods have also been used to measure $H_0$, including Megamasers \citep{2013ApJ...767..155K,2013ApJ...767..154R,2015ApJ...800...26K,2019ApJ...886L..27R}, gravitational waves \citep{2017Natur.551...85A,2018Natur.561..355M,2019NatAs...3..940H}, fast radio bursts \citep{2022MNRAS.511..662H,2022MNRAS.515L...1W,2022MNRAS.516.4862J,2023ApJ...946L..49L,2022arXiv221213433Z}, baryon acoustic oscillations \citep{2018ApJ...853..119A,2021A&A...647A..38B}, Type II SNe \citep{2022MNRAS.514.4620D}, ages of old astrophysical objects \citep{2022JHEAp..36...27V,2022ApJ...928..165W}, and others \citep{2022LRR....25....6M}. However, a satisfactory answer to this tension has not been found yet.

Another independent approach for $H_0$ measurements is to focus on time delays between multiple images of strong gravitational lenses.
It was pointed out in the 1960s that observations
of time delays between multiple images of gravitationally
lensed SNe can be used to estimate $H_0$ independently of the distance ladder~\citep{1964MNRAS.128..307R}.
However, it is only recently that we have been able to measure the Hubble constant accurately enough to use it to potentially resolve the Hubble tension.

Due to the scarcity of lensed SNe, lensed quasars~(QSOs) have been mostly used to constrain the Hubble constant.
For example, in \citet{2020MNRAS.498.1420W}, the $H_0$ Lenses in COSMOGRAIL's Wellspring~(H0LiCOW) collaboration estimated $H_0$ as $H_0=73.3_{-1.8}^{+1.7} \mathrm{~km} \mathrm{~s}^{-1} \mathrm{Mpc}^{-1}$ based on analysis of the time delays of six lensed QSOs with the simple mass distribution of lensing galaxies such as an elliptical power-law or stars plus standard dark matter halos.
In \citet{2020A&A...643A.165B}, the time-delay cosmography~(TDCOSMO) team obtained a constraint of $H_0=74.5_{-6.1}^{+5.6} \mathrm{~km} \mathrm{~s}^{-1} \mathrm{Mpc}^{-1}$, by considering a more flexible radial mass distribution and breaking the degeneracy in mass models with stellar kinematics.
While galaxy-scale lensed QSOs are used in the H0LiCOW and TDCOSMO analyses, recently, in \citet{2023Sci...380.1322K}, SN Refsdal, the first multiply-lensed SN with time delay measurements, is finally used to obtain the constraint on the Hubble constant of $H_0=64.8_{-4.3}^{+4.4} \mathrm{~km} \mathrm{~s}^{-1} \mathrm{Mpc}^{-1}$ \citep[see also][for the detailed analysis of the mass model dependence]{2024arXiv240213476L}. In addition to the difference in the source population, SN Refsdal was strongly lensed by a massive cluster of galaxies rather than a single galaxy and, hence, is highly complementary to the constraints from H0LiCOW and TDCOSMO.  Besides, another measurement of $H_0$ from the gravitationally lensed SN H0pe due to a massive cluster of galaxies, $H_0=75.7_{-5.5}^{+8.1} \mathrm{~km} \mathrm{~s}^{-1} \mathrm{Mpc}^{-1}$, was reported very recently \citep{2024arXiv240318902P}.

The measurement of $H_0$ using time-delay lensings will advance significantly in the coming decade thanks to several large-scale galaxy survey projects, including Euclid and Rubin Observatory Legacy Survey of Space and Time~(LSST).
\citet[OM10 hereafter]{2010MNRAS.405.2579O} was among the first to make a comprehensive prediction of event rates of galaxy-scale lensed SNe and QSOs, such as those used in the H0LiCOW and TDCOSMOS analysis, in various wide-field cadenced imaging surveys including LSST.
According to \cta{2010MNRAS.405.2579O}, the LSST should find more than 8000 galaxy-scale lensed QSOs, some 3000 of which will have well-measured time delays. The LSST should also find some 130 lensed SNe during the 10-year survey duration.
\cta{2010MNRAS.405.2579O} also computed the available precision on the Hubble constant from the LSST survey and obtained $\sigma(h)=0.017$ as the predicted marginalized $68$ percent confidence intervals, where $h\equiv H_0/100~\mathrm{km} \mathrm{~s}^{-1} \mathrm{Mpc}^{-1}$.
We note that there are some minor updates in the calculation of \cta{2010MNRAS.405.2579O}, as presented in \citet{2018MNRAS.480.3842O, 2023MNRAS.520.3305L,2023ApJ...943...38S}. Henceforth, we refer to the updated \cta{2010MNRAS.405.2579O} calculation as OM10+.

Strong gravitational lenses are predominantly produced by single massive galaxies. This is one of the reasons why the analysis of TDCOSMO and H0LiCOW, as well as \cta{2010MNRAS.405.2579O}, have focused on galaxy-scale strong lenses.
However, several cluster-scale lensed QSO systems have already been found, which recently attracts increasing attention due to the Hubble tension problem.
\citet{2021MNRAS.501..784D} used the cluster-scale lensed QSO \text {SDSS J1004+4112}~\citep{2003Natur.426..810I, 2004ApJ...605...78O, 2005PASJ...57L...7I, 2008ApJ...676..761F} together with 7 galaxy-scale lensed QSOs to obtain
$H_0=71.8_{-3.3}^{+3.9} \mathrm{~km} \mathrm{~s}^{-1} \mathrm{Mpc}^{-1}$.
\citet{2023ApJ...959..134N} focused more specifically on cluster-scale lensed QSOs by studying three cluster-lensed QSOs \text {SDSS J1004+4112, SDSS J1029+2623}~\citep{2006ApJ...653L..97I, 2008ApJ...676L...1O, 2013ApJ...764..186F}, and \text {SDSS J2222+2745}~\citep{2013ApJ...773..146D, 2015ApJ...813...67D}, to produce a
combined measurement of the Hubble constant as $H_0=74.1 \pm 8.0 \mathrm{~km} \mathrm{~s}^{-1} \mathrm{Mpc}^{-1}$. 
Very recently, \text {SDSS J1004+4112} has been reanalyzed by \citet{2023arXiv230914776M} using a free-form lens model and by \citet{2023PhRvD.108h3532L} using 16 different lens models, with results of $H_0=74_{-13}^{+9} \mathrm{~km} \mathrm{~s}^{-1} \mathrm{Mpc}^{-1}$ and $H_0=67.5_{-8.9}^{+14.5} \mathrm{~km} \mathrm{~s}^{-1} \mathrm{Mpc}^{-1}$, respectively.

In this work, we present detailed predictions of the numbers
of gravitationally lensed QSOs and SNe produced for a wide mass range of lens objects, including both galaxy- and cluster-scale lenses, with a particular focus on the LSST.
Specifically, we adopt the so-called halo model \citep[see e.g.,][for a review]{2023OJAp....6E..39A} approach adopting a compound lens model that consists of the dark matter halo and galaxy (stellar mass) components.
This approach differs from the one in \cta{2010MNRAS.405.2579O} and OM10+ that focuses only on galaxy-scale strong lenses.
They assume the singular isothermal ellipsoid~(SIE) model and calculate the lensing properties from the line-of-sight projected stellar velocity dispersion of lens galaxies obtained from the distributions of the velocity dispersions, the so-called velocity function, empirically derived in the local universe by the Sloan Digital Sky Survey~(SDSS).
While the SIMCT pipeline presented in \citet{2016MNRAS.455.1191M} can also generate mock lenses at galaxy- as well as group- and cluster-scales, their approach differs from the halo model approach in this paper in that they use a hybrid approach of injecting mock lensed sources in foreground deflectors selected from a real or mock survey. 

In this work, equipped with the halo model, we generate mock catalogs of lensed QSOs and SNe, especially focusing on time delay and highly magnified systems, which would be observed in LSST's 10-year long observing run. Thanks to the halo model approach, our mock catalogs cover a wide image separation range. We then discuss the properties of lens objects and images for the mocks while comparing the mock catalogs with those of OM10+.
Our mock catalog generation code, \texttt{Strong Lensing Halo model-based mock catalogs~(SL-Hammocks)}, will be made publicly available on GitHub.
Although we here only focus on the LSST survey, the \texttt{SL-Hammocks} code can be applied to arbitrary ongoing and future time-domain optical imaging surveys.
Our work and the \texttt{SL-Hammocks} code will be helpful for future astrophysical and cosmological applications of lensed QSO and SN systems that have been/will be discovered.

In addition to mock catalogs of multiply lens systems as those presented in OM10+, we construct mock catalogs of highly magnified systems, for which both single and multiple image systems are included. Since the contribution from group-scale halos is expected to be important for predicting the abundance of highly magnified systems \citep{2005ApJ...621..559K}, we expect that our halo model approach enables a reliable prediction on the abundance and properties of such systems. The mock catalogs of highly magnified systems will be useful for e.g., assessing the magnification bias on observations of high-redshift QSOs and SNe in the LSST survey.

This paper is organized as follows. We describe our halo model-based deflector model in Sec.~\ref{sec: lensmodel}. We check the validity of our model to populate galaxies in halos by comparing various observables predicted by our model with observational data in Sec.~\ref{sec: original_pop}. We then generate mock catalogs for lensed QSOs and SNe with our deflector model. The results of our mock catalogs are presented and analyzed in
Sec.~\ref{sec: lensmock}. 
We compare our mock catalog with known lensed QSO samples in Sec.~\ref{sec: mock_sdss}.
We summarize the main conclusion in Sec.~\ref{sec: conclusion}.

Throughout this paper, we assume a flat $\Lambda$CDM cosmology and fix the cosmological parameters to the Planck 2018 best-fits~\citep{2020A&A...641A...6P}, specifically $h=0.677$, $\Omega_{\mathrm{b}}h^2=0.0224, \omega_{\mathrm{cdm}}=0.120, \ln \left(10^{10} \mathcal{A}_{\zeta}\right)=3.05$, and $n_{\mathrm{s}}=0.967$.

\section{Deflector model}\label{sec: lensmodel}
This section addresses the halo-model-based deflector model, which distinguishes our new mock catalog from that of \cta{2010MNRAS.405.2579O}. In this model, we generate deflector populations from dark matter halos and subhalos.
We employ a compound mass model where the mass distribution of individual deflectors is represented by a superposition of dark matter and stellar components. The advantage of this compound model over the simple SIE model used in \cta{2010MNRAS.405.2579O} is that it can naturally account for the transition in the density profile from the galaxy- to the cluster-mass scale, reflecting the structure formation through dark matter and baryon~\citep[e.g.,][]{2001ApJ...559..531K,2006MNRAS.367.1241O,2012ApJ...749...38M}.
While \cta{2010MNRAS.405.2579O} provides the distribution of image separations for galaxy-scale lenses expected to be found in the LSST, this study extends the scope to include not only galaxy-scale lenses but also group- and cluster-scale lenses.
Since early-type galaxies, particularly elliptical galaxies, are known to account for approximately $80\%$ of the total lensing probability \citep{1984ApJ...284....1T, 1992ApJ...393....3F, 1996ApJ...466..638K,  2007MNRAS.379.1195M}, we focus solely on elliptical galaxies as the stellar components of deflectors throughout this paper.

In the following subsections, we describe the population and mass distribution for each component of our deflector model.

\subsection{Host dark matter halos}
First, we present the population and mass distribution for distinct dark matter halos (henceforth referred to as host halos to distinguish them from subhalos that are introduced later), which is a basic unit of our deflector model.

To calculate the population of host halos, we adopt the halo mass function presented in \citet{2008ApJ...688..709T}. This model was the first to be calibrated for the halo mass defined from the spherical overdensity. We use the so-called virial mass for the halo mass throughout this paper. We compute the halo mass function for the virial mass using the \texttt{COLOSSUS} package \citep{2018ApJS..239...35D}. 
Strictly speaking, this halo mass in the halo mass function is the total mass of the halo, $M_{\mr{tot}}$, including the host halo mass, $M_{\mr{hh}}$, and the total subhalos' mass, $\sum_{i}M^{i}_{\mr{sh}}$, where $M^{i}_{\mr{sh}}$ is the $i$-th subhalo mass. In this study, we investigate strong gravitational lenses due to subhalos in addition to the host halo, as will be explained in detail later. We adopt $M_\mr{tot}$ as the mass of the host halo when we investigate the strong gravitational lensing phenomena of the host halo. On the other hand, we adopt $M_\mr{tot}-M_{\mr{sub}}^{i}$ as the mass of the host halo when calculating the strong gravitational lenses caused by the $i$-th subhalo. This would be a reasonable assumption since the subhalos exist along the smooth mass distribution of the host halo, as explained below.

For the radial density profile of host halos, we adopt the commonly-used Navarro-Frenk-White \citep[NFW;][]{1996ApJ...462..563N} profile
\eq{\label{eq: def_bfw}
\rho_{\mathrm{hh}}(r)=\frac{\rho_{\mathrm{s, hh}}}{\left(r / r_{\mathrm{s, hh}}\right)\left(1+r / r_{\mathrm{s, hh}}\right)^2},
}
where $\rho_{\mathrm{s, hh}}$ is the characteristic density, which is related to the host halo mass $M_{\mathrm{hh}}$ as
\eq{
\rho_{\mathrm{s, hh}}=\frac{M_{\mathrm{hh}}}{4\pi r_{\mathrm{s, hh}}^3 m_{\mathrm{nfw}}(c_{\mathrm{hh}})},
}
\eq{
m_{\mathrm{nfw}}(c)=\int_0^c \frac{r}{(1+r)^2} d r=\ln (1+c)-\frac{c}{1+c},
}
and $r_{\mathrm{s, hh}}$ and $c_{\mathrm{hh}}$ are the scale radius and the concentration parameter, respectively, which are related with each other using a virial radius $r_{\mathrm{vir, hh}}$ as $r_{\mathrm{s, hh}} = r_{\mathrm{vir, hh}}/c_{\mathrm{hh}}$.
For the mean concentration parameter $\bar{c}_{\mathrm{hh}}$, we adopt a fitting form of the mass–concentration relation for the virial overdensity presented by \citet{2015ApJ...799..108D} with updates of fitting parameters by \citet{2019ApJ...871..168D}, which is also evaluated using the \texttt{COLOSSUS} package.
We also include the scatter of the concentration parameter described by a lognormal distribution
\eq{\label{eq: concent_hh_w_scatter}
p(x)=\frac{1}{\sqrt{2 \pi} \sigma_{\ln x}} \exp \left[-\frac{(\ln x-\ln \bar{x})^2}{2 \sigma_{\ln x}^2}\right],
}
with the standard deviation $\sigma_{\ln c_{\mathrm{hh}}}=0.33$~\citep{2007MNRAS.378...55M}.
We note that throughout the paper we ignore any modification of the NFW profile due to e.g., the adiabatic contraction.

In this paper, we introduce the ellipticity for the projected surface mass density profile of host halos. 
Throughout this paper, it is introduced by replacing the distance $r$ in the NFW density profile Eq.~\eqref{eq: def_bfw} projected along the line-of-sight as
\eq{\label{eq: r_to_v}
r \rightarrow v \equiv \sqrt{\frac{\tilde{x}^2}{(1-e)}+(1-e) \tilde{y}^2},
}
where the new coordinates $\tilde{x}$ and $\tilde{y}$ are defined by
\eq{
\begin{aligned}
\tilde{x} & =x \cos \phi+y \sin \phi, \\
\tilde{y} & =-x \sin \phi+y \cos \phi,
\end{aligned}
}
with the position angle $\phi$ indicating the position angle of the ellipse measured counterclockwise from the $y$-axis. 

For host halos, we apply the transformation in Eq.~\eqref{eq: r_to_v} with their ellipticity, $e_{\mr{hh}}$, and the position angle, $\phi_{\mr{hh}}$. 
The ellipticity $e_{\mr{hh}}$ of each halo is assigned following a truncated normal distribution, which is defined by
\eq{\label{eq: truncated_normal_dist}
\mathcal{F}(x; \bar{x}, \sigma, a, b)=\frac{1}{\sigma} \frac{\mathcal{N}\left(\frac{x-\bar{x}}{\sigma}\right)}{\Phi\left(\frac{b-\bar{x}}{\sigma}\right)-\Phi\left(\frac{a-\bar{x}}{\sigma}\right)},
}
where $\mathcal{N}(x)$ shows the standard normal distribution, and $\Phi(x)=(1+\operatorname{erf}(x / \sqrt{2}))/2$ is the cumulative distribution function.
We set the mean of $e_{\mr{hh}}$ as a function of halo mass $M_{\mathrm{hh}}$ from \citet{2020MNRAS.496.2591O} as
\eq{\label{eq: mean_elip_hh}
\bar{e}_{\mathrm{hh}}(M_{\mathrm{hh}}) = \mathrm{max}\left[0.23,~0.094\log_{10}(M_{\mathrm{hh}})-0.95\right],
}
and the variance $\sigma_{e_{\mathrm{hh}}}=0.13$. We also set the lower cutoff $a_{e_{\mathrm{hh}}}=0.0$ and the upper cutoff $b_{e_{\mathrm{hh}}}=0.9$.
The value of the position angle $\phi_{\mr{hh}}$ is assigned according to a uniform random distribution between $-180$ to $180$ in units of degrees.

\subsection{Central galaxies in host halos}\label{subsec: central_gal}
\begin{figure}
    \centering
    \includegraphics[width=\linewidth]{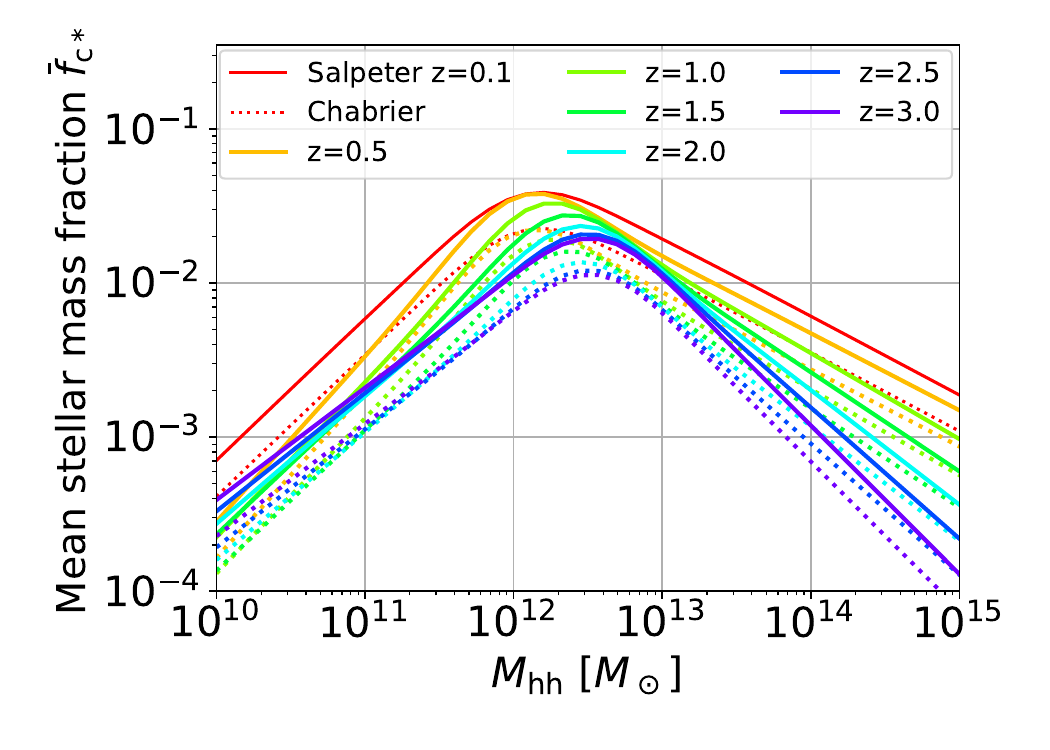}
    \caption{Mean stellar mass fractions for central galaxies. The solid lines show the fractions in our fiducial Salpeter IMF case, while the dotted lines represent the ones in the Chabrier IMF case. We note that the star formation efficiency peaks at round masses slightly smaller than $10^{12}M_\odot$ \citep{2013ApJ...762L..31B}.}
    \label{fig: fcen_imf}
\end{figure}

Next, we consider central galaxies in host halos.
We assume that each host halo has one central galaxy at its center with a stellar mass of $M_{\mathrm{c*}}=f_{\rm{c*}}M_{\mr{hh}}$. Throughout the paper we ignore any offset of centers between galaxies and dark matter halos, because $\Lambda$CDM simulations confirm that central galaxies are located almost always very near the centers of host halos with small offsets~\citep[e.g.,][]{2024OJAp....7E..65R}.
In practice, we should calculate $M_\mr{hh}$ by subtracting the stellar component of the central galaxy from the mass of the host halo. Since the total mass of the stellar component in the central galaxy is typically lower than $10\%$ of one of the host halo, we, however, ignore the stellar mass budget when calculating the host halo mass. 
To determine the mean value of the mass fraction, $\bar{f}_{\rm{c*}}$,
we use the fitting formula of stellar mass–halo mass relation presented in Appendix~J of \citet{2019MNRAS.488.3143B}. While in \citet{2019MNRAS.488.3143B} the Chabrier initial mass function \citep[Chabrier IMF;][]{2003PASP..115..763C} is used in deriving the fitting formula, we generalize this formula for other IMFs by multiplying a factor correcting for the difference of the mass-to-light ratios between the Chabrier and other IMFs, $f_{\mr{IMF}}$, by the stellar mass-halo mass relation. In this work, we adopt the Salpeter IMF~\citep{1955ApJ...121..161S}, which corresponds to $f_{\mr{IMF}}=1.715$, as our fiducial choice of the IMF, because the Salpeter IMF is generally favored in strong lensing observations \citep[e.g.,][]{2010ApJ...709.1195T,2014MNRAS.439.2494O,2015ApJ...800...94S}, albeit with some exceptions \citep[e.g.][]{2013MNRAS.434.1964S,2015MNRAS.449.3441S,2019A&A...630A..71S}.
We show the mean stellar mass fraction for both the Salpeter IMF and Chabrier IMF cases in Fig.~\ref{fig: fcen_imf}.

We also include the scatter in the stellar mass–halo mass relation, which is assumed to follow the log-normal distribution presented in Eq.~\eqref{eq: concent_hh_w_scatter} with $\sigma_{\ln M_{\mathrm{c*}}} =0.2$~\citep{2022ApJ...928...28G}.

For the density profile of central galaxies, we adopt the elliptical Hernquist profile~\citep{1990ApJ...356..359H}
\eq{
\rho_{\mathrm{c*}}(v_{\mathrm{c*}})=\frac{f_{\mathrm{c*}} M_{\mr{hh}}}{2 \pi\left(v_{\mathrm{c*}} / r_{\mathrm{b, cen}}\right)\left(v_{\mathrm{c*}}+r_{\mathrm{b, cen}}\right)^3},
}
where $v_{\mr{c*}}\equiv v(x,y,e_{\mr{c*}},\phi_{\mr{c*}})$ given in Eq.~\eqref{eq: r_to_v}, $e_{\mr{c*}}$ is the ellipticity, $\phi_{\mr{c*}}$ is the position angle, $r_\mathrm{b, cen}=0.551 r_{\mathrm{e, cen}}$, and $r_{\mathrm{e, cen}}$ is the so-called effective radius of the central galaxy.
For the mean value of the effective radius $\bar{r}_{e}$, we adopt the stellar size-mass relation for quiescent galaxies presented in \citet[vdW24 hereafter]{2024ApJ...960...53V}, which is based on observational results of the James Webb Space Telescope (JWST).
We employ the following fitting functional form for the stellar mass/size relation~\citep{2003MNRAS.343..978S, 2015MNRAS.447.2603L, 2021MNRAS.506..928N}
\eq{\label{eq: reff}
\log_{10}\lr{\frac{r_e}{h^{-1}\mr{kpc}}}&=\Gamma + \alpha\log_{10}\left(\frac{M_{\mathrm{c*}}}{\Ms}\right) \\
&\ + (\beta-\alpha)\log_{10}\left(1+10^{\log_{10}\lr{\frac{M_{\mathrm{c*}}}{\Ms}}-\delta}\right)\\
&\ -\omega\log_{\mr{10}}\lr{\frac{1+z}{1+z_{\mathrm{data}}}},
}
with fitting parameters of $(\Gamma , \alpha, \beta, \delta, \omega)$.
It should be noted that this fitting function is expressed by a stellar mass based on the Chabrier IMF.
Adopting the data for $0.5<z<1.0$ shown in Table 5 of \cta{2024ApJ...960...53V}, we obtain $(\Gamma, \alpha, \beta, \delta)=( 0.58, -0.067,  1.14,  10.82)$. 
We set $\omega=1.72$ for massive galaxies~($M_{\mathrm{c*}}>10^{10.5} \Ms$) according to \cta{2024ApJ...960...53V}.
For galaxies with masses lower than $10^{10.5} \Ms$, \cta{2024ApJ...960...53V} suggest that they have an approximately constant effective radius regardless of their mass. Thus, we set $r_e$ of such galaxies to $r_e(10^{10.5}M_\odot)$ calculated by Eq.~\eqref{eq: reff}.
We also include the scatter around the mean effective radius assuming the lognormal distribution in Eq.~\eqref{eq: concent_hh_w_scatter} with the dispersion of
\eq{\label{eq: sigma_Re}
\sigma_{\ln r_e}=\frac{\ln r_e^{(84)}-\ln r_e^{(16)}}{2},
}
where the values of the $16\%$th and $84\%$th percentiles, $r_e^{(84)}$ and $r_e^{(16)}$, respectively, are obtained from Table 5 of \cta{2024ApJ...960...53V}. Specifically, we calculate the scatter in Eq.~\eqref{eq: sigma_Re} using the fitting parameters $(\Gamma, \alpha, \beta, \delta) = (0.64, -0.054, 1.02, 10.80)$ for $r_e^{(84)}$ and $(0.77, -0.11,  1.19, 10.69)$ for $r_e^{(16)}$ at $z=z_{\mathrm{data}}$. We also assume that the redshift evolution can be neglected.

For the ellipticity $e_{\mathrm{c*}}$, we assume a truncated normal distribution in Eq.~\eqref{eq: truncated_normal_dist}, following \citet{2008AJ....135..512O}, with the mean of $0.3$ and the dispersion of $0.16$, that is consistent with the observed ellipticity distributions of early-type galaxies obtained by the SDSS measurements~\citep{1989A&A...217...35B,1993A&A...279...75S,1995MNRAS.276.1341J,2001AJ....121.2431R,2002ApJ...570..610A,2003ApJ...594..225S}. The distribution is truncated at $e_{\mathrm{c*}} = 0$ and $e_{\mathrm{c*}} = 0.9$.
Considering that the position angles of central galaxies often correlate with those of host halos, we set the position angle of each central galaxy $\phi_{\mr{c*}}$ to follow a Gaussian distribution centered on the position angle of its host halo ($\phi_{\mathrm{hh}}$), with a dispersion of $35.4~\mathrm{deg}$, as suggested by \citet{2009ApJ...694..214O}.

Since our model of the stellar mass-halo mass relation and the stellar mass-galaxy size relation does not assume any distinction between galaxy- and cluster-scale halos, galaxy properties change smoothly as a function of the halo mass from galaxy- to cluster-scale halos.

\subsection{Subhalos}\label{sec: subhalo}
Another important difference between this work and \cta{2010MNRAS.405.2579O} is that we also consider subhalos.
Here, we describe our model of subhalos and satellite galaxies.

To calculate the population of subhalos, we adopt a simple analytic model proposed in \citet{2020ApJ...901...58O} for the subhalo mass function.
The model is based on the extended Press-Schechter theory~\citep{1991ApJ...379..440B, 1991MNRAS.248..332B, 1993MNRAS.262..627L}, but additionally includes the two important effects of tidal stripping and dynamical friction.
The subhalo mass function at a given $z$, taking account of these effects, can be written as
\eq{\label{eq: combined_dnsh_dm}
\frac{d N_\mathrm{sh}}{d M_{\mathrm{sh}}}=f_{\mathrm{df}} \frac{d N_{\mathrm{sh}}}{d M_{\mathrm{f}}} \frac{d M_{\mathrm{f}}}{d M_{\mathrm{sh}}},
}
where $M_{\mathrm{sh}}$ is the subhalo mass at $z$, $M_{\mathrm{f}}$ denotes the subhalo mass at their accretion time, $f_{\mathrm{df}}$ is a suppression factor due to the dynamical friction, and $d N_{\mathrm{sh}}/d M_{\mathrm{f}}$ denotes the number distribution of progenitors with mass $M_{\mathrm{f}}$ at redshift $z_{\mathrm{f}}$ for a host halo with mass $M_{\mathrm{hh}}$ and redshift $z$, which is predicted by the extended Press-Schechter theory as
\eq{\label{eq: dndmf_EPS}
\frac{d N_{\mathrm{sh}}}{d M_{\mathrm{f}}}=\frac{M_{\mathrm{hh}}}{M_{\mathrm{f}}} P\left(M_{\mathrm{f}}, z_{\mathrm{f}} \mid M_{\mathrm{hh}}, z\right),
}
where $P\left(M_{\mathrm{f}}, z_{\mathrm{f}} \mid M_{\mathrm{hh}}, z\right)$ is the mean fraction of the total mass for $M_{\mathrm{f}}$,
\eq{
P\left(M_{\mathrm{f}}, z_{\mathrm{f}} \mid M_{\mathrm{hh}}, z\right)=\frac{1}{\sqrt{2 \pi}} \frac{\Delta \omega}{\Delta S^{3 / 2}} \exp \left(-\frac{\Delta \omega^2}{2 \Delta S}\right)\left|\frac{d \Delta S}{d M_{\mathrm{f}}}\right|,
}
with
\eq{
\Delta \omega=\delta_{\mathrm{c}}\left(z_{\mathrm{f}}\right)-\delta_{\mathrm{c}}(z) \text { and } \Delta S=\sigma^2\left(M_{\mathrm{f}}\right)-\sigma^2(M_{\mathrm{hh}}),
}
and $\delta_{\mathrm{c}}(z)=(3 / 20)(12 \pi)^{2 / 3}\left\{\Omega_{\mathrm{m}}(z)\right\}^{0.0055} / D_{+}(z)$ with the linear growth factor $D_+(z)$. 
To evaluate Eq.~\eqref{eq: dndmf_EPS}, we assume that all subhalos accrete at their median formation time instead of considering their assembly history. Following \citet{2007MNRAS.376..977G}, we calculate the median formation time by solving
\eq{
\delta_{\mathrm{c}}\left(z_{\mathrm{f}}\right)=\delta_{\mathrm{c}}(z)+\frac{0.974}{\sqrt{q}} \sqrt{\sigma^2\left(f_{\mathrm{f}} M\right)-\sigma^2(M)},
}
where $q=0.707$ and $f_{\mathrm{f}}=0.5$. 

The factor $d M_{\mathrm{f}}/d M_{\mathrm{sh}}$ in Eq.~\eqref{eq: combined_dnsh_dm} represents the effect of mass loss due to the tidal stripping.
We calculate this factor and the suppression factor $f_{\mathrm{df}}$ following \citet{2020ApJ...901...58O}.

We assume that the spatial distribution of subhalos within their host halo follows the mass distribution of the host halo that is modeled by the elliptical NFW profile. The radial distribution function of subhalos with masses in the range $[M_{\mr{f}}, M_{\mr{f}}+\Delta M_{\mr{f}}]$ at their accretion time is given by
\eq{
\frac{d^2N_{\mr{sh}}}{dM_{\mr{f}}dr}  =  \frac{4\pi r^2\rho_{\mr{hh}}(M_{\mr{hh}},z)}{M_{\mr{hh}}}\frac{dN_{\mr{sh}}}{dM_{\mr{f}}} = \frac{r}{m_{\mr{NFW}}(c_\mr{hh})(r_\mr{s,hh}+r)^2}\frac{dN_{\mr{sh}}}{dM_{\mr{f}}}.
}
In practice, when we spatially distribute subhalos, we make use of the inverse transform sampling method with the cumulative distribution function, 
\eq{
P_{\mr{sh}}(x_{\mr{hh}},M_{\mr{f}}) \equiv  \int_{0}^{x_{\mr{hh}}r_{\mr{vir,hh}}} \frac{d^2N_{\mr{sh}}}{dM_{\mr{f}}dr^\prime} dr^\prime = \frac{m_{\mr{NFW}}(c_\mr{hh}x_{\mr{hh}})}{m_{\mr{NFW}}(c_\mr{hh})}\frac{dN_{\mr{sh}}}{dM_{\mr{f}}},
}
where $x_{\mr{hh}}\equiv r/r_{\mr{vir,hh}}$.
After spatially distributing subhalos within their host halo, we project them along the line-of-sight.

For the density profile of subhalos, the truncated NFW profile is often adopted, considering tidal stripping effects. In fact, the mass loss due to tidal stripping in Eq.~\eqref{eq: combined_dnsh_dm} is also calculated with an assumption of the truncated NFW profile of subhalos.
However, while the tidal stripping is taken into account in calculating the subhalo mass function, for numerical reasons, in calculating lensing properties of subhalos we adopt the conventional elliptical NFW profile without truncation for the mass distribution of subhalos as well.
Instead, when calculating lensing properties of subhalos, we use the masses of subhalos before tidal stripping mass loss effects, i.e., $M_{\mathrm{f}}$, rather than their true masses after the tidal stripping.
This approach can be justified by the fact that tidal stripping and projection effects strip away the matter only outside of the subhalos, and their mass distribution near the center, which is important for strong lensing, is not changed.
We calculate ellipticity $e_{\mathrm{sh}}$ and position angle $\phi_{\mathrm{sh}}$ of subhalos in the same manner as host halos.

For the mean concentration parameter of subhalos $\bar{c}_{\mathrm{sh}}$, we use a fitting function of the mean mass-concentration relation specialized for subhalos, presented in \citet{2020MNRAS.492.3662I}.
Since this fitting formula is applicable only for subhalos with masses smaller than $10^{12} \Ms$, we adopt the larger of the concentration parameter $c^{\mr{IA}}$ calculated from \citet{2020MNRAS.492.3662I} or the concentration parameter $c^{\mr{DKJ}}$ derived from \citet{2015ApJ...799..108D} and \citet{2019ApJ...871..168D}.
We confirm that this transition occurs at around $M_{\mr{sh}}=10^{12} \Ms$.
Additionally, we multiply the concentration parameter by the ratio of the virial radius of the masses at accretion time and at $z$.
This correction corresponds to assuming that the scale radius does not change before and after mass loss due to tidal stripping effects.
To summarize, we calculate the mean concentration parameter of subhalos as
\eq{\label{eq: mean_csh}
\bar{c}_{\mathrm{sh}}(M_{\mathrm{sh}},z) =  \text{max}\left[c^{\mr{IA}}(M_{\mathrm{sh}},z) , c^{\mr{DKJ}}(M_{\mathrm{sh}},z) \right]\times \frac{r_{\mr{vir}}(M_{\mr{f}},z)}{r_{\mr{vir}}(M_{\mr{sh}},z)}.
}
It should be noted that $c_{\mathrm{sh}}$ is defined in terms of the virial radius, often referred to as $c_{\mathrm{vir}}$, while the mean mass-concentration relation presented in \citet{2020MNRAS.492.3662I} is for the one based on a radius defined at $200$ times the critical density, often denoted as $c_{\mathrm{200c}}$. 
Therefore, although not explicitly stated, we use the following conversion assuming the NFW profile,
\eq{
c_{\mathrm{vir}} = \left(\frac{200}{\Delta_{\mathrm{vir}} \Omega_{\mathrm{m}}(z)}\right)^{\frac{1}{3}}\left(\frac{H(z)}{H_0}\right)^{-\frac{2}{3}}c_{\mathrm{200c}},
}
where $\Delta_{\mathrm{vir}}$ is the virial overdensity \citep{10.1143/PTP.97.49}.
We include the scatter in the same manner as for $c_{\mathrm{hh}}$, which is described by Eq.~\eqref{eq: concent_hh_w_scatter} with the standard deviation $\sigma_{\ln c_{\mathrm{sh}}}=0.33$.
We also assign the ellipticity $e_{\mr{sh}}$ and the position angle $\phi_{\mr{sh}}$ of subhalos in the same manner as for host halos.

Similarly to central galaxies, we assume that each subhalo has one satellite galaxy at its center. 
We calculate properties of satellite galaxies, such as the density profile, the effective radius, the ellipticity, and the position angle, in the same manner as for central galaxies, which are described in Sec.~\ref{subsec: central_gal}, except for stellar masses of satellite galaxies.
The stellar masses of satellite galaxies are derived using the stellar-mass-halo-mass relation for satellite galaxies which is provided in Appendix~J of \citet{2019MNRAS.488.3143B} with masses of subhalos at their accretion time as well as considering the differences of the stellar IMF choice in the same way as for central galaxies.

When considering the lens equation for these subhalos, we always include the contribution of host halos and central galaxies to the lens potential. We do so by considering the same mass distributions as those used for computing their lensing properties, namely the elliptical NFW density profile for host halos and the elliptical Hernquist profile for central galaxies.

\subsection{External shear}
In addition to the deflector model mentioned above, we include external shear to account for the effect of the matter fluctuation along the line of sight. 
In \cta{2010MNRAS.405.2579O}, external shear $\gamma_{\mathrm{ext}}$ is assumed to follow a log-normal distribution with a mean of $0.05$ and dispersion of $0.2~\mathrm{dex}$. 
In OM10+, it is assumed that both $\gamma_{\mathrm{ext}, 1}$ and $\gamma_{\mathrm{ext}, 2}$, appeared in $\gamma_{\mathrm{ext}} = \sqrt{\gamma_{\mathrm{ext}, 1}^2 + \gamma_{\mathrm{ext}, 2}^2}$, follow a truncated normal distribution as
\eq{\label{eq: truncated_normal_dist_gamma_ext}
\mathcal{F}(x;\sigma_{\gamma_{\mathrm{ext}}}) = \frac{\frac{1}{\sigma_{\gamma_{\mathrm{ext}}}} \mathcal{N}\left(x/\sigma_{\gamma_{\mathrm{ext}}}\right)}{\Phi\left(0.5\right)-\Phi\left(-0.5\right)},
}
with the dispersion of $\sigma_{\gamma_{\mathrm{ext}}}$ which is estimated from the nonlinear power spectrum of \citet{2012ApJ...761..152T} as a function of $z$ and fitted by a simple function
\eq{\label{eq: sigma_gamma_ext}
\sigma_{\gamma_{\mathrm{ext}}} =
\left\{
\begin{array}{ll}
    0.023 z & \quad z<1.0\\
    0.023 + 0.032 \ln(z) & \quad \text{otherwise}.
\end{array}
\right.
}
This mode of external shear is confirmed to better match the ray-tracing simulation result \citep{2022AJ....163..139Y}. 

In this work, we adopt this improved model in Eqs.~\eqref{eq: truncated_normal_dist_gamma_ext} and \eqref{eq: sigma_gamma_ext}, including the redshift dependence for assigning the external shear.
Although external convergence may also be important, particularly for time delays \citep{2005MNRAS.364.1451O}, we do not include it explicitly in our calculation.

We note that \citet{2021ApJ...915....4L} analyze the contribution of the lens quadrupole to observed quad lens QSOs to suggest that about half of the systems are dominated by external shear rather than lens ellipticity, which is in tension with the prediction of \cta{2010MNRAS.405.2579O} as they predict that most of the lens systems are ellipticity dominated. Since the external shear model in this paper is updated from the model used in \cta{2010MNRAS.405.2579O} to include the redshift evolution, it is of interest to revisit this analysis using our new mock catalog.

\subsection{Lensing calculations}
The calculation of gravitational lensing effects is performed using the public code \texttt{glafic} \citep{2010PASJ...62.1017O, 2021PASP..133g4504O}.
For each host halo and subhalo, we solve the lens equation to derive multiple images, as well as magnifications and time delays for individual images. Here, for numerical reasons, we do not calculate the lens equation with all deflectors, including a host halo and the subhalos, simultaneously, but break them down into individual deflector systems. To summarize, the lensing calculation of each host halo involves three deflector components: the elliptical NFW profile for the host halo, the elliptical Hernquist profile for the central galaxy, and external shear, while the lensing calculation of each subhalo involves five deflector components: the elliptical NFW profile for the host halo, the elliptical Hernquist profile for the central galaxy, the elliptical NFW profile for the subhalo, the elliptical Hernquist profile for the satellite galaxy, and external shear. While we expect some correlation of external shear between a halo and a subhalo within the halo or between subhalos within a halo, we ignore such correlation for simplicity.
This approximation not to calculate the lens equation with all deflectors would be validated for two reasons. Firstly, as described in Sec.~\ref{sec: lensmodel}, we adopt $M_\mr{tot}-M^{i}_{\mr{sub}}$ as the mass of the host halo when performing lensing calculations for $i$-th subhalos of mass $M^{i}_{\mr{sub}}$. Therefore,  influences from other subhalos are collectively included as the influence from the host halo. Secondly, it would be very rare to have several subhalos located within a scale of about their Einstein radii.

We summarize the notation introduced in Sec.~\ref{sec: lensmodel} in Table~\ref{tab: notation}.

\begin{table*}[htbp]
\begin{center}
\caption{Definitions of properties of deflectors, including host halos, subhalos, central galaxies, satellite galaxies, and external shear, appeared in our deflector model described in Sec.~\ref{sec: lensmodel}. 
}
\begin{adjustbox}{width=\linewidth,center}
\begin{tabular}{ cccccc } 
\hline \hline
Symbol $X$ & Explanation & Mean $\bar{X}$ & Variance $\sigma_{X\mathrm{or}\ln(X)} $ & Distribution\\
\hline
$M_{\mr{hh}}$ & mass of host halo in units of $\Ms$& --  & -- & \citet{2008ApJ...688..709T}\\ 
$M_{\mr{tot}}$ & total mass of halo in units of $\Ms$& --  & -- & \citet{2008ApJ...688..709T}\\ 
$M_{\mr{sh}}$ & mass of subhalo in units of $\Ms$& --  & -- & ~\citet{1991ApJ...379..440B, 1991MNRAS.248..332B, 1993MNRAS.262..627L}\\ 
$M_{\mr{h}}$ & mass  of subhalo at the accretion time in units of $\Ms$& --  & -- & \citet{2020ApJ...901...58O}\\ 
$c_\mr{h}$ & concentration parameter for host halos & DKJ & 0.33 & Log-normal dist. (Eq.~\eqref{eq: concent_hh_w_scatter})\\ 
$c_\mr{sh}$ & concentration parameter for subhalos & Eq.~\eqref{eq: mean_csh} & 0.33 & Log-normal dist. (Eq.~\eqref{eq: concent_hh_w_scatter})\\ 
$e_\mr{h}$ & ellipticity for host halos & Eq.~\eqref{eq: mean_elip_hh} & 0.13 & Truncated normal dist. (see below Eq.~\eqref{eq: mean_elip_hh})\\ 
$e_\mr{sh}$ & ellipticity for subhalos & Eq.~\eqref{eq: mean_elip_hh} & 0.13 & Truncated normal dist. (see below Eq.~\eqref{eq: mean_elip_hh})\\ 
$\phi_\mr{h}$ & position angle for host halos & -- & -- & Uniform in $[-180, 180]$ deg\\ 
$\phi_\mr{sh}$ & position angle for subhalos & -- & -- & Uniform in $[-180, 180]$ deg\\ 
$M_{\mr{c*}}$ & stellar mass of central galaxy in units of $\Ms$& \cite{2019MNRAS.488.3143B} & 0.20 & Log-normal dist.\\ 
$M_{\mr{s*}}$ & stellar mass of satellite galaxy in units of $\Ms$& \cite{2019MNRAS.488.3143B} & 0.20 & Log-normal dist.\\ 
$r_e$ & effective radius of central/satellite galaxy in units of $h^{-1}\mathrm{kpc}$& Eq.~\eqref{eq: reff}  & Eq.~\eqref{eq: sigma_Re} & Log-normal dist.\\ 
$e_\mr{c*}$ & ellipticity for central galaxies & 0.30 & 0.16 & Truncated normal dist. (see below Eq.~\eqref{eq: sigma_Re})\\ 
$e_\mr{s*}$ & ellipticity for satellite galaxies & 0.30 & 0.16 & Truncated normal dist. (see below Eq.~\eqref{eq: sigma_Re})\\ 
$\phi_\mr{c*}$ & position angle for central galaxies & $\phi_\mr{h}$ & 35.4 & Normal dist.\\ 
$\phi_\mr{s*}$ & position angle for satellite galaxies & $\phi_\mr{sh}$ & 35.4 & Normal dist.\\ 
$\gamma_\mr{ext, 1}$ & external shear component & 0 & Eq.~\eqref{eq: sigma_gamma_ext} & Truncated normal dist. (Eq.~\eqref{eq: truncated_normal_dist_gamma_ext})\\ 
$\gamma_\mr{ext, 2}$ & external shear component  & 0 & Eq.~\eqref{eq: sigma_gamma_ext} & Truncated normal dist. (Eq.~\eqref{eq: truncated_normal_dist_gamma_ext})\\ 
\hline
\label{tab: notation}
\end{tabular}
\end{adjustbox}
\end{center}
\end{table*}

\section{Validation of the galaxy model}\label{sec: original_pop}

In order to create a realistic mock catalog of strong gravitational lenses, it is essential to check whether our galaxy model within halos is reliable.
In this section, we discuss the validity of our model in populating galaxies in halos by comparing various observables predicted by our model with observational data. To be more specific, we calculate quantities such as the velocity dispersion function and the stellar mass fundamental plane~(MFP) of galaxies for our lens model described in Sec.~\ref{sec: lensmodel}. In this section, we adopt a Salpeter IMF for our fiducial computation.

\subsection{Velocity dispersion function}

First, we discuss the velocity dispersion function of galaxies. Throughout this paper, we consider a luminosity-weighted averaged value of the line-of-sight velocity dispersion within the effective radius $r_{e}$, which we simply call the velocity dispersion and denote $\sigma_{e}$.

We calculate velocity dispersions of galaxies in our compound lens model using the \texttt{GalKin} package in the \texttt{Lenstronomy} python library \citep{2018PDU....22..189B, 2021JOSS....6.3283B}. 
In order to derive the velocity dispersion function of galaxies in our model, we first specify the field of view to populate halos and galaxies and then create a mock catalog of halos and galaxies in that region for each redshift bin as described in Sec.~\ref{sec: lensmodel}.
We include not only main halos and central galaxies but also subhalos and satellite galaxies. In the latter case, we adopt the masses at their accretion time for computing velocity dispersions because we use those masses for calculating lensing properties of subhalos (see the discussion in Sec.~\ref{sec: subhalo} for more details). We then use the \texttt{GalKin} package to compute the velocity dispersion for each galaxy in the mock catalog, assuming the NFW and Hernquist profiles as density profiles of dark matter and stellar mass components, respectively. Here, the effect of the ellipticity is ignored because, as explained later, velocity dispersions are obtained from observations after circularization.

We compare the velocity dispersion function predicted by our model with that from OM10+, which is based on \citet{2010MNRAS.404.2087B}.
In \citet{2010MNRAS.404.2087B}, the velocity dispersion function derived from the SDSS Data Release 6 is fitted to the Schechter function \citep{2003ApJ...594..225S} as 
\eq{\label{eq: vdf_sigma_e8}
\varphi_{\mathrm{loc}}(\sigma_{e/8})=\varphi_*\left(\frac{\sigma_{e/8}}{\sigma_{*}}\right)^\alpha \exp \left[-\left(\frac{\sigma_{e/8}}{\sigma_{*}}\right)^\beta\right] \frac{\beta}{\Gamma(\alpha / \beta)} \frac{1}{\sigma_{e/8}},
}
with $\varphi_{*}=2.099 \times 10^{-2}(h / 0.7)^3 \mathrm{Mpc}^{-3}, \sigma_*=113.78 \mathrm{~km} \mathrm{~s}^{-1}$, $\alpha=0.94$, and $\beta=1.85$.
OM10+ also takes account of the redshift evolution of the velocity dispersion function, mainly reflecting an evolution in the mass-to-light ratio of galaxies \citep{2005ApJ...622L...5T,2007ApJ...655...30V,2010A&A...524A...6S,2010ApJ...724..714H,2014ApJ...793L..31V,2015ApJ...806..122Z}, as
\eq{\label{eq: vdf_sigma_e8_z}
\varphi(\sigma_{e/8}, z)=\varphi_{\text {loc}}(\sigma_{e/8}) \frac{\left.\varphi_{\text {hyd }}(\sigma_{e/8}, z\right)}{\varphi_{\text {hyd }}(\sigma_{e/8}, 0)}.
}
Here $\varphi_{\text {hyd }}(\sigma, z)$ is the velocity dispersion function presented in \citet{2015MNRAS.454.2770T}, which is derived from the Illustris simulation \citep{2014MNRAS.444.1518V,2014Natur.509..177V,2014MNRAS.445..175G}. We caution that Eqs.~(\ref{eq: vdf_sigma_e8}) and (\ref{eq: vdf_sigma_e8_z}) are used only for the validation test conducted in this Section, and not used for our main calculations to generate mock catalogs based on the halo model approach.

It should be noted that this velocity dispersion function is for $\sigma_{e/8}$, i.e., a luminosity-weighted velocity dispersion averaged within the effective radius $r_{e}/8$, rather than within $r_{e}$\footnote{Conventionally, this luminosity-weighted velocity dispersion within $r=r_{e}/8$ has often been adopted, historically based on the aperture size of previous observations. In this study, since we are considering strong gravitational lensing effects and the Einstein radii are typically comparable to $r=r_{e}$ \citep[e.g.,][]{2014MNRAS.439.2494O}, the luminosity-weighted velocity dispersion averaged within $r=r_{e}$ is more relevant than $r=r_{e}/8$.}.
Hence, we convert the velocity dispersion function of $\sigma_{e/8}$ to that of $\sigma_{e}$ using the commonly-used relation for elliptical galaxies, which is derived from 40 early-type galaxies of the SAURON integral field spectroscopy data in \citet{2006MNRAS.366.1126C},
\eq{\label{eq: convert_sig_e8_to_sig_e}
\frac{\sigma_1}{\sigma_2}=\left(\frac{r_1}{r_2}\right)^{-0.066},
}
where $\sigma_1$ and $\sigma_2$ are velocity dispersions measured in apertures of radii $r_1$ and $r_2$, respectively. In the current situation, we multiply $8^{-0.066}\approx 0.87$ by $\sigma_{e/8}$ obtained from Eq.~\eqref{eq: vdf_sigma_e8} to convert it to $\sigma_{e}$.

\begin{figure*}[t]
    \begin{tabular}{ccc}
       \begin{minipage}{.31\linewidth}
        \centering
        \includegraphics[width=\hsize]{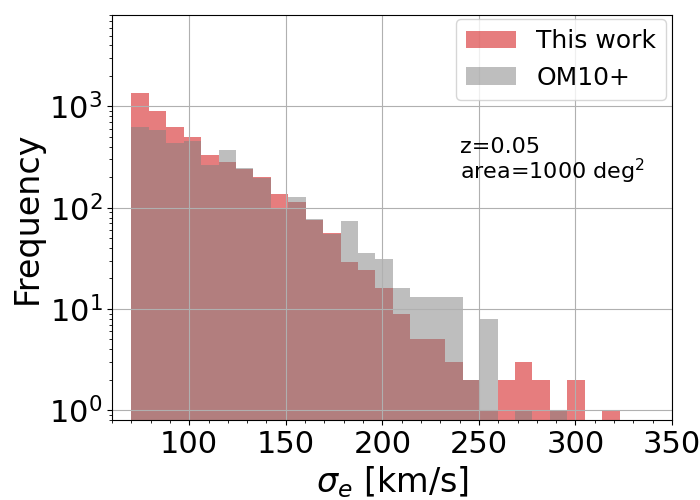}
      \end{minipage} &
      \begin{minipage}{.31\linewidth}
        \centering
        \includegraphics[width=\hsize]{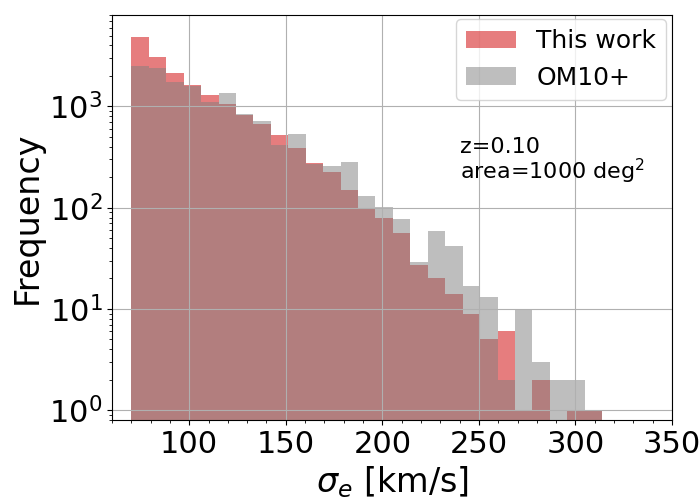}
      \end{minipage} &
   
      \begin{minipage}{.31\linewidth}
        \centering
        \includegraphics[width=\hsize]{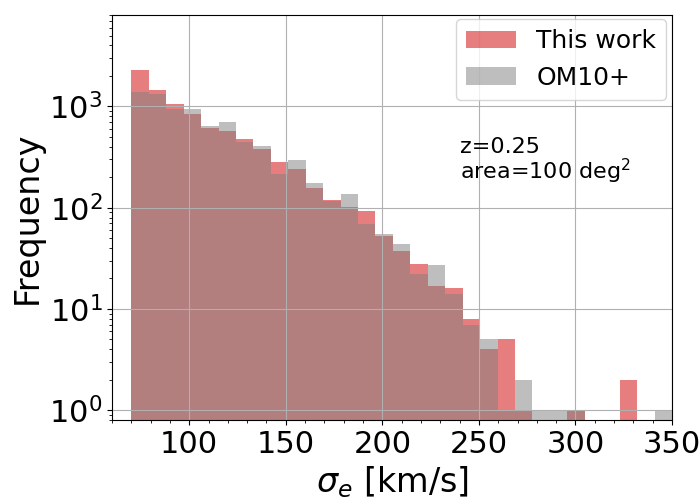}
      \end{minipage}\\
      \begin{minipage}{.31\linewidth}
        \centering
        \includegraphics[width=\hsize]{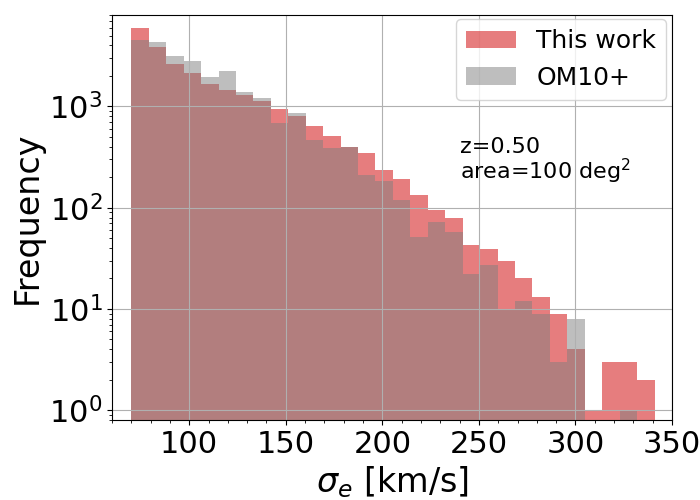}
    \end{minipage} &
   
      \begin{minipage}{.31\linewidth}
        \centering
        \includegraphics[width=\hsize]{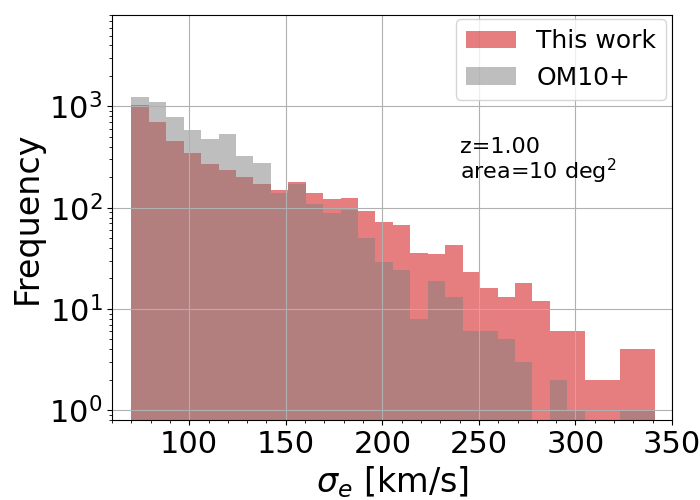}
      \end{minipage} &
      \begin{minipage}{.31\linewidth}
        \centering
        \includegraphics[width=\hsize]{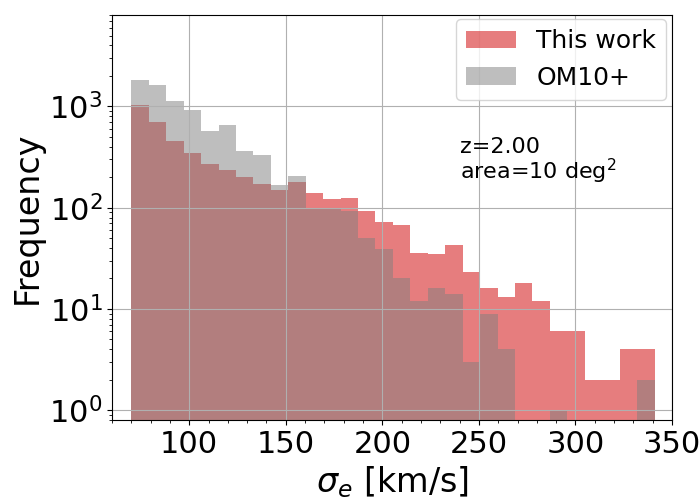}
    \end{minipage}
    \end{tabular}
     \caption{Comparison between the velocity dispersion function derived with our model described in Sec.~\ref{sec: lensmodel}~(red) and that presented in \citet{2018MNRAS.480.3842O}~(grey). 
     The vertical axis shows the number of galaxies in each redshift bin with the width of $\Delta z=0.01$ for each velocity dispersion bin. Here, we set the area written on each panel to derive the number.
     \textit{Top left}: $z=0.05$, \textit{Top center}: $z=0.1$, \textit{Top right}: $z=0.25$, \textit{Bottom left}: $z=0.5$, \textit{Bottom center}: $z=1.0$, \textit{Bottom right}: $z=2.0$.
     }
     \label{fig: veldisp_func}
  \end{figure*}

Fig.~\ref{fig: veldisp_func} shows the comparison between the velocity dispersion function derived from our model and that of OM10+ calculated from Eq.~\eqref{eq: vdf_sigma_e8_z} for several redshifts.
For numerical reasons, we set a lower limit on the mass of halos to be calculated,  $M_{\mathrm{hh}}\gtrsim 10^{11}~\Ms$ (or $M_{\mathrm{sh}}\gtrsim 10^{10}~\Ms$ for subhalos), which may lead to the underestimation of the number of galaxies with low velocity dispersions in our model.
We find that both the velocity dispersion functions agree well for low redshifts ($z\lesssim 0.5$), which are main redshifts of lensing objects for lensed QSOs and SNe.
In contrast, we find some mismatch at high redshifts ($z \gtrsim 0.5$), indicating that the redshift evolution in our model deviates from that in the Illustris simulation. To be more quantitative, we conduct the Kolmogorov-Smirnov test focusing on the velocity dispersion function at high velocity dispersions $v\geq 200 \mathrm{~km} \mathrm{~s}^{-1}$ to find that p-values are larger than 0.002 for $z\leq 0.25$ and are smaller than $2\times 10^{-5}$ for $z\geq 0.5$, indicating that the mismatch of two velocity dispersion functions is indeed significant at high redshifts. Although the Illustris simulations were calibrated to match a set of global galaxy properties, they predicted galaxy sizes about twice as large as those observed \citep{2014MNRAS.445..175G}. In the updated IllustrisTNG simulations incorporating AGN feedback, the predicted galaxy sizes are significantly reduced \citep{2018MNRAS.473.4077P, 2018MNRAS.474.3976G}. Since there is a negative correlation between the galaxy size and the velocity dispersion for given stellar and halo masses, it is possible that the original Illustris simulations predict significantly smaller velocity dispersion at such high redshifts. As described in Sec.~\ref{sec: lensmodel}, our model adopts galaxy sizes directly measured from observations. Therefore, we expect our model to predict the redshift evolution of the velocity dispersion function more accurately and reliably than OM10+.

\subsection{Effective radii and velocity dispersions at intermediate redshift}

\begin{figure*}[t]
    \begin{tabular}{cc}
       \begin{minipage}{.5\linewidth}
        \centering
        \includegraphics[width=\hsize]{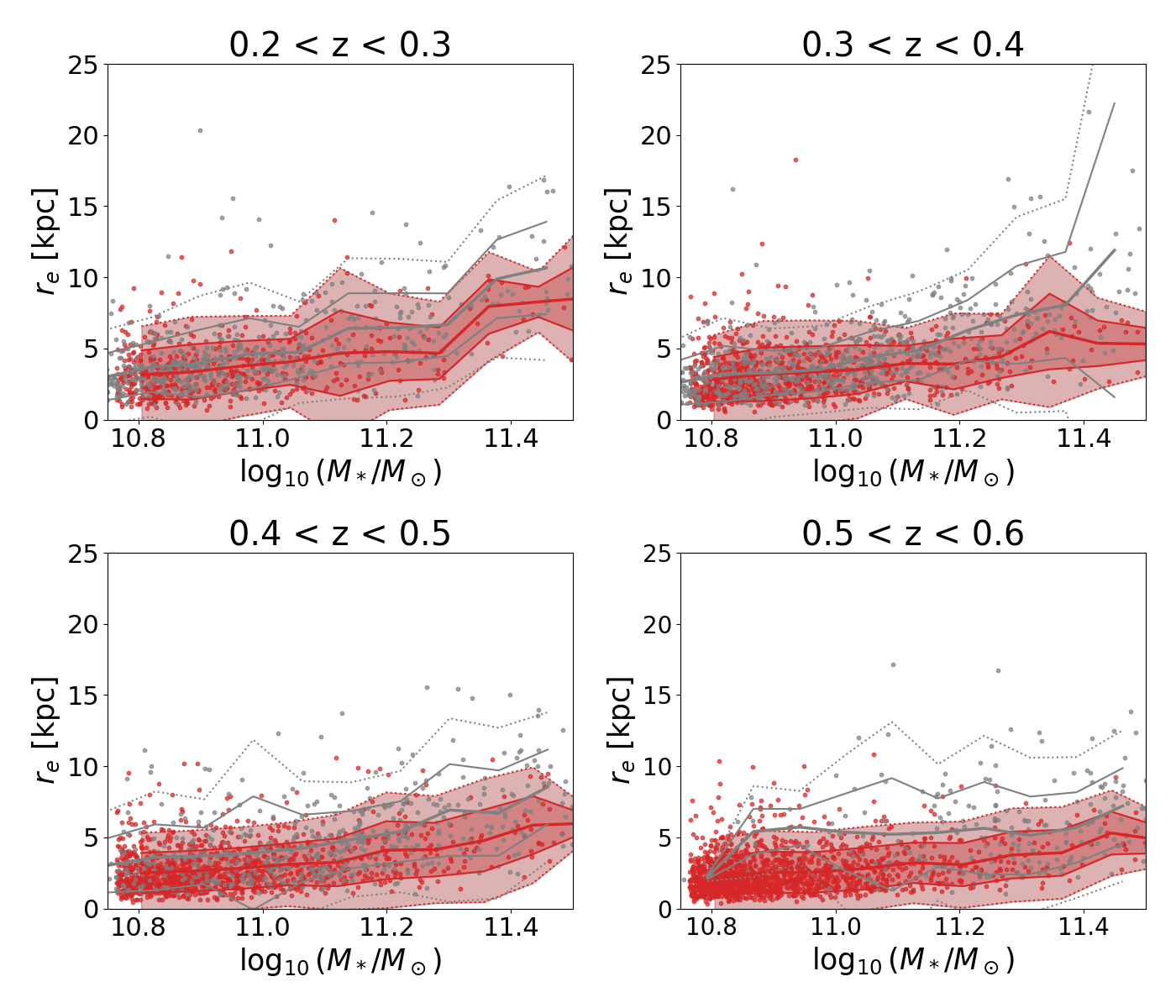}
      \end{minipage}
   
      \begin{minipage}{.5\linewidth}
        \centering
        \includegraphics[width=\hsize]{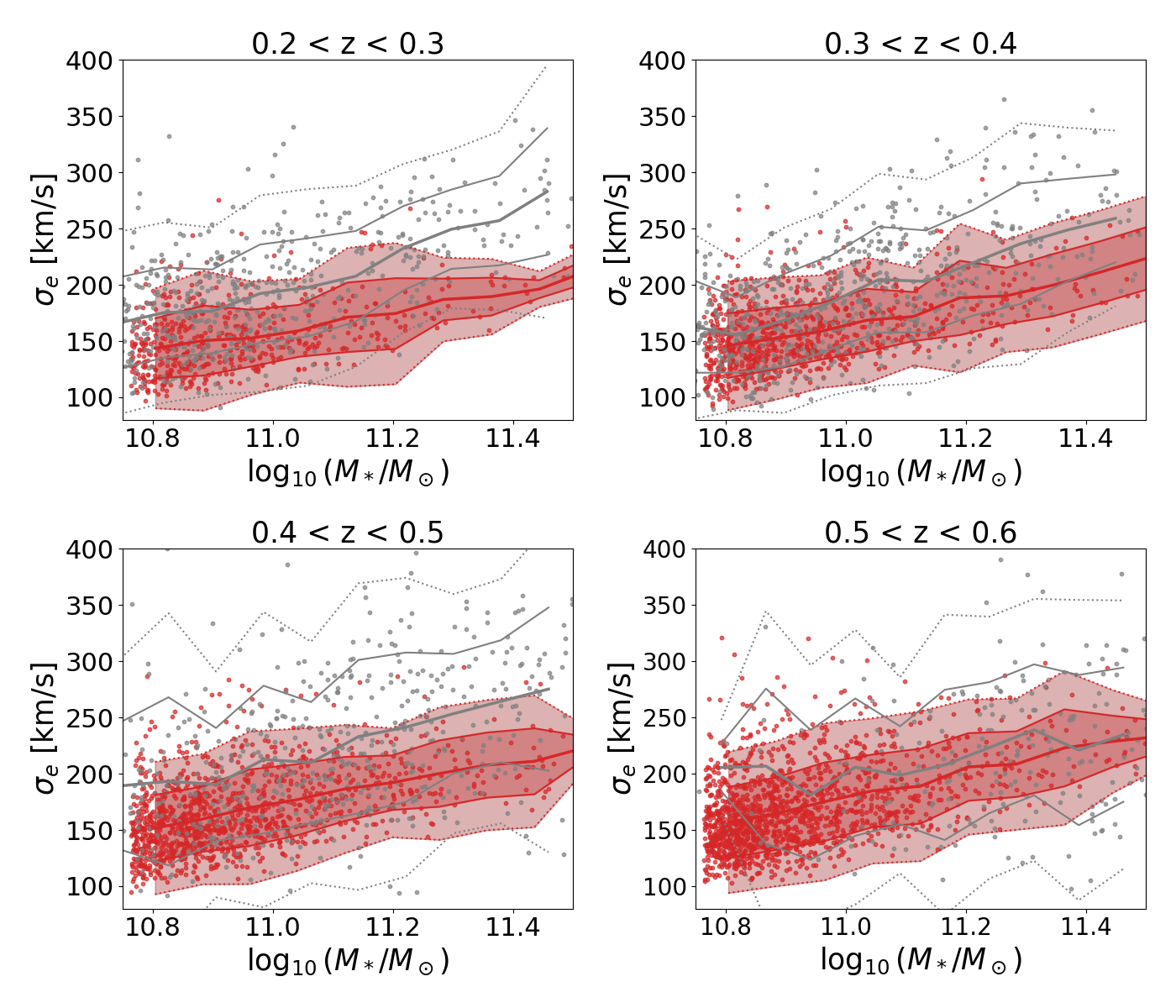}
      \end{minipage}
    \end{tabular}
        \caption{Comparison between our model with observation data of SHELS F2 at intermediate redshifts, $z\lesssim 0.6$. 
        Only the galaxy samples with stellar masses of $M_* \gtrsim 10^{10.7}~\Ms$ are plotted.
        The four panels on the left side show effective radii as a function of stellar mass at $0.2<z<0.3$, $0.3<z<0.4$, $0.4<z<0.5$, and $0.5<z<0.6$, while the four panels on the right represent those of velocity dispersions. The thick solid lines show the center lines for the galaxy samples.
        The dark red and light red regions show the $68\%$ and $95\%$ areas for our galaxy samples, while the thin grey solid and dotted lines show the $68\%$ and $95\%$ areas for SHELS F2 galaxy samples. We set the area of $4.2~\mathrm{deg}^2$ to create a mock galaxy sample, matching the area covered by the SHELS F2 catalog. Note that we do not plot the error bars for the effective radii and velocity dispersions for the SHELS F2 data.
        }
    \label{fig: reff_sig_e_evolve}
\end{figure*}

As mentioned above, our model of the effective radius of galaxies is based on measurements with JWST for galaxies at redshift $0.5<z<2.5$.
For the intermediate redshift range of $0.2\lesssim z\lesssim 0.6$, there is a galaxy catalog from the SHELS survey of the F2 field of
the Deep Lens Survey \citep{2006ApJ...643..128W,2014ApJS..213...35G,2022ApJ...929...61D}.
As an additional validity check, we compare effective radii and velocity dispersions for early-type galaxies at $0.2\lesssim z\lesssim 0.6$ in that galaxy catalog with those predicted by our model. 
The four panels on the left side of Fig.~\ref{fig: reff_sig_e_evolve} show the effective radii of galaxies as a function of stellar mass, focusing on massive galaxies with masses of $M_*\gtrsim 10^{10.7}~M_\odot$ at $0.2<z<0.3$, $0.3<z<0.4$, $0.4<z<0.5$, and $0.5<z<0.6$. The right panel displays the corresponding velocity dispersions.
For both effective radii and velocity dispersions, the SHELS F2 catalog galaxies, especially more massive galaxies than $10^{11}~M_\odot$, appear to have somewhat larger value than our mock galaxies, which is, however in agreement at the $2 \sigma$ level.

\subsection{Intrinsic scatter of velocity dispersions}

Next, we discuss the intrinsic scatter of velocity dispersions, $\sigma_{e}$, for a given stellar mass of galaxies, which is also important in order to create a realistic mock catalog.
For instance, ignoring the intrinsic scatter typically is suggested to lead to an underestimation of the slope of the $M_*$-$\sigma_{e}$ relation \citep[see, e.g.,][]{2010ApJ...724..511A}.
Using the mock catalog obtained for the comparison of the velocity dispersion function, we can estimate the scatter of $\sigma_{e}$ at given $M_*$ and $z$.
In the case of our model, the intrinsic scatter originates from three sources,  (1) the scatter of the stellar-mass-halo-mass relation \citep{2022ApJ...928...28G}, (2) the scatter of the halo mass-concentration parameter relation \citep{2007MNRAS.378...55M}, and (3) the scatter of the galaxy size-mass relation \cpa{2024ApJ...960...53V}. We note that \cta{2024ApJ...960...53V} claims that measurement uncertainties in observations are smaller than the intrinsic scatter of $M_*-r_{e}$ relation. Therefore, we assume that the scatter obtained from our model is also dominated by the intrinsic scatter.

\citet{2020MNRAS.498.1101C} suggest that the intrinsic scatter of the $M_*$-$\sigma_{e}$ relation is about $0.075_{-0.003}^{+0.003}~\mathrm{dex}$ at given $M_*$ for early-type galaxies with masses $M_*\gtrsim 3.0\times 10^{10}~\Ms$ in the SDSS sample.
They also investigate the redshift evolution of the intrinsic scatter using the sample of the massive early-type galaxies at higher redshifts, $z\lesssim 2$, composed of 26 galaxies drawn from the LRIS sample presented in \citet{2014ApJ...783..117B}, 56 galaxies from \citet{2013ApJ...771...85V}, four galaxies from \citet{2015A&A...573A.110G}, and 24 galaxies from \citet{2017ApJ...834...18B}.
As a result, it is found that the redshift evolution of the scatter is weak in terms of the Bayesian factor comparison.

\begin{figure*}[t]
    \begin{tabular}{ccc}
       \begin{minipage}{.32\linewidth}
        \centering
        \includegraphics[width=\hsize]{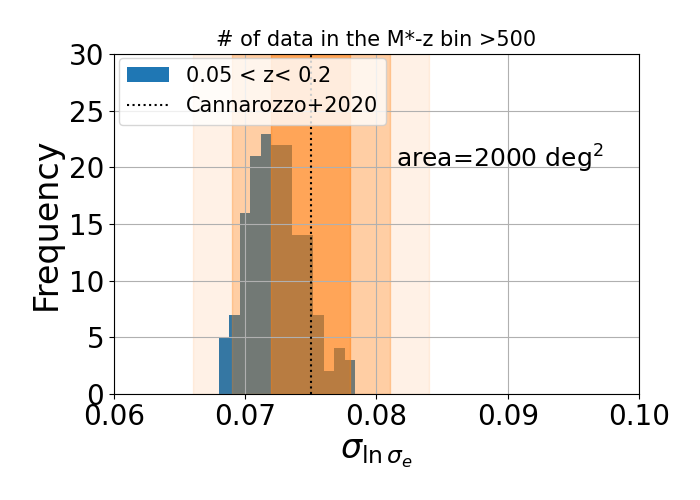}
      \end{minipage}
   
      \begin{minipage}{.32\linewidth}
        \centering
        \includegraphics[width=\hsize]{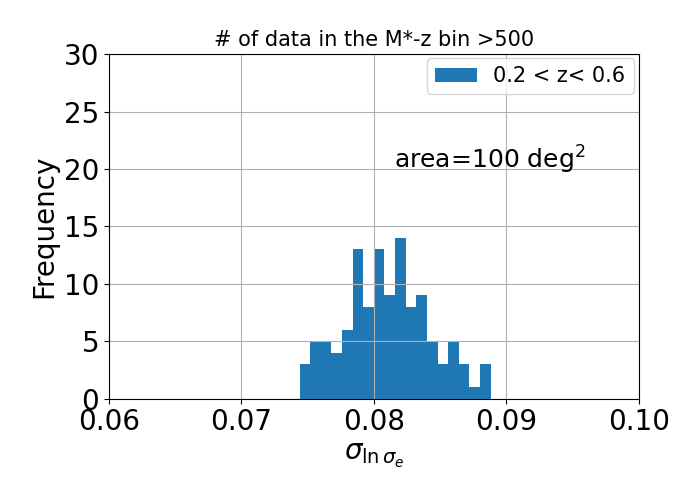}
      \end{minipage}
   
      \begin{minipage}{.32\linewidth}
        \centering
        \includegraphics[width=\hsize]{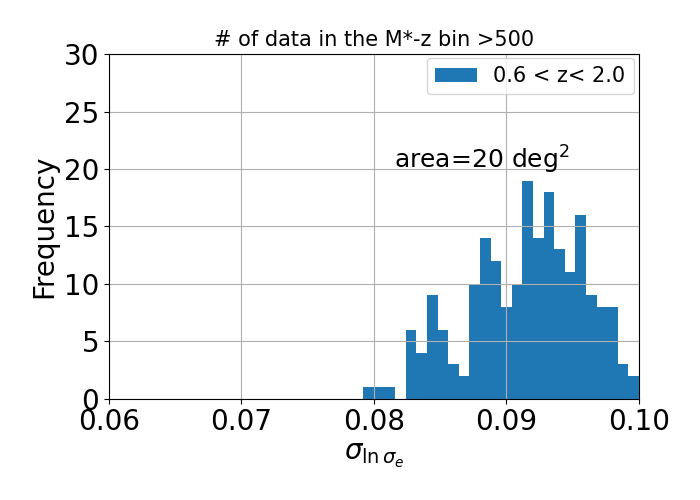}
      \end{minipage}
    \end{tabular}
    \caption{The scatter of $\sigma_{e}$ for mock galaxies at given $z$ and $M_*$ calculated with our model. We compute the scatter only for mock galaxies with $10.5<\log_{10}(M_*/\Ms)<11.7$ at $z<2$. We set $2000~\mathrm{deg}^2$, $100~\mathrm{deg}^2$, and $20~\mathrm{deg}^2$ to create mock galaxy samples at $0.05<z<0.2$~(left), $0.2<z<0.6$~(middle), and $0.6<z<2.0$~(right), respectively. The calculation method of the scatter is as follows. We first divide our mock galaxies into ($z, M_*$)-bins. We then calculate the scatter of velocity dispersion for the sample of galaxies in each bin. We finally plot these histograms only using those bins that hold more than $\mathcal{O}(10^2)$ galaxy samples to ensure that the variance is properly computed. Here we set $(\Delta z,\Delta\log_{10}(M*/\Ms))=(0.01,0.02)$ for the galaxy sample at $0.05<z<0.2$, $(0.02,0.02)$ at $0.2<z<0.6$, and $(0.1,0.02)$ at $0.6<z<2.0$ for a numerical reason. 
     The grey dotted line appearing only in the left panel represents a best-fit value obtained for the SDSS galaxy sample at $0.05<z<0.2$ in \citet{2020MNRAS.498.1101C}, $\sigma_{\ln \sigma_e}= 0.075^{+ 0.003}_{- 0.003} ~\mathrm{dex}$.
     The orange-colored regions represent the $\pm 1\sigma$, $\pm 2\sigma$, and $\pm 3\sigma$ regions, in order from darkest to lightest.}
     \label{fig: scatter_sigma_e}
\end{figure*}

We compute the intrinsic scatter in a stellar mass range of $10.5 < \log_{10}(M_*/\Ms) < 11.7$ at $z < 2$. First, we divide our mock galaxies within this mass range into ($z, M_*$)-bins. We then calculate the scatter of velocity dispersion for the galaxy samples in each bin. The histograms of the scatter are plotted in Fig.~\ref{fig: scatter_sigma_e}, divided into three redshift ranges: $0.05 < z < 0.2$, $0.2 < z < 0.6$, and $0.6 < z < 2.0$. For the galaxy samples, we set $(\Delta z, \Delta\log_{10}(M*/\Ms))=(0.01,0.02)$ for $0.05 < z < 0.2$, $(0.02,0.02)$ for $0.2 < z < 0.6$, and $(0.1,0.02)$ for $0.6 < z < 2.0$ for numerical reasons.
It should be noted that the histograms in Fig.~\ref{fig: scatter_sigma_e} include only the scatter from bins containing a sufficient number of galaxy samples, $N_{\mathrm{gal}}^{(z, M_*)-\mathrm{bin}} > \mathcal{O}(10^2)$. In the left panel of Fig.~\ref{fig: scatter_sigma_e}, we also plot the result of \citet{2020MNRAS.498.1101C} for SDSS samples containing galaxies at $0.05 < z < 0.2$ for comparison.
We find that our model predicts the intrinsic scatter of about $0.072^{+0.002}_{-0.002}$~dex at $0.05< z < 0.2$, which agrees with the result of \citet{2020MNRAS.498.1101C} at $\sim 1\sigma$ level.
We also find that our model does not exhibit a strong redshift dependence on the intrinsic scatter. Since \citet{2020MNRAS.498.1101C} suggest the weak redshift evolution of the scatter, both results appear to be consistent.

\subsection{Stellar mass fundamental plane}

Finally, we check the MFP for early-type galaxies. 
A broad range of early-type galaxies has been known to exhibit a tight relation between their sizes, velocity dispersions, and surface brightnesses, which is the so-called fundamental plane. When the surface luminosity is replaced by the surface mass, it is then called the MFP, which is also known as a tight relation for early-type galaxies.
These are considered to reflect the virial equilibrium of galaxies.
This empirical relation provides a good indicator for assessing the validity of our mock catalog.

In our mock, we calculate the stellar mass surface density as
\eq{
\Sigma_*=\frac{M_*}{2 \pi r_e^2},
}
in units of $\Ms~\mr{kpc}^{-2}$, ignoring the potential mass-to-light gradients. 
The MFP can be described by
\eq{\label{eq: mfp_general}
a \log_{10} \left(\Sigma_*\right)+b \log_{10} (\sigma_{e})+c \log_{10} \left(r_{e}\right)+d=0,
}
with $\sigma_{e}$ in units of $\mathrm{km/s}$ and $r_{e}$ in units of $\mathrm{kpc}$.
\citet[Z16 hereafter]{2016ApJ...821..101Z} provide the best-fit values for $(a,b,c,d)$ in Eq.~\eqref{eq: mfp_general} as $(-0.84,1.63,-1.00,4.42)$. 
It is known that this plane is almost invariant as a function of redshift, although the zero point may slightly evolve with redshift.

\begin{figure}
    \centering
    \includegraphics[width=\linewidth,clip]{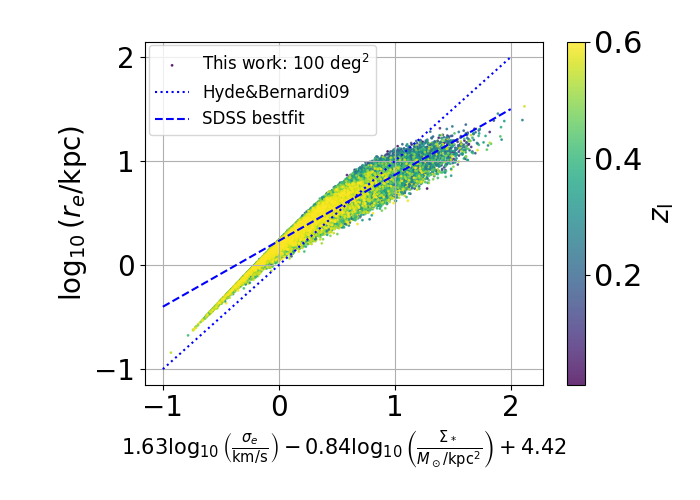}
    \caption{Our mock galaxy catalog is plotted on the MFP. The MFP is described by Eq.~\eqref{eq: mfp_general} with parameters of $(a,b,c,d) = (-0.84,1.63,-1.00,4.42)$, which is obtained for quiescent galaxies in \citet{2009MNRAS.396.1171H}. The blue dashed line displays the best-fit line for the SDSS sample of quiescent galaxies obtained in \cta{2016ApJ...821..101Z}. 
    }
    \label{fig: galaxy_fp}
\end{figure}

Figure~\ref{fig: galaxy_fp} shows our mock galaxy sample with masses larger than $3\times 10^{10}M_\odot$ projected onto the MFP defined by Eq.~\eqref{eq: mfp_general} with the parameters of \cta{2016ApJ...821..101Z}. 
We find that our galaxy population is well distributed on this MFP.
The root-mean-square orthogonal scatter in the MFP for our sample is $0.12~\mathrm{dex}$, almost independently of redshift.
Fig.~7 of \cta{2016ApJ...821..101Z} shows the SDSS galaxy samples at $z<0.1$ and the hCOSMOS samples at $z<0.6$ on the MFP, which are in good agreement with our result in Fig.~\ref{fig: galaxy_fp}. The root-mean-square  orthogonal scatters in the MFP for the SDSS and hCOSMOS sample are $0.17 ~\mathrm{dex}$ and $0.24 ~\mathrm{dex}$, including not only the intrinsic scatter but also measurement errors, respectively. 
It should be noted that our mock galaxy population in Fig.~\ref{fig: galaxy_fp}  is not filtered by any observational selection function, such as the one from which the MFP relation is taken.
Also, the sharp galaxy population with small $r_e$ on the MFP in Fig.~\ref{fig: galaxy_fp} comes from the cutoff of the stellar mass to create this galaxy population.

\section{Create a mock catalog of time-variable strong lenses}\label{sec: lensmock}

In Sec.~\ref{sec: original_pop}, we confirm that our model of halos and galaxies reproduces the observations well.
In this section, we adopt this model to create mock catalogs of strong lenses. 
These catalogs are intended for use in assessing the feasibility of various science projects for a given survey.

\begin{table*}[htbp]
\begin{center}
\caption{Properties of mock catalog in light of the time-domain survey of LSST.}
\begin{tabular}{lccccccccc} \hline
\hline Survey & $\Omega$ & $i_{\text {lim}}$ & $i_{\text {lim}}$ & $i_{\text {lim}}$ & $\theta_{\mathrm{PSF}}$ & cadence & $t_{\text {season}}$ & $t_{\text {survey}}$ & $t_{\text {eff}}$ \\ 
 & $\left(\mathrm{deg}^2\right)$ & (visit) & (yr) & (total) & (arcsec) & (d) & (month) & (yr) & (yr) \\
LSST & 20000 & 23.3 & 24.9 & 26.2 & 0.75 & 5 & 3 & 10 &2.5\\
\hline
\end{tabular}
\end{center}
\end{table*}

As a specific example, this paper presents a mock catalog of lensed QSOs and SNe generated using our model, tailored for the baseline survey planned by LSST. We adopt the same source models of QSOs and SNe as OM10+, and update the deflector model that is detailed in Sec.~\ref{sec: lensmodel}. As in \cta{2010MNRAS.405.2579O}, we consider five types of SNe, Ia, Ib/c, IIP, IIL, and IIn. Therefore, any difference between results in this paper and those in OM10+ should originate from the difference of their lens models rather than the source models. We set criteria specific to LSST, where the survey limiting magnitude in the $i$-band is $m_{\mr{i,lim}}=23.3$, and the flux ratio between the brightest and the second brightest images for double lenses should be $f_{\mr{lim}}<0.1$\footnote{Changing the flux ratio limit from $0.1$ to $0.01$ increases the number of lensed events by approximately $12\%$.}. This is because asymmetric double images with large flux differences can be very challenging to detect due to dynamic range problems produced by the brighter image as well as the obscuration of the fainter images by lensing galaxies. Following \cta{2010MNRAS.405.2579O}, we use the magnification of the fainter image (excluding the central image) for double lens systems and that of the third brightest image for the other multiple image systems for computing the magnification bias.

\cta{2010MNRAS.405.2579O} sets a limit on image separation as $\Delta\theta_{\mr{min}}=(2/3)\theta_{\mr{PSF}}$, where $\theta_{\mr{PSF}}$ is the typical seeing full width at half-maximum (FWHM) for a given survey (e.g., $\theta_{\mr{PSF}}=0''.75$ for LSST).
On the other hand, several approaches to search for small-image separation strong lenses with apparently unresolved multiple images have been proposed and demonstrated \citep{2014Sci...344..396Q,2017Sci...356..291G,2017ApJ...834L...5G,2023NatAs...7.1098G,2021NatAs...5..569S,2024MNRAS.531.3509A,2024arXiv240415389B}. Therefore, in this work, we do not impose such a lower limit on the image separation when creating the mock catalog, as adopted in OM10+.

Since the brightness of SNe changes drastically with time, the definition of the magnitude limit to detect lensed SNe is non-trivial.
We follow \cta{2010MNRAS.405.2579O} and set the condition in which lensed SNe will be detected if their peak magnitude is more than $0.7$ mag brighter than the magnitude limit for each visit.
Additionally, we consider the effective survey duration $t_{\mr{eff}}$ to predict the total number of lensed SNe. In the case of LSST, we assume that the effective survey period is about $3$ months each year out of 10 years of the observation, i.e., $t_{\mr{eff}}=2.5~\mr{yr}$, following \cta{2010MNRAS.405.2579O}. We note that this effective survey period is a conservative estimate. In reality, a longer observation period is planned, which could result in the discovery of more lensed SNe. However, we set this effective survey period for the comparison with the OM10+ results.

We largely follow \cta{2010MNRAS.405.2579O} for the specific procedure to generate mock catalogs, which we described below. For a given survey area and depth, we first generate a list of sources according to the adopted source luminosity function. To take account of the magnification bias, we actually adopt a 3~mag deeper limit than the magnitude limit of the survey for generating the source catalog. Each source in the sample contains information on only the redshift and apparent magnitude without any lensing magnification, and hence we ignore spatial correlations between sources. As a function of the halo mass and redshift, we also pre-compute the size of a square boundary in the source plane such that the magnification at the boundary is $\mu=1.5$ along the major axis direction, assuming an elliptical halo and galaxy with an extreme ellipticity of $0.8$, in order to be conservative. We then generate a list of halos and subhalos according the adopted halo mass function and subhalo mass function (see Sec.~\ref{sec: lensmodel}), again ignoring any spatial correlations between halos as well as between subhalos in a halo. For each halo and subhalo, we assign an ellipticity, position angle, and stellar component, adopting the model detailed in Sec.~\ref{sec: lensmodel}. For each halo and subhalo, we randomly distribute source objects from the pre-computed source catalog within a square region in the source plane whose size is pre-computed as mentioned above, assuming the Poisson distribution. For each realization of sources, we solve the lens equation using \texttt{glafic}, and record only events with multiple images, or with magnifications larger than some threshold even for single-image systems, which are used for the analysis in Sec.~\ref{sec: high_mag}. We also check the detection criteria for the magnitude limit and the flux ratio mentioned above to construct the final sample of the mock catalog. 

The catalog contains all the necessary information for each mock lens, including the properties of lenses and sources, image positions, magnifications, time delays between image pairs.
In practice, the catalog is five times (QSOs) or ten times (SNe) oversampled in order to reduce the shot noise.
We caution that our calculations use the Salpeter IMF~($f_{\mathrm{IMF}} = 1.715$) for the fiducial calculation. 
For comparison, mock catalogs are also produced with the Chabrier IMF~($f_{\mathrm{IMF}} = 1.0$). However, the results in the Chabrier IMF case are only mentioned if the behavior differs significantly from that of the Salpeter IMF case.
Additionally, we note that our mock catalog does not consider lens objects with halo masses below {$10^{10}~\Ms$} (or {$10^9~\Ms$} for subhalos) for a numerical reason. Therefore, it may underestimate the number of lenses with small image separation or short time delay, which will be discussed again later. 

When a subhalo is located close to the center of its host halo in projected space, however, it can be located within the critical curve of the host halo. Strong lensing by such a subhalo near the host halo center should not be counted as a strong lens event produced by an isolated subhalo. We remove such lenses based on the convergence values of the host halo and the central galaxy, as explained below.

We analytically compute the total environmental convergence $\kappa_{\mr{env}}$ as 
\eq{\label{eq: convergence_hh_cen}
\kappa_{\mr{env}}(x_{\mathrm{sh}}, y_{\mathrm{sh}}) = \kappa_{\mr{env, hh}} + \kappa_{\mr{env, cen}},
}
where $\kappa_{\mr{env, hh}}$ and $\kappa_{\mr{env, cen}}$ are the environmental convergence due to the host halo and the central galaxy, respectively.
Since the host halo follows the elliptical NFW profile, $\kappa_{\mr{env, hh}}$ is given by \citep{1996A&A...313..697B}
\eq{
\kappa_{\mr{env, hh}}(u_{\mr{sh, hh}})=\frac{4 \rho_s r_s}{\Sigma_{\text {crit }}}\frac{1}{2\left(u_{\mr{sh, hh}}^2-1\right)}[1-G(u_{\mr{sh, hh}})],
}
where $u_{\mr{sh, hh}}\equiv v(x_{\mathrm{sh}}, y_{\mathrm{sh}}, e_{\mathrm{hh}}, \phi_{\mathrm{hh}})/r_{\mathrm{s, hh}}$ with the new coordinate $v$ in Eq.~\eqref{eq: r_to_v} that includes the ellipticity and the position angle for the host halo, $\Sigma_{\mathrm{crit }}$ is the critical surface density, and 
\eq{
G(u)= \begin{cases}\frac{1}{\sqrt{1-u^2}} \operatorname{arctanh} \sqrt{1-u^2} & (u<1) \\ \frac{1}{\sqrt{u^2-1}} \arctan \sqrt{u^2-1} & (u>1)\end{cases}.
}
On the other hand, since the central galaxy follows the elliptical Hernquist profile, $\kappa_{\mr{env, cen}}$ can be written by \citep{2001astro.ph..2341K}
\eq{
\kappa_{\mr{env, cen}}&(u_{\mathrm{Hern}})\\
&\ =\frac{M_*}{2 \pi r_b^2 \Sigma_{\text {crit }}}\frac{1}{\left(u_{\mathrm{Hern}}^2-1\right)^2}\left[-3+\left(2+u_{\mathrm{Hern}}^2\right) F(u_{\mathrm{Hern}})\right],
}
with $u_{\mathrm{Hern}}\equiv v(x_{\mathrm{sh}}, y_{\mathrm{sh}}, e_{\mathrm{c*}}, \phi_{\mathrm{c*}})/r_{\mathrm{b, cen}}$.

For each subhalo, we calculate the total external convergence and retain only subhalos with $\kappa_{\mr{env}}<0.4$ in our calculation. We emphasize that this analytically derived total external convergence is used only for removing subhalos near the host halo center, and actual calculations for lenses produced by subhalos are numerically conducted by including their host halo and central galaxies. 

In addition, our numerical calculation setup may result in counting multiple images produced by the host halo in the lensing calculation for the subhalos.
In order to avoid counting these host halo images as duplicates, we set another condition for lensing calculations concerning the subhalos that images that are closer to the host halo than the subhalo are to be removed.
In what follows, we discuss the properties of our mock lens in comparison with the results of OM10+.

\subsection{Mock catalog of lensed QSOs}\label{sec: mock_multi_qso}
First, we examine properties of mock catalogs for lensed QSOs with multiple images.
The total number in the fiducial mock catalog is {$4907$} for lensed QSOs, which is similar to but is somewhat {larger} than the result of $4286$ from OM10+. 
Of these, {$1297$} lenses, or approximately {$25\%$} of all lenses, are primarily produced by subhalos. The fraction of lensing by subhalos in our mock catalog is broadly consistent with a previous analytic estimate \citep{2006MNRAS.367.1241O}.
We find that the mock catalog with the Chabrier IMF instead of the Salpeter IMF includes {$2732$} lensed QSOs, which is significantly smaller compared with the case assuming the fiducial Salpeter IMF. Of these, {$800$} lenses are primarily produced by subhalos, indicating that the ratio is comparable to that in the fiducial mock catalog.
When we add a criterion that the maximum image separation, which is defined by the maximum angular separation between any pair of multiple images, should be larger than $\Delta\theta_{\mr{min}}=(2/3)\theta_{\mr{PSF}}$, the total numbers of lensed QSOs reduces to {$4223$} in the fiducial case, {$2179$} in the Chabrier IMF case, and $3013$ in the case of OM10+. 
Since the number of lenses predicted by OM10+ has been shown to be approximately consistent with observations \citep[see e.g.,][]{2023MNRAS.520.3305L}, our result
should also be consistent with observations.
We explicitly check the consistency with observations in Sec.~\ref{sec: mock_sdss}.
The above results regarding the total number of lenses are summarized in Table~\ref{tab: lensed_mocks}.

The large difference of predicted numbers of strongly lensed QSOs between Salpeter and Chabrier IMFs is explained as follows. First, the stellar mass is not a direct observable, but rather we infer it from the observed luminosity and color of a galaxy assuming the stellar IMF. Therefore, the stellar mass-halo mass relation that is constrained from the abundance and clustering of galaxies depends on the choice of the IMF, as discussed in Sec.~\ref{subsec: central_gal}. This indicates that a typical luminosity of a central galaxy for a given halo is well constrained from the observation, while a typical stellar mass of the central galaxy is subject to the uncertainty of the choice of the stellar IMF. Since the bottom-heavy Salpeter IMF predicts the higher mass-to-light ratio than the Chabrier IMF, the typical stellar mass of the central galaxy for a given halo is higher for the Salpeter IMF than for the Chabrier IMF. The higher stellar mass naturally results in the higher surface mass density at the center of the halo, and hence the larger lensing cross section. This well explains the reason why our fiducial model with the Salpeter IMF predicts roughly a factor of two larger number of lensed QSOs than the Chabrier IMF case. 

Basically, our catalogs consist of three- or five-image lenses in accordance with the odd number theorem~\citep[see, e.g.,][]{1981ApJ...244L...1B}. However, in most cases, the images near the centers of lensing galaxies are strongly demagnified and unobservable. We refer to three- and five-image lenses as double (two-image) and quadruple (quad; four-image) lenses to keep the notation consistent with the one in OM10+, which basically predicts double or quad lenses from the SIE plus external shear model.
We find that the fraction of quad lenses in the fiducial~(Chabrier) model is approximately $10\%$~({$13\%$}), while the one in OM10+ is about $15\%$, which is slightly higher than our results.
Note that these quad fractions are lower than the current observed values, about $14\%$~\citep[see e.g.,][]{2023MNRAS.520.3305L}, which is understood from the viewpoint that the current sample of known lenses is more strongly affected by magnification bias, and hence has higher quad fractions, compared to the lenses expected from LSST. We will discuss the detail in Sec.~\ref{sec: mock_sdss}.

\begin{table*}[htbp]
\begin{center}
\caption{The expected numbers of lensed QSOs and SNe with a view to LSST survey. We calculate the expected number with and without a criterion of the minimum image separation $\Delta\theta>(2/3)\theta_{\mr{PSF}}$. The uncertainty of the expected number represents the standard deviation of the lens counts across different realizations of mock catalogs, corresponding to a $68\%$ confidence interval.}\label{tab: lensed_mocks}
\begin{tabular}{llccccc}
\hline \hline &  & \multicolumn{3}{c}{ Multiple images } & \multicolumn{2}{c}{ Highly magnified systems} \\
Source & IMF/Model &$N_{\text {multi}}$ ($\Delta\theta>(2/3)\theta_{\mr{PSF}}$) & $N_{\text {quad.}}$ & $N_{\text {sub}}$ & $N_{\text {high}}|_{\mu>3}$ & $N_{\text {high}}|_{\mu>10}$ \\
QSOs & OM10+ & $4286\pm 28$ ($3013\pm 35$) & $625\pm 15$ & --- & --- &  --- \\
QSOs & Salpeter & {$4907\pm 53$} ({$4223\pm 62$}) & {$514\pm 25$} & {$1297\pm 29$} & {$7459\pm 96$} &  {$764\pm 19$} \\
QSOs & Chabrier & {$2732\pm 43$} ({$2179\pm 34$}) & {$362\pm 19$} & {$800\pm 25$} & {$6422\pm 67$} &  {$629\pm 19$} \\
SNe & OM10+ & $248\pm 18$ ($178\pm 13$) & $69\pm 10$ & --- & --- &  --- \\
SNe & Salpeter & {$288\pm 17$} ({$234\pm 12$}) & {$69\pm 7$} & {$85\pm 10$} & {$1098\pm 19$} &  {$230\pm 8$} \\
SNe & Chabrier & {$169\pm 18$} ({$127\pm 15$}) & {$39\pm 6$} & {$54\pm 9$} & {$822\pm 30$} & {$170\pm 11$} \\
\hline
\end{tabular}
\end{center}
\end{table*}

\begin{figure*}[htbp]
    \begin{tabular}{cc}
       \begin{minipage}[t]{.48\linewidth}
        \centering
        \includegraphics[width=\hsize]{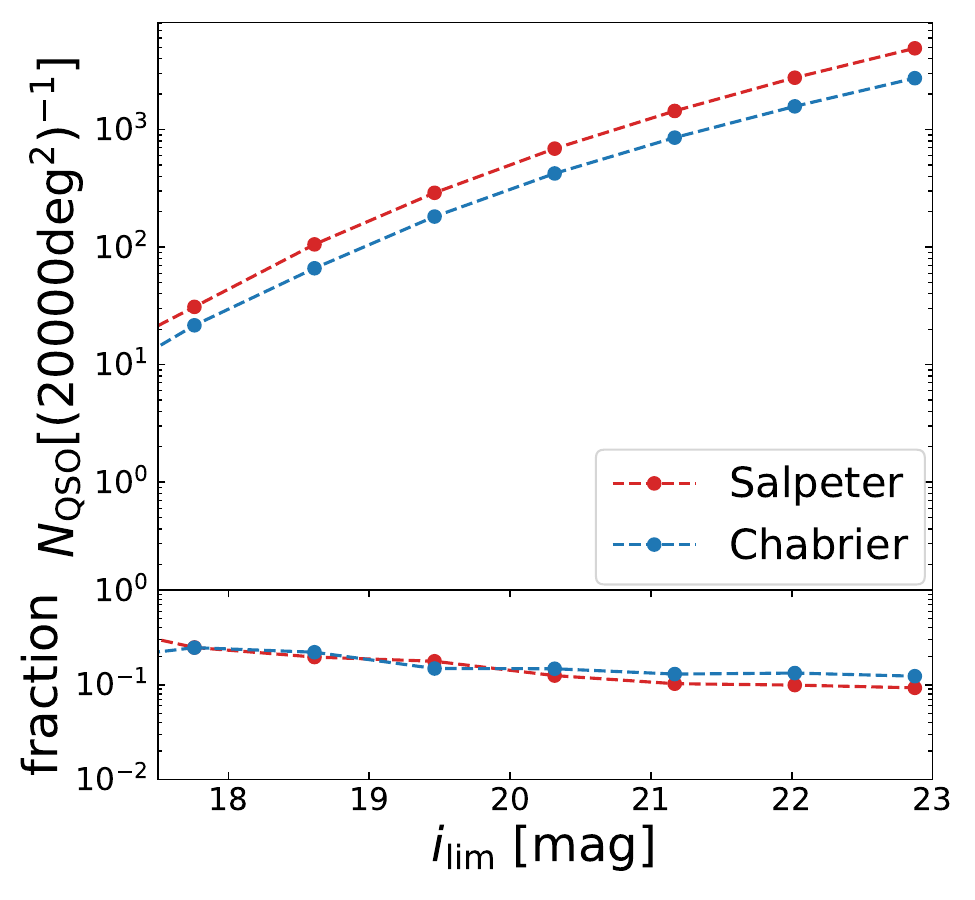}
      \end{minipage} &
      \begin{minipage}[t]{.48\linewidth}
        \centering
        \includegraphics[width=\hsize]{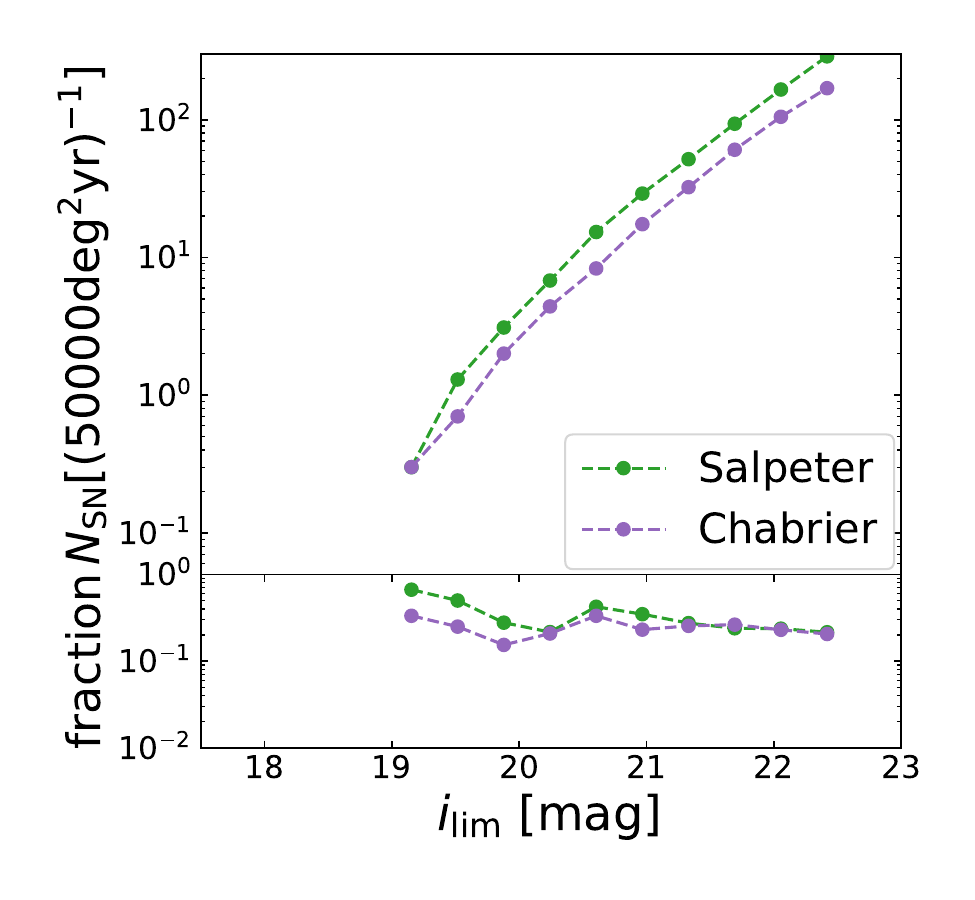}
      \end{minipage}
  \end{tabular}
  \caption{The expected number of lensed QSOs (left panel) and lensed SNe (right panel) as a function of the $i$-band limiting magnitude, $i_\mathrm{lim}$. Here, we assume the survey area of $20,000~\mathrm{deg^2}$, and the effective survey period of $2.5$-yr for lensed SNe. The lower panels show the ratio of quad (four-image) lenses to the total number of lenses as a function of $i_\mathrm{lim}$. The red and green lines show the results from our fiducial Salpeter IMF mock catalogs for lensed QSOs and SNe, respectively, while the blue and purple lines represent those from the Chabrier IMF mock catalogs. Quad-lensed SNe in the Chabrier IMF mock catalogs is very rare at $i_\mathrm{lim}<20$, and as a result the fraction becomes $0$ or $1$ in our calculation.
  }
  \label{fig: dN_mobs_cum}
\end{figure*}

Here, although the discussion in this paper is primarily tailored for future LSST observations, we explore how the number of observable strong lenses increases with the survey depth. In the left panel of Fig.~\ref{fig: dN_mobs_cum}, we plot the number of lensed QSOs over an area of $20,000~\mathrm{deg}^2$ for both our fiducial Salpeter IMF and Chabrier IMF mock catalogs as a function of the $i$-band limiting magnitude, $i_\mathrm{lim}$. The result for the Salpeter IMF is consistent with that of \cta{2010MNRAS.405.2579O} (Fig.~3), showing a shallow slope and indicating that survey area is more crucial than survey depth for discovering many lensed QSOs. The result for the Chabrier IMF is similar to that of the Salpeter IMF, but showing the smaller number of lenses. The double- and quad-fractions are also in agreement with the results of \cta{2010MNRAS.405.2579O}, showing a gradual decrease with increasing $i_\mathrm{lim}$
The right panel depicts the results of lensed SNe for both our fiducial Salpeter IMF and Chabrier IMF mock catalogs over an area of $20,000~\mathrm{deg}^2$ with the effective survey period of $2.5$-yr. The slope of the number counts is steeper than the one for the lensed QSOs, suggesting that the survey depth is relatively more important for lensed SNe. The steeper dependence of the number of strong lenses on the limiting magnitudes for lensed SNe originates from lower source redshifts ($z \lesssim 1$) of lensed SNe where the gravitational lensing probability is more sensitive to the source redshift. As for \cta{2010MNRAS.405.2579O}, we find that the fraction of quad lens systems tends to be higher than the one for the lensed QSOs, due to the very large magnification bias associated with the bright lenses, and it changes rapidly as a function of the limiting magnitude.

\begin{figure*}[htbp]
    \begin{tabular}{cc}
       \begin{minipage}[t]{.48\linewidth}
        \centering
        \includegraphics[width=\hsize]{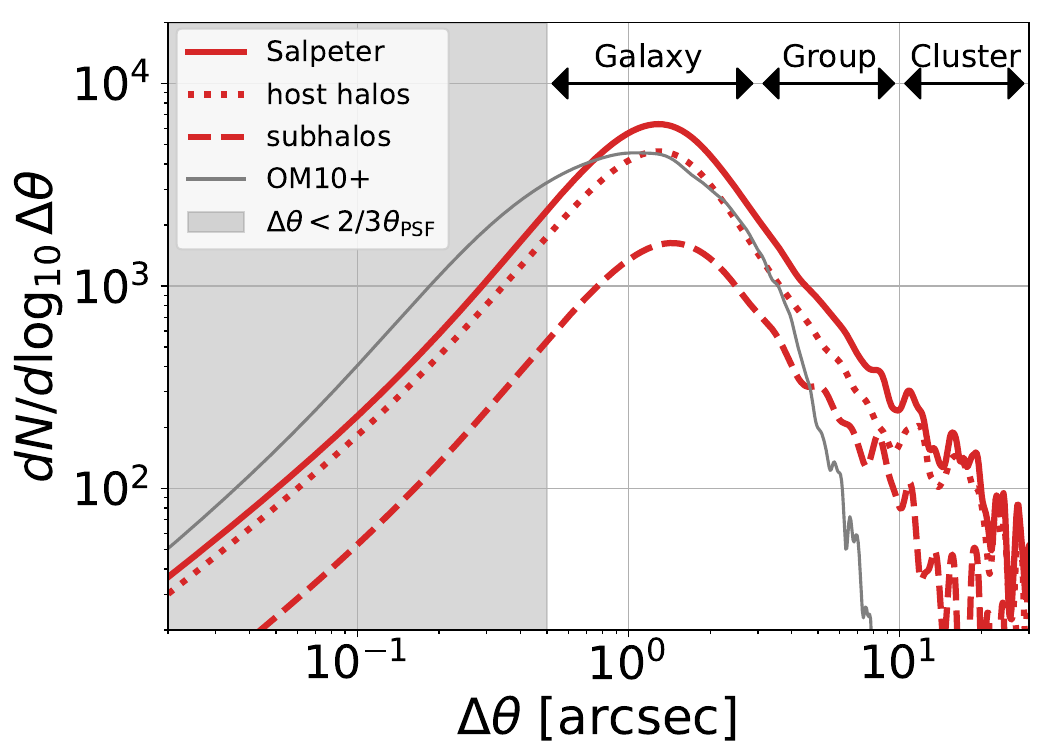}
      \end{minipage} &
      \begin{minipage}[t]{.48\linewidth}
        \centering
        \includegraphics[width=\hsize]{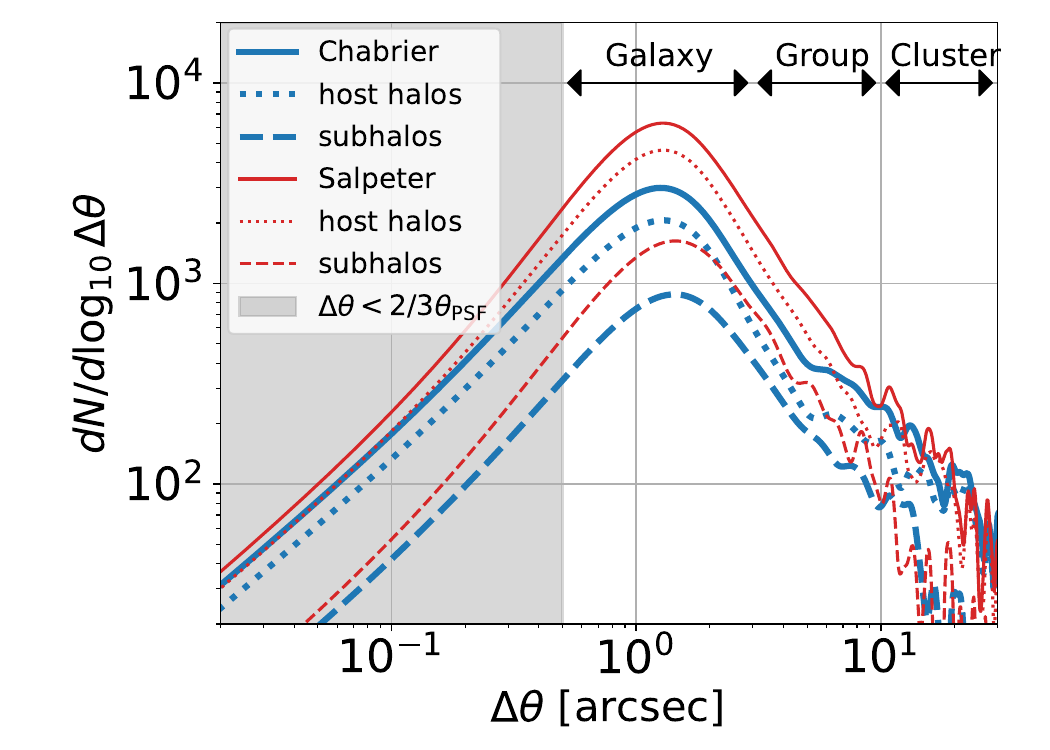}
      \end{minipage}
  \end{tabular}
  \caption{The number distribution of lensed QSOs as a function of maximum image separations $\Delta\theta$ between multiple images. In the left panel, we plot the results from our fiducial Salpeter IMF mock catalogs with red lines, comparing them to the OM10+ results in grey. The solid red line represents all lenses in the mock catalogs, while the results for lenses caused by host halos (subhalos) are shown with dotted (dashed) lines. In the right panel, we show the results from the Chabrier IMF mock catalogs with blue lines, comparing them to our fiducial Salpeter IMF mock catalogs. The grey-shaded region indicates the limitations of LSST in resolving multiple images. At the top of each panel, we also include approximate lens-object scales corresponding to the image separation scales for reference.
  }
  \label{fig: dtheta_dist_qso}
\end{figure*}

To check the observability of multiple images, it is essential to investigate the distribution of image separations between multiple images. 
Fig.~\ref{fig: dtheta_dist_qso} shows the distribution functions of maximum image separation between multiple images in our mock lenses.
We find that the distribution function from our mock catalogs has a different peak from OM10+, and several mock lenses display a large image separation, $\Delta\theta>10~\mathrm{arcsec}$, mainly produced by galaxy groups or clusters, unlike the result of OM10+.
This is one of the major results obtained by adopting the compound lens model described in Sec.~\ref{sec: lensmodel}.
Note that OM10+ adopts the SIE lens model, focuses on massive galaxies as lens objects, and does not predict the lenses produced by more massive objects such as galaxy groups and clusters.
For example, our fiducial model predicts that LSST 10-year observations will find about 80 lensed QSOs with the maximum image separations larger than $10~\mathrm{arcsec}$. We also find that the overall number of lenses in the Chabrier IMF case is lower than in the Salpeter IMF case, except for cluster-scale lenses with maximum image separations larger than 10 arcsec for which dark matter rather than the stellar mass dominates the lens potential.
Furthermore, there are notable differences compared with OM10+, even on the small image separation side.
A caveat is that since we set a lower limit on the mass of halos to be calculated as {$M_{\mathrm{hh}}>10^{10}~\Ms$} for a numerical reason, we may underestimate lenses with small image separations. 
We confirm that the number of lenses with image separations of $\Delta \theta\geq (2/3)\theta_{\mr{PSF}}$ is not affected by this lower limit on the halo mass.

\begin{figure}
    \centering
    \includegraphics[width=\linewidth]{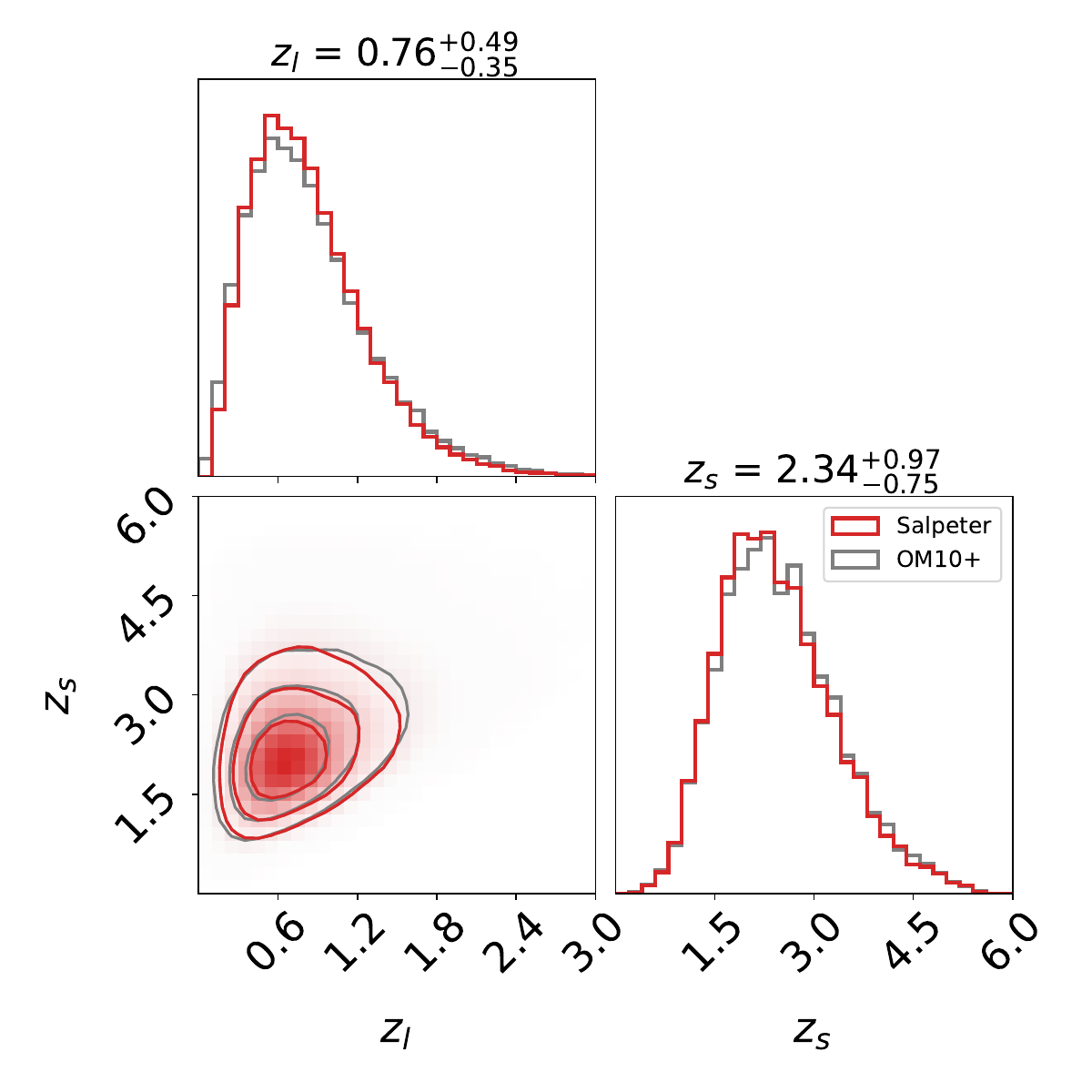}
  \caption{Predicted joint distribution of lens and source redshifts in our fiducial lensed QSO mock catalogs. The top panel shows the distributions for the lens redshift, while the right panel displays those for the source redshift. The central panel shows contour plots at $25\%$, $50\%$, and $75\%$. Results for our fiducial lensed QSOs are plotted with red lines, while grey lines represent OM10+ results for comparison. The titles of the top and right panels display the median ($50\%$ quantile) with upper and lower errors corresponding to the $16\%$ and $84\%$ quantiles, respectively, for our results.
  }
  \label{fig: zl_zs_dist_qso}
\end{figure}

Next we discuss the statistical properties of our mock catalogs for lensed QSOs in more detail, in accordance with the results in Table~\ref{tab: lensed_mocks} and Fig.~\ref{fig: dtheta_dist_qso}.
We plot the redshift distributions of mock lenses in Fig.~\ref{fig: zl_zs_dist_qso}.
The left panels in Fig.~\ref{fig: zl_zs_dist_qso} show the source redshift distributions of the mock lenses, while the right panels display the lens redshift distributions.
The upper panels represent the redshift distributions for the lensed QSOs of our fiducial mock catalogs.
Since the same distribution of sources is adopted as in OM10+, and the redshift distribution of the lens objects is not significantly changed, our mock catalogs show similar distributions for both the source and lens redshifts.
There also appears to be no difference in distribution between lenses produced by subhalos and those by host halos.

\begin{figure*}[ht]
    \begin{tabular}{ccc}
       \begin{minipage}[t]{.315\linewidth}
        \centering
        \includegraphics[width=\hsize]{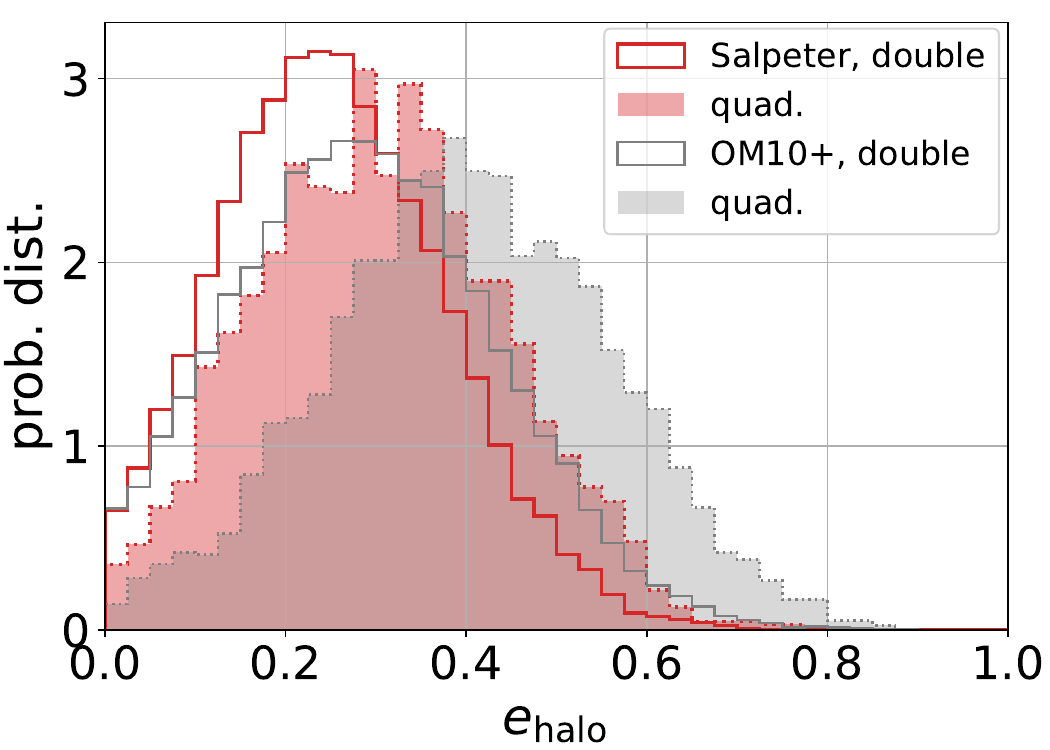}
      \end{minipage} &
      \begin{minipage}[t]{.315\linewidth}
        \centering
        \includegraphics[width=\hsize]{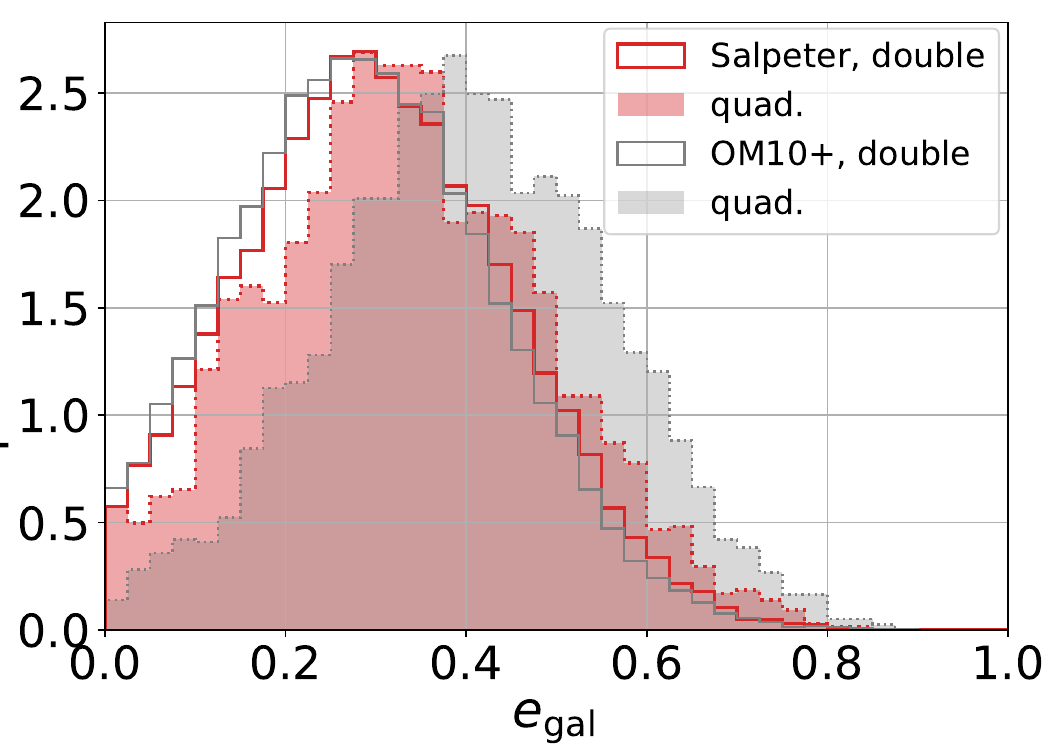}
      \end{minipage} &
      \begin{minipage}[t]{.315\linewidth}
        \centering
        \includegraphics[width=\hsize]{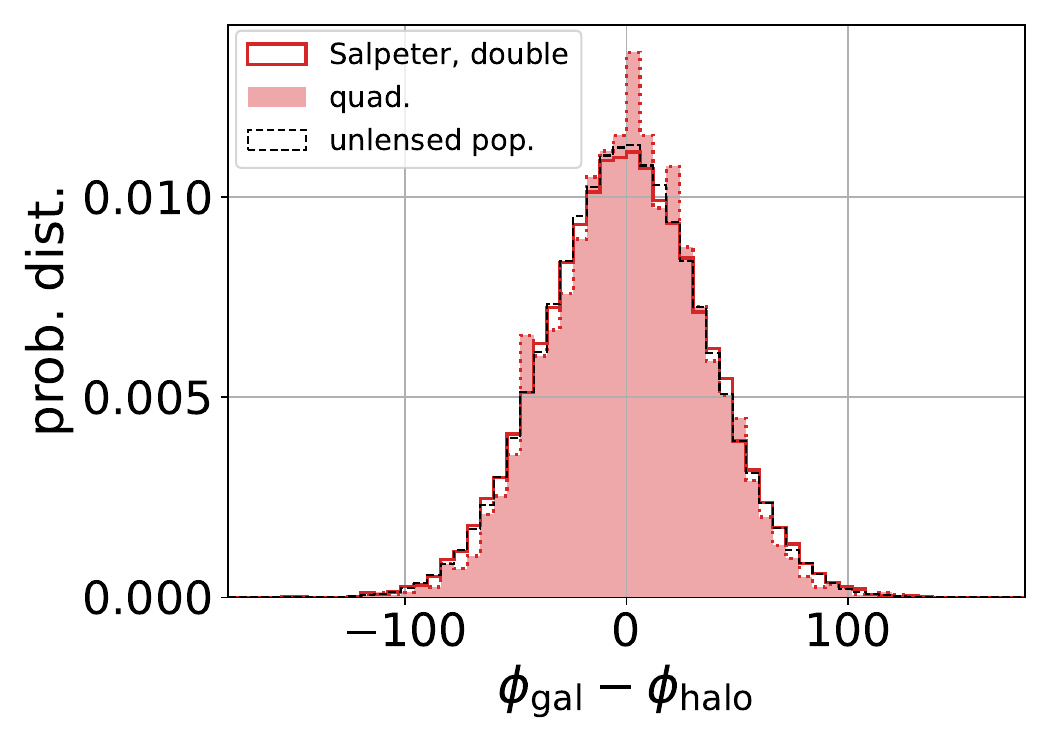}
      \end{minipage}
  \end{tabular}
  \caption{The probability distributions of the lens halo~(left panel) and galaxy ellipticity (middle panel) for double (solid) and quad (dotted and shaded) lenses obtained from our fiducial mock lens catalog. For the left and middle panels, we also plot the results of OM10+ in grey and the results from the unlensed original population as mentioned in Sec.~\ref{sec: original_pop} in black for comparison. The right panel shows the probability distribution of $\phi_{\mr{gal}}-\phi_{\mr{halo}}$, where $\phi_{\mr{gal}}$ and $\phi_{\mr{halo}}$ are position angles of the lens galaxy and halo. We also plot the results of $\phi_{\mr{gal}}-\phi_{\mr{halo}}$ from the unlensed original population in black.
  }
  \label{fig: ehalo_egal_dist_qso}
\end{figure*}

An advantage of mock catalogs is that we can easily check how distributions of lens and source parameters differ from those of unlensed populations. As specific examples, Fig.~\ref{fig: ehalo_egal_dist_qso} shows probability distributions of the lens ellipticity and the alignment between the lens galaxy and lens halo separately for double and quad lenses. 
We find that the lensing halos and galaxies of quad lens systems typically have higher ellipticity than those of double lenses, although not as pronounced as in quad lens systems in OM10+.
Our result is broadly consistent with previous studies~\citep[see, e.g.,][]{Rozo:2007bj}.

\begin{figure*}[htbp]
    \begin{tabular}{ccc}
       \begin{minipage}[t]{.32\linewidth}
        \centering
        \includegraphics[width=\hsize]{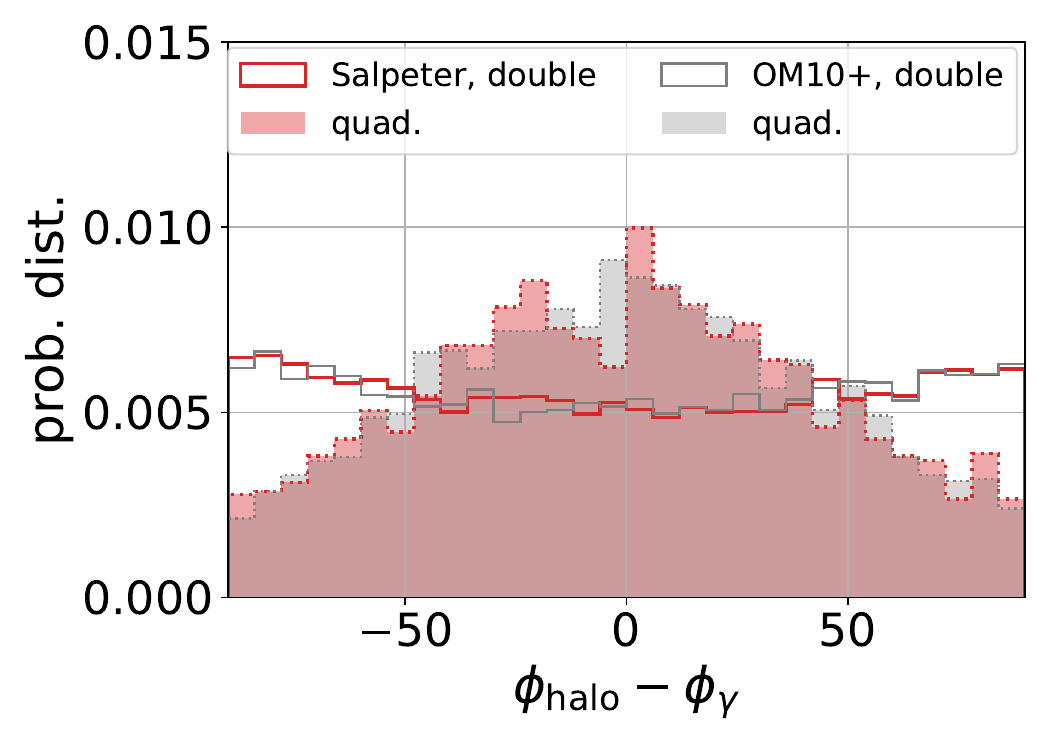}
      \end{minipage} &
      \begin{minipage}[t]{.32\linewidth}
        \centering
        \includegraphics[width=\hsize]{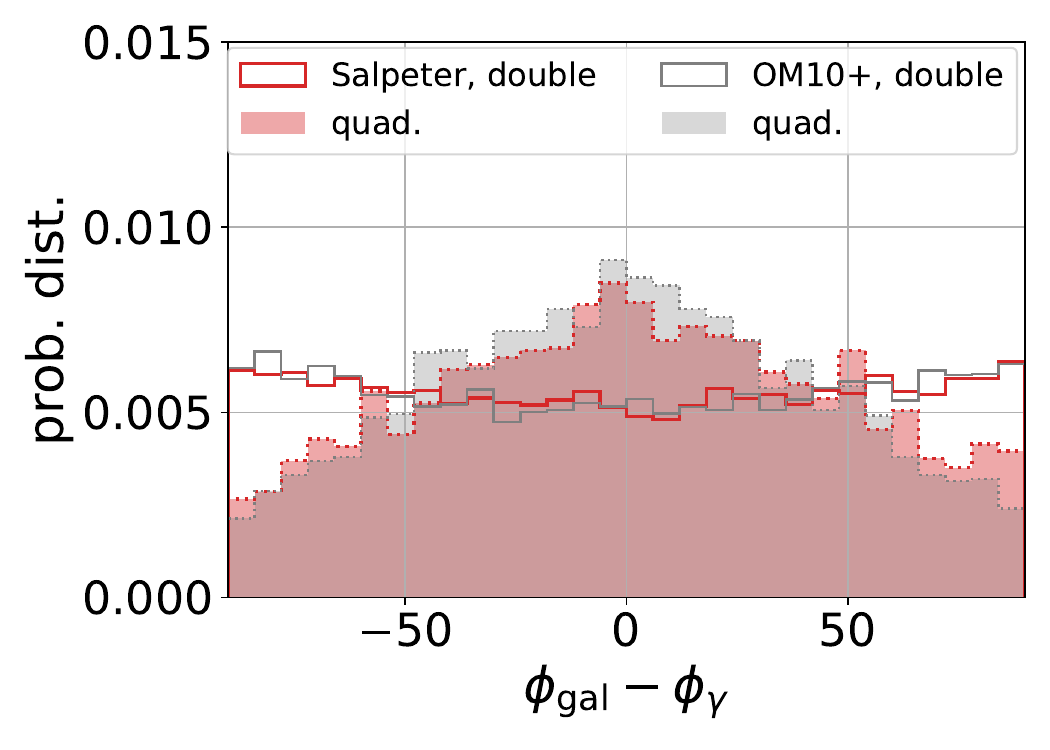}
      \end{minipage} &
      \begin{minipage}[t]{.32\linewidth}
        \centering
        \includegraphics[width=\hsize]{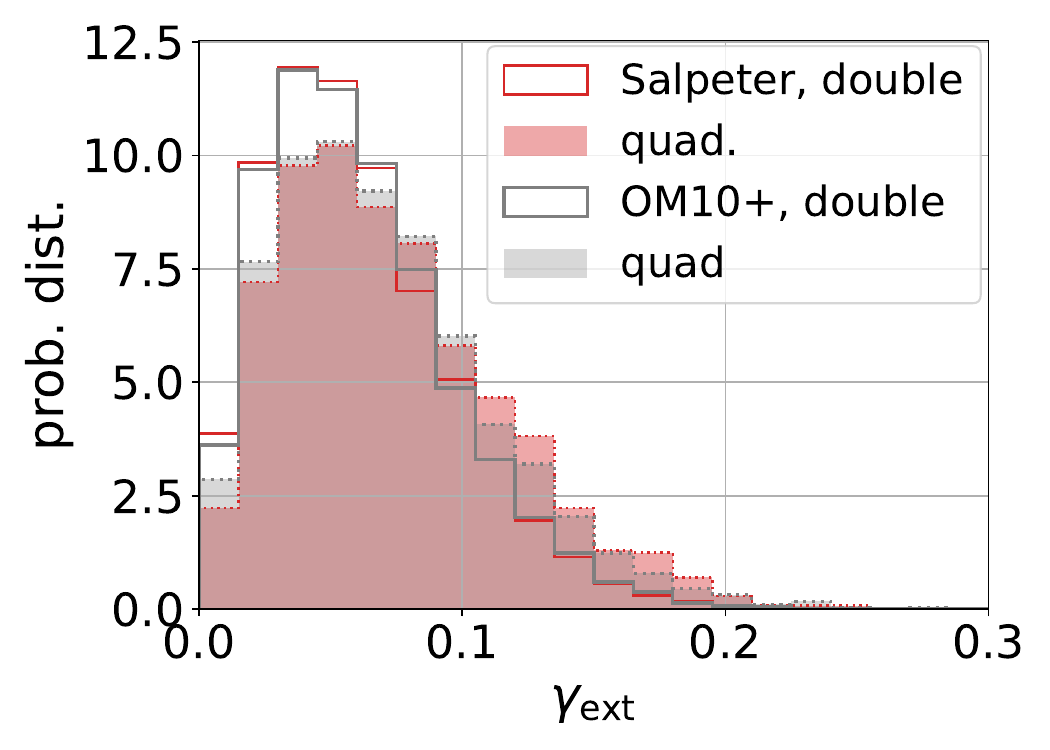}
      \end{minipage}
  \end{tabular}
  \caption{The probability distribution of $\phi_{\mr{halo}}-\phi_{\gamma}$~(left panel), $\phi_{\mr{gal}}-\phi_{\gamma}$~(middle panel), and external shear $\gamma_{\mr{ext}}$ (right panel) for double (solid) and quad (dotted and shaded) lenses obtained from our fiducial mock lens catalog. Here $\phi_{\mr{halo}}$, $\phi_{\mr{gal}}$, and $\phi_{\gamma}$ are position angles of halo, galaxy, and external shear in units of degree, respectively. We also plot the results of OM10+ for comparison. In the left and middle panels, the distribution for the unlensed population is a horizontal line, because we assume no correlation between position angles of ellipticities and external shear.}
  \label{fig: phalo_pgal_ppert_dist_qso}
\end{figure*}

The probability distributions of the position angle of lens halos and galaxies against the position angle of external shear are shown in the left and middle panels in Fig.~\ref{fig: phalo_pgal_ppert_dist_qso}. We also plot the distribution of external shear itself in the right panel of Fig.~\ref{fig: phalo_pgal_ppert_dist_qso}.
It appears that the probability distribution in quad lenses shows slight tails toward the higher values compared to the one in double lenses. We conduct the Kolmogorov-Smirnov test at $\gamma_{\mathrm{ext}}$ to confirm that the difference of the tails is significant with the p-value of {$3\times 10^{-10}$}.
We also find that the relative orientation of lens halos and galaxies against external shear is biased for both cases of double and quad lenses, albeit weakly, compared to OM10+.
That is, for double lenses, the direction of external shear is more likely to be aligned with the minor axis of the lens halo and lens galaxy, whereas, for quad lenses, the direction of external shear is more likely to be aligned with the major axis.
This bias can be understood through the degeneracy between the ellipticity of the lens object and the external shear~\citep[see, e.g.,][]{1997ApJ...482..604K}.
However, our mock catalogs exhibit a more complex bias than OM10+ due to the compound lens model as described in Sec.~\ref{sec: lensmodel}.
It may be related with their differences in the distributions of lens ellipticities discussed in Fig.~\ref{fig: ehalo_egal_dist_qso}.

\begin{figure}[htbp]
    \centering
    \includegraphics[width=\hsize]{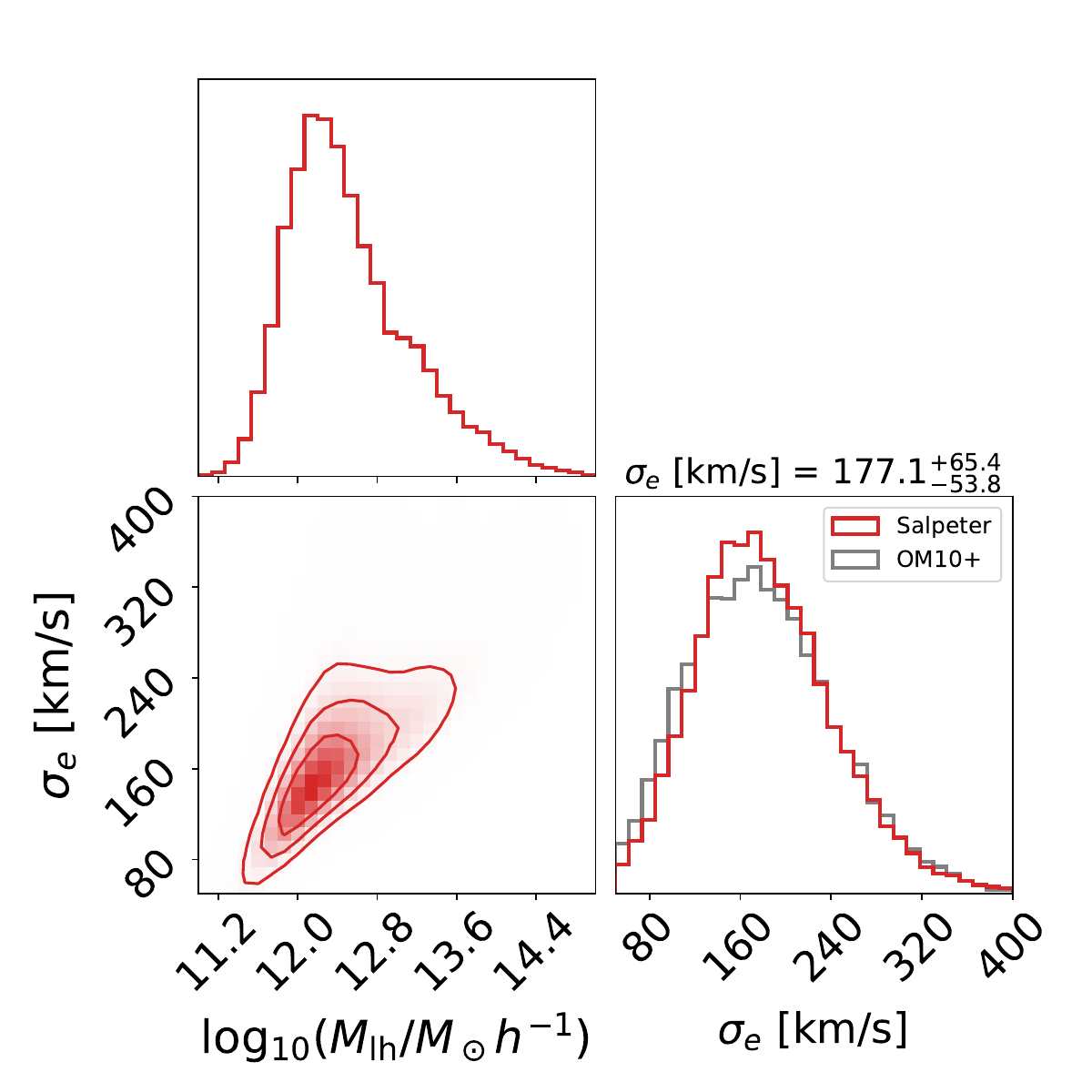}
  \caption{The joint distribution of the velocity dispersion and lens halo mass, as defined in Eq.~\eqref{eq: mlens_halo}, in our fiducial lensed QSO mock catalogs. The top panel shows the distribution of the lens halo mass, while the right panel displays that of the velocity dispersion. The central panel shows contour plots at $25\%$, $50\%$, and $75\%$. The titles of the top and right panels display the median ($50\%$ quantile) with upper and lower errors corresponding to the $16\%$ and $84\%$ quantiles, respectively. For the histogram of the velocity dispersion, the result for OM10+ is overplotted by a grey line for comparison.
  }
  \label{fig: velp_dist_qso}
\end{figure}

One of the quantities that are related to the maximum image separation is the velocity dispersion of galaxies, which is displayed in Fig.~\ref{fig: velp_dist_qso} in the same manner as in Fig.~\ref{fig: zl_zs_dist_qso}.
We remind the reader that OM10+ adopts the velocity dispersion function for $\sigma_{e/8}$ based on \cite{2018MNRAS.480.3842O}, which is shown in Eq.~\eqref{eq: vdf_sigma_e8} with the redshift evolution in Eq.~\eqref{eq: vdf_sigma_e8_z}. 
We convert $\sigma_{e/8}$ to $\sigma_{e}$ using Eq.~\eqref{eq: convert_sig_e8_to_sig_e} to fairly compare both results.
The distributions of maximum image separation and velocity dispersion show similar shapes because of their tight relation, e.g., $\Delta \theta\propto \sigma_e^2$ in the SIS case.
As noted above, the reason why lenses with larger $\sigma_e$ are expected compared to OM10+ is that our lens model is able to represent galaxy groups and galaxy clusters.
In addition, we may underestimate the number of lenses at small velocity dispersions due to the cutoff of small-mass halos.

\begin{figure}
    \centering
    \includegraphics[width=\linewidth]{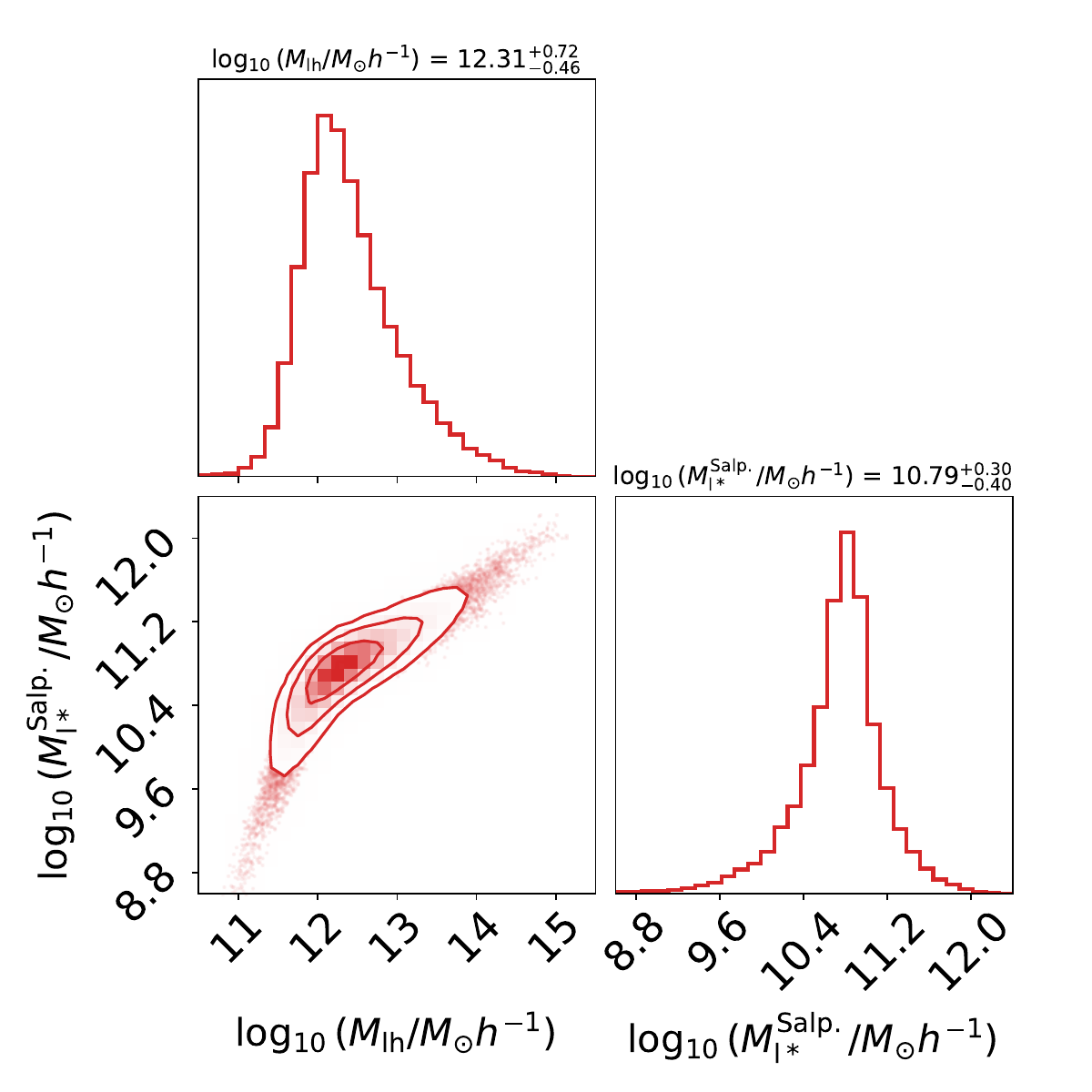}
  \caption{The joint distribution of the lens halo mass and the lens galaxy mass, as defined in Eqs.~\eqref{eq: mlens_halo} and \eqref{eq: mlens_gal}, respectively, in our fiducial lensed QSO mock catalogs. The superscript ``Salp.'' indicates that the lens galaxy mass is given in terms of the Salpeter IMF. The top panel shows the distribution for the lens halo mass, while the right panel displays that for the lens galaxy mass. The central panel shows contour plots at $25\%$, $50\%$, and $75\%$. The titles of the top and right panels display the median ($50\%$ quantile) with upper and lower errors corresponding to the $16\%$ and $84\%$ quantiles, respectively.
  }
  \label{fig: m_lens_halo_gal}
\end{figure}

The masses of lens halos and galaxies are also strongly correlated with maximum image separations, which are displayed in Fig.~\ref{fig: m_lens_halo_gal}.
Here, we define the lens halo mass as
\eq{\label{eq: mlens_halo}
M_{\text {lh}}=\left\{\begin{array}{lr}
M_{\mathrm{f}} & \text { (lenses by subhalos) } \\
M_{\mathrm{hh}} & \text { (otherwise) },
\end{array}\right.
}
and the lens galaxy mass as
\eq{\label{eq: mlens_gal}
M_{\text {l*}}=\left\{\begin{array}{lr}
M_{\mathrm{s*}} & \text { (lenses by subhalos) } \\
M_{\mathrm{c*}} & \text { (otherwise) },
\end{array}\right.
}
Although the result in the Chabrier IMF case is not shown, the difference between the Salpeter and Chabrier IMF cases is only the shift of the lens galaxy mass by $f_{\mathrm{IMF}}=1.715$.
This indicates that the difference in the stellar mass–halo mass relation due to the choice of IMF changes the total number of lenses but not the properties of the lens mass distribution.

\subsection{Mock catalog of lensed SNe}\label{sec: mock_multi_sne}
Next, we discuss the features of mock catalogs for
lensed SNe with multiple images. As mentioned before, we set the effective survey period of LSST to be $t_{\mr{eff}}=2.5~\mr{yr}$ to generate the mock catalogs.
The total number in the fiducial mock catalog is {$288$} for lensed SNe, which is somewhat {larger} than the result of $248$ from OM10+. Of these, {$85$} lenses, or approximately {$30\%$} of all lenses, are primarily produced by subhalos. On the other hand, the mock catalog with the Chabrier IMF includes {$169$} lenses, and {$32\%$} of these lenses are produced by subhalos.
When we add a criterion of maximum image separation as $\Delta\theta\leq (2/3)\theta_{\mr{PSF}}$, the total number of observed lenses reduces to {$234$} in the fiducial case, {$127$} in the Chabrier IMF case, which should be compared with $178$ in the case of OM10+. 
Note that approximately one-third of the lensed SNe is of Type Ia, which is consistent with OM10+.

As in the case of lensed QSOs, our catalogs of lensed SNe basically consist of double~(two images and one faint central image) and quad~(four images and one faint central image) lenses. The fraction of quad lenses in the fiducial~(Chabrier) model is approximately {$24\%$}~($23\%$), while the one in OM10+ is about $28\%$.

It should be noted that the expected number of lensed SNe (especially Type Ia) in future surveys have been investigated in previous studies other than OM10+ \citep[e.g.,][]{2019MNRAS.487.3342W,2019ApJS..243....6G, 2021ApJ...908..190P,2023MNRAS.526.4296S,2024arXiv240710470D}. Basically, our results are consistent with these previous studies, but with some differences due to different selection criteria of lenses. For example, in \citet{2023MNRAS.526.4296S}, the peak brightness of multiply lensed Type Ia SNe are counted as detectable lenses, while we adopt a selection criterion of 0.7 magnitude brighter than the limiting magnitude of the survey for the peak brightness of either the faintest image in a double lens system or the third brightest image in a quadruple lens system, as described in Sec.~\ref{sec: lensmock}. \citet{2019ApJS..243....6G} performed pixel-level simulations of lensed SNe considering the LSST observing strategy and following selection criteria based on photometric classification by spectral template fitting.

\begin{figure*}[htbp]
    \begin{tabular}{cc}
      \begin{minipage}[t]{.48\linewidth}
        \centering
        \includegraphics[width=\hsize]{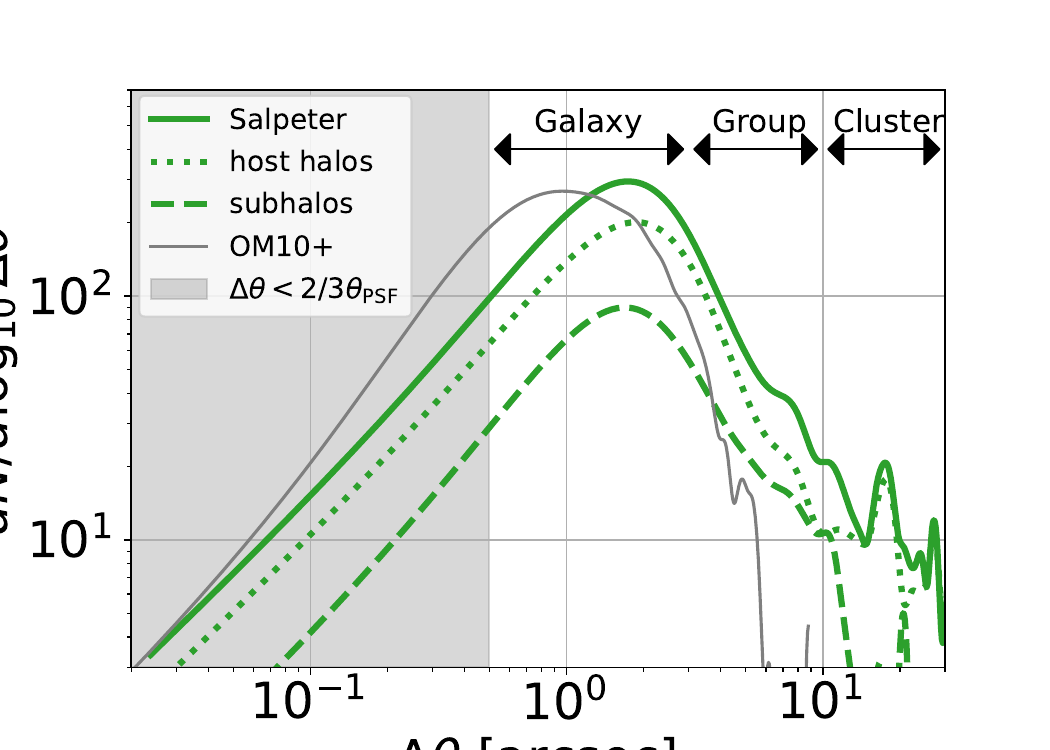}
      \end{minipage} &
      \begin{minipage}[t]{.48\linewidth}
        \centering
        \includegraphics[width=\hsize]{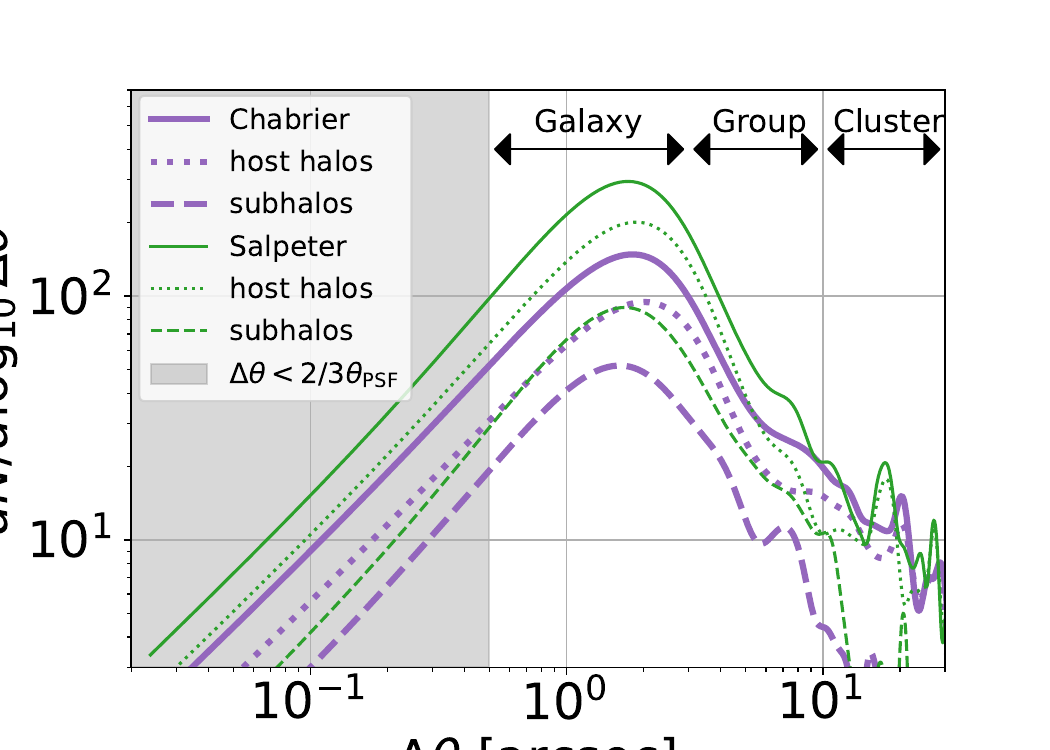}
      \end{minipage}
  \end{tabular}
  \caption{Same as Fig.~\ref{fig: dtheta_dist_qso}, but the distribution of the maximum image separation of our mock catalogs for lensed SNe are shown. The green lines show the results in our fiducial Salpeter IMF mock catalogs, while the purple lines represent the ones in the Chabrier IMF mock catalogs.}
  \label{fig: dtheta_dist_sn}
\end{figure*}

The distribution of the maximum image separation between multiple images is plotted in Fig.~\ref{fig: dtheta_dist_sn} in the same manner as in Fig.~\ref{fig: dtheta_dist_qso}.
As with the lensed QSOs, lenses with large image separations are expected from galaxy groups or clusters. For example, LSST 10-yr observations will find about seven lensed SNe with maximum image separations larger than $10~\mathrm{arcsec}$.
We also find the overall shift of the distribution toward the lower maximum image separation in the case of the Chabrier IMF compared to the fiducial case as well as the underestimation of lenses with small image separations as compared with OM10+, as found in Fig.~\ref{fig: dtheta_dist_qso}.

\begin{figure}[htbp]
    \centering
    \includegraphics[width=\linewidth]{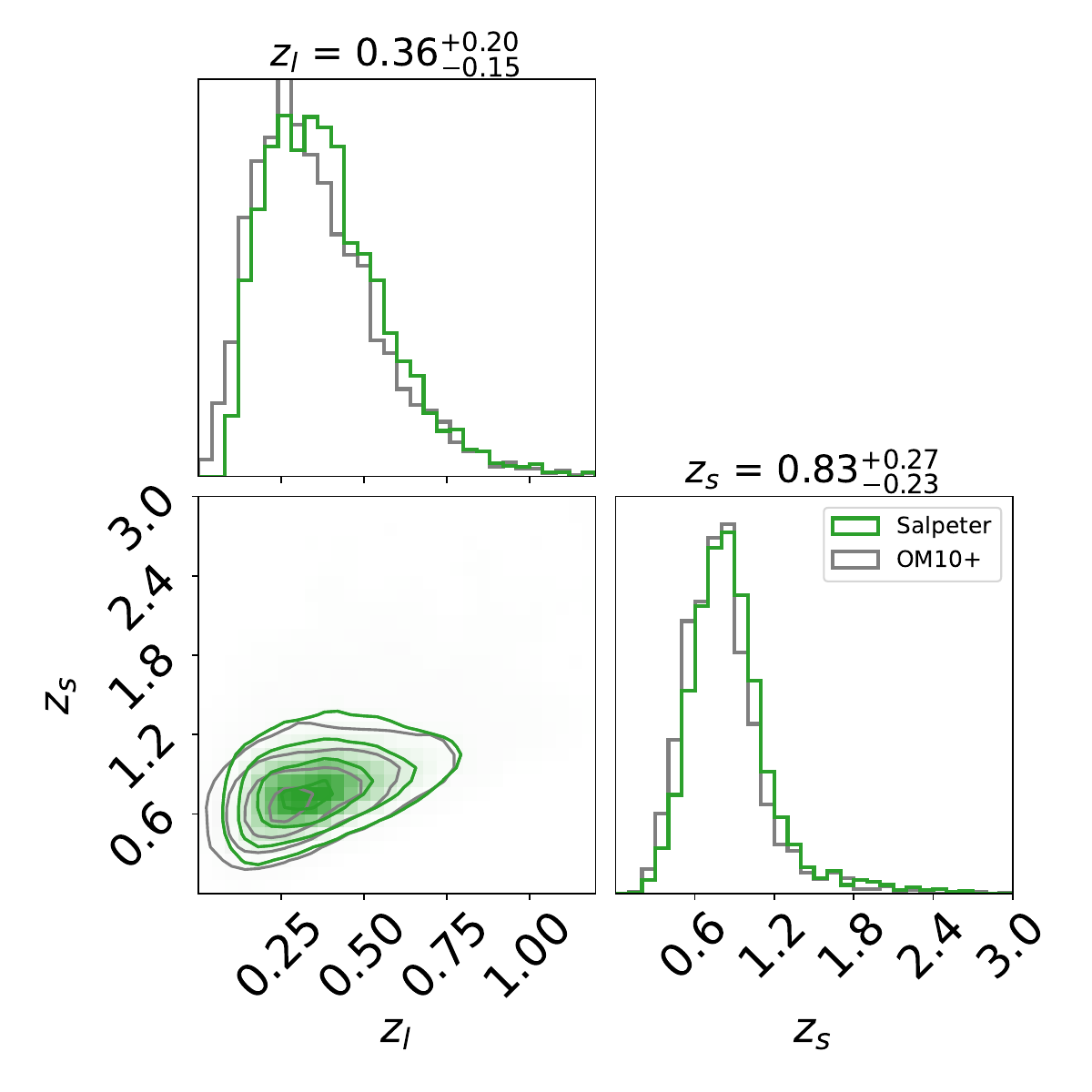}
  \caption{Same as Fig.~\ref{fig: zl_zs_dist_qso}, but the redshift distributions in our fiducial lensed SN mock catalogs are shown. }
  \label{fig: zl_zs_dist_sn}
\end{figure}

We plot the redshift distributions of the lensed SNe mock catalogs in Fig.~\ref{fig: zl_zs_dist_sn} in the same manner as in Fig.~\ref{fig: zl_zs_dist_qso}.
Similarly, for lensed QSOs, since the same distribution of sources is adopted as in OM10+, and the redshift distribution of the lens objects is not significantly changed, our mock catalog shows similar redshift distributions for both the source and lens redshifts.
Source and lens redshifts of lensed SNe are typically much lower than those for lensed QSOs, simply because SNe are typically much fainter than QSOs.
There also appears to be no difference in distributions between lenses produced by subhalos and those by host halos and central galaxies.
We do not show distributions of the ellipticity, position angle, velocity dispersion, external shear, and mass distributions for lensed SNe mock catalogs because of their similarity of the statistical properties to those in lensed QSOs.

\subsection{Distributions of time delays for lensed QSOs and SNe}
Predicting distributions of time delays is crucial when assessing the measurability of time delays from future time-domain surveys.
Here, we discuss the detailed properties of all multiple images, especially for double and quad lenses.

\begin{figure*}[htbp]
        \centering
        \includegraphics[width=\hsize]{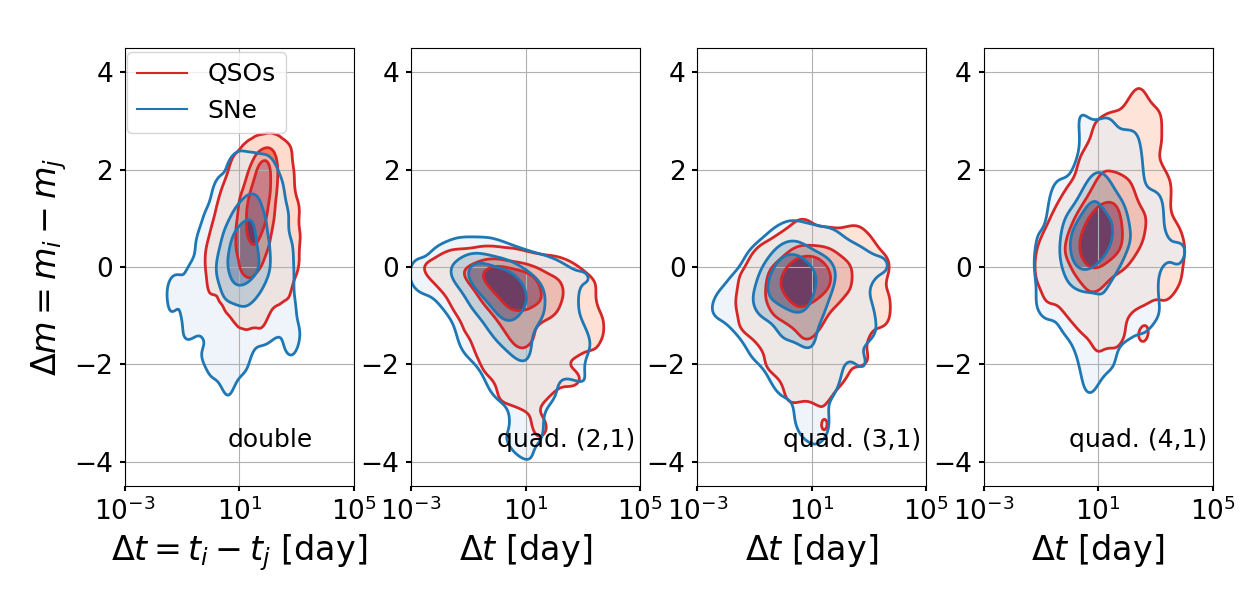}
    \caption{Predicted distributions of time delays and magnitude differences for double~(left) and quad~(three panels on the right) lenses in our fiducial lensed QSO and SN mock catalogs. The red-shaded regions show contour plots at $25\%$, $50\%$, and $75\%$ for lensed QSOs, while the green-shaded regions represent those for lensed SNe. The numbers in brackets in the bottom right-hand corner indicate the set of labels for the images being plotted. Results for each image pair are plotted against image 1, with images named in order of arrival: image 1 arrives first, followed by image 2, and so on.}
  \label{fig: dt_dmag_nim3_5}
\end{figure*}

The left panel in Fig.~\ref{fig: dt_dmag_nim3_5} shows the distribution of time delays and magnitude differences for double lenses 
in our fiducial mock for lensed QSOs and SNe.
In most cases for double lenses, the brighter images arrive prior to the fainter ones, which is the same as the expectations of basic spherical mass models, although the arrival order is inverted in some cases.
Based on this, 
we use the magnification of the second brightest image when calculating the magnification bias.
Thus, the detection of the first arriving images is ensured in either case.
We also find that lensed SNe have shorter time delays on average than lensed QSOs partly due to their lower lens/source redshifts and the stronger effect of the magnification bias.

The distributions of time delays and magnitude differences for quad lenses derived from our fiducial mock catalogs are displayed in Fig.~\ref{fig: dt_dmag_nim3_5}.
Quad lenses exhibit additional complexity due to the presence of more independent image pairs compared with double lenses.
However, in the three panels on the right of Fig.~\ref{fig: dt_dmag_nim3_5}, we clearly see some general trends as follows. The first arrival image tends to be fainter than the second and third arrival images, i.e., $\Delta m_{21} < 0$ and $\Delta m_{31} < 0$, and also tends to be brighter than the fourth image,
i.e., $\Delta m_{41} > 0$.
Hence for quad lenses, it is anticipated that the third brightest image is most likely to arrive first, which is in agreement with \cta{2010MNRAS.405.2579O} and supports our choice to make use of the third brightest image for the calculation of the magnification bias.
This selection ensures the successful detection of the first three images to arrive.
No significant differences are found in this distribution for quad lenses between lensed QSOs and SNe.

\subsection{Distributions of highly magnified QSOs and SNe}\label{sec: high_mag}
In addition to mock catalogs of multiply imaged QSOs and SNe, we construct mock catalogs of highly magnified systems, which are defined by lens systems with their maximum magnifications larger than a threshold value. With this definition, highly magnified systems include both single and multiple image systems. Such mock catalogs will be useful for investigating magnification bias in searching for high-redshift objects~\citep[e.g.,][]{2005ApJ...621..559K}. Recently, \citet{2023AJ....165..191Y} reported that J0025–0145 is a lensing system that is not multiply imaged but has significant magnification, which is referred to as an intermediate lensing system. Investigating the statistics of highly magnified systems also helps in discovering such intermediate lensing systems. 

We employ two thresholds for highly magnified systems, one with the maximum magnification larger than 3 and the other with the maximum magnification larger than 10. 
For lensed QSOs, we find that the number of highly magnified systems is about {$7459$} ($\mu>3$) and {$764$} ($\mu>10$) for the fiducial Salpeter IMF case and approximately {$6422$} ($\mu>3$) and {$629$} ($\mu>10$) for the Chabrier IMF case.
For lensed SNe, the number of highly magnified systems is about {$1098$} ($\mu>3$) and {$230$} ($\mu>10$) for the fiducial Salpeter IMF mock and approximately {$822$} ($\mu>3$) and {$170$} ($\mu>10$) for the Chabrier IMF.

\begin{figure}
    \centering
    \includegraphics[width=\linewidth]{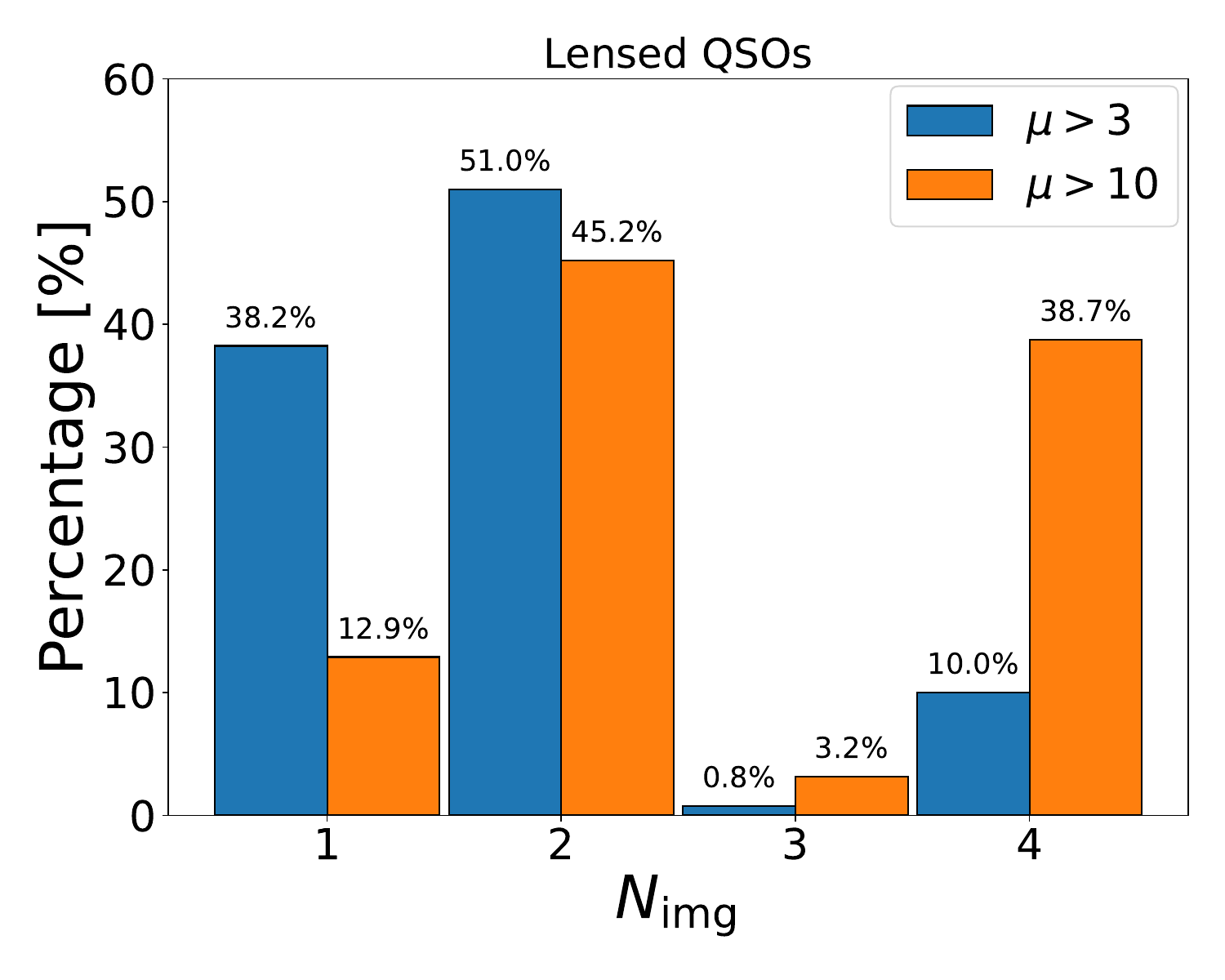}
    \caption{Fraction of highly magnified systems for lensed QSOs with one, two, three, or four images. Note that the demagnified image near the centre of the lens object is not counted here. 
}
    \label{fig: highmag_nimg}
\end{figure}

We plot the fraction of the number of images for our fiducial lensed QSOs in Fig.~\ref{fig: highmag_nimg}. For $\mu>3$, the majority of systems are two- and single-image  (i.e., intermediate lens) systems, while in $\mu>10$, the majority are two- and four-image systems.
\citet{2023AJ....165..191Y} suggests that J0025–0145 has a high magnification of $\mu \gg 1$ despite several systematic uncertainties, including the errors in the supermassive black hole mass measurement and bolometric correction. According to our fiducial mock catalogs, the probability of finding an intermediate lensing system with such a high magnification, $\mu>10$, is not negligible, around {$13\%$}, and it is predicted that there are about {100} intermediate lensing systems in LSST. This result is broadly consistent with the earlier analysis by \cite{2005ApJ...621..559K}. We also find similar results for the lensed SNe, which are not shown here.

We next discuss properties of lens objects causing these highly magnified systems. Such discussions have recently been advanced primarily in the context of gravitationally lensed gravitational waves.
There is still disagreement in the literature as to which lens populations are most important. 
For instance, \cite{2008MNRAS.386.1845H} and \cite{2018MNRAS.475.3823S} concluded that galaxy clusters are the most important lens population using ray tracing results from cosmological $N$-body simulations. 
Meanwhile, \cite{2020MNRAS.495.3727R} showed that massive galaxies are the most important lens population by performing a similar test using cosmological hydrodynamical simulations.

\begin{figure*}
    \begin{tabular}{cc}
       \begin{minipage}[t]{.5\linewidth}
        \centering
        \includegraphics[width=\hsize]{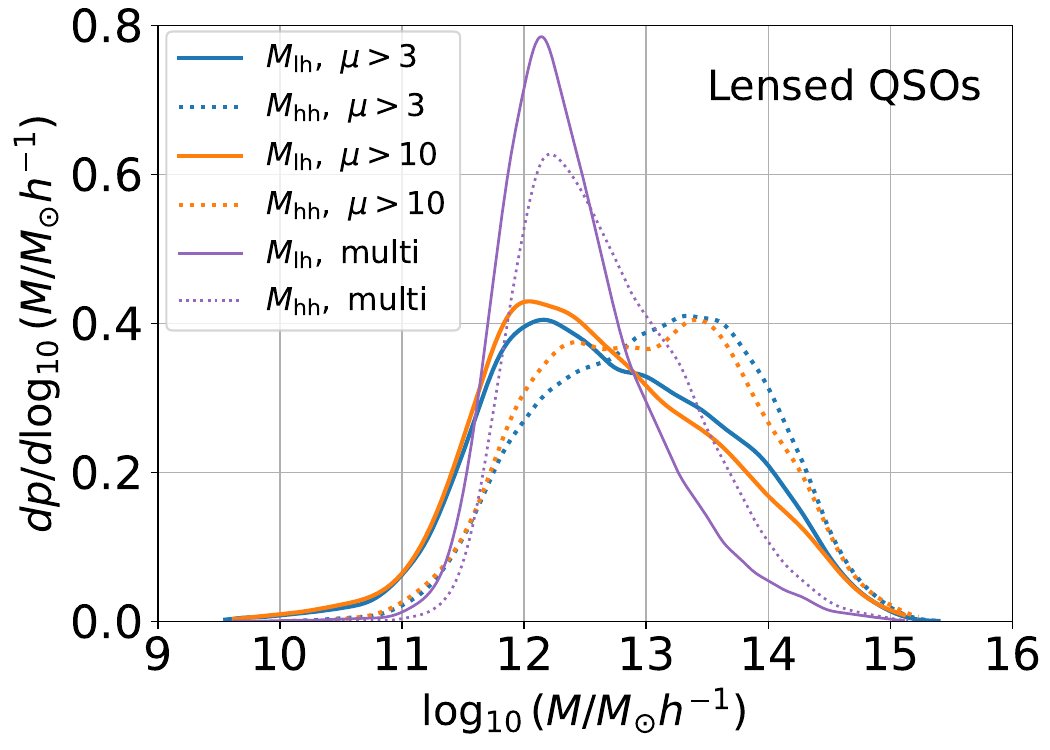}
      \end{minipage} &
      \begin{minipage}[t]{.5\linewidth}
        \centering
        \includegraphics[width=\hsize]{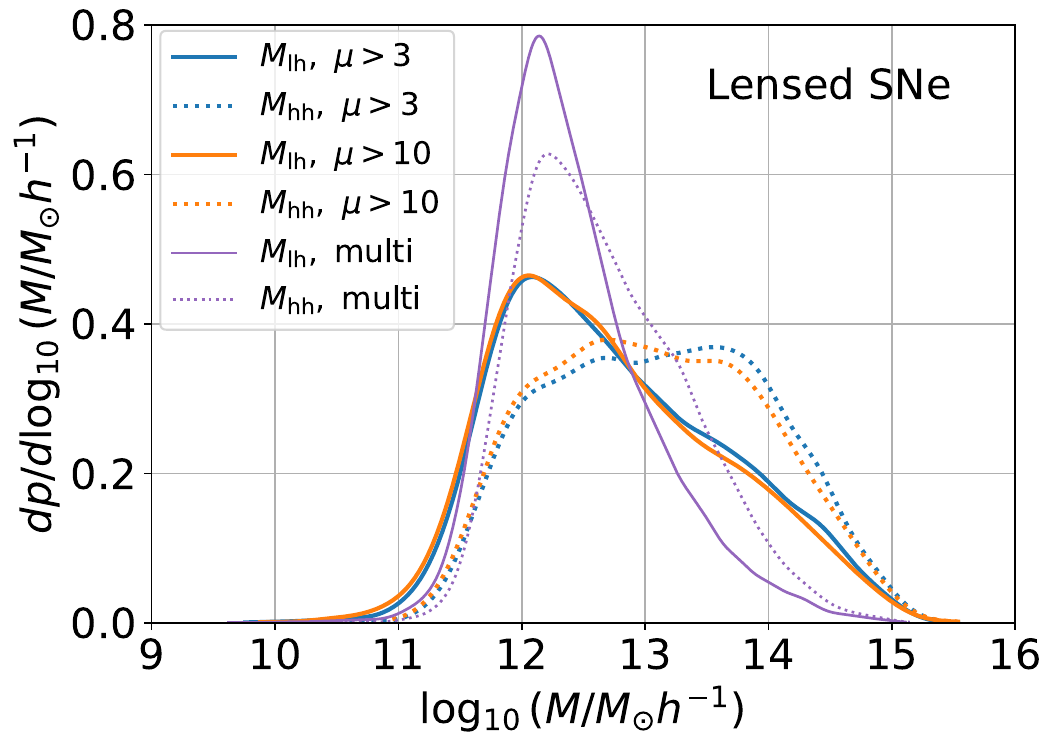}
      \end{minipage}
    \centering
    \end{tabular}
  \label{fig: highly_magnified_events_qso}
  \caption{The lens mass distributions for highly magnified systems. The left panel shows results from our fiducial lensed QSO mock catalogs, while the right panel presents those from our fiducial lensed SN mock catalogs. The solid lines represent the distribution of lens halo mass, as defined in Eq.~\eqref{eq: mlens_halo}. The dotted lines depict the host halo mass distribution. The blue and orange lines correspond to lenses with $\mu>3$ and $\mu>10$, respectively. For comparison, we also show lens mass distributions for multiple image systems presented in Sec.~\ref{sec: mock_multi_qso} and \ref{sec: mock_multi_sne} with thin purple lines.
  }
\end{figure*}

The lens mass distributions of these highly magnified systems in our fiducial mock catalogs are displayed in Fig.~\ref{fig: highly_magnified_events_qso}.
For lensed QSOs, we find that while highly magnified systems are often triggered by massive galaxies with $M\approx 10^{12-13}\Ms$, they are also frequently caused by subhalos, such as member galaxies in the outskirts of galaxy groups and clusters with $M\approx 10^{14-15}\Ms$. This is due to the additional magnification effect from environmental convergence within massive groups or clusters. This indicates that, although typical lenses of highly magnified systems are galaxy-scale lenses, galaxies in the outskirts of galaxy groups and clusters also contribute significantly. The distinct contributions of central galaxies in galaxy-scale halos and satellite galaxies in the outskirt of cluster-scale halos, in combination with the fact that magnifications of satellite galaxies are boosted due to the environmental convergence and shear, can presumably explain the two peaks of distributions of $M_{\mathrm{hh}}$ in Fig.~\ref{fig: highly_magnified_events_qso}. However, we also find that the lensing probability of massive galaxies with $M\approx 10^{12-13}~\Ms$ tends to be slightly more dominant when the magnification criterion is increased from 3 to 10. For lensed SNe, the results are similar to those for lensed QSOs, but there is a stronger tendency for highly magnified systems to be caused by massive galaxies because of the stronger effect of the magnification bias. We note that, as also shown in Fig.~\ref{fig: highly_magnified_events_qso}, lens mass distributions for multiple image systems have one prominent peak at $M\approx 10^{12}\Ms$, suggesting that lensing galaxies of multiply imaged QSOs and SNe are typically isolated massive galaxies in galaxy-scale halos.

The reason why we do not see as much difference in the amount of highly magnified systems in the fiducial Salpeter IMF and Chabrier IMF case as we do in the lensed mock catalogs can be explained as follows.
While it is the mass and surface density inside and near the Einstein radius that is important for the multiple-image lenses, mass distributions at the outer radii become important when predicting (moderately) high-magnification lenses, especially in the single-image case.
In this case, the interior mass is no longer dominated by stellar components, and the difference in the IMF becomes unimportant.
In fact, the difference in the number of lenses with $\mu>3$ between the fiducial Salpeter IMF and the Chabrier IMF cases comes almost entirely from the difference in the number of highly magnified multiple-image lenses.

\subsection{Stellar mass fractions for lensed QSOs and SNe} 
Another important quantity related to strong lenses is the stellar mass fraction
\eq{
f_*(\bm{\theta})\equiv \frac{\kappa_{\mathrm{lens,gal}}(\bm{\theta})}{\kappa_{\mr{tot}}(\bm{\theta})},
}
where $\kappa_{\mathrm{lens, gal}}(\bm{\theta})$ and $\kappa_{\mr{tot}}(\bm{\theta})$ are the convergence from the primary lens galaxy, as defined in Eq.~\eqref{eq: mlens_gal}, and the total convergence at the image position of $\bm{\theta}$, respectively.
While in this paper we do not consider the microlensing effect, the stellar mass fractions of individual images are quite important when adding microlensing to our mock catalog \citep{2024SSRv..220...14V}.

\begin{figure}[htbp]
    \centering
    \includegraphics[width=\linewidth,clip]{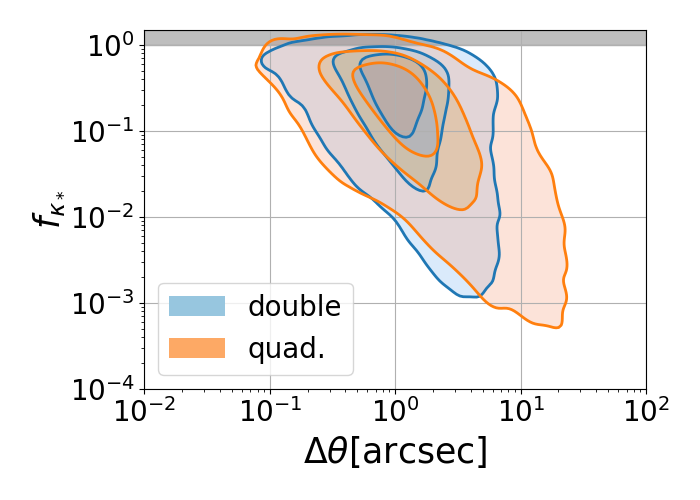}
    \caption{The stellar mass fractions in our mock catalogs for lensed QSOs as a function of maximum image separation. The blue-shaded regions show contour plots at $25\%$, $50\%$, and $75\%$ for double lens mock catalogs, while the orange-shaded regions represent those for quad lens mock catalogs. The grey-shaded area indicates the excluded parameter region where $f_{\kappa*}> 1$.}
    \label{fig: kapst_fraction_qso}
\end{figure}

The stellar convergence fraction in our mock catalog for lensed QSOs is plotted in Fig.~\ref{fig: kapst_fraction_qso}.
In this plot we remove central demagnified images that always have quite high stellar mass fractions due to their proximity to central galaxies.
As expected, we find an inverse correlation between the maximum image separation and the stellar convergence fraction, indicating that the contribution of dark matter becomes more dominant as the mass scale of the lens object increases. We also find that a quad lens tends to show a higher stellar convergence fraction than a double lens. 

No result is shown for the stellar mass fraction for lensed SNe as it differs little from that for lensed QSOs shown in Fig.~\ref{fig: kapst_fraction_qso}.

\section{Comparing with currently known samples of lensed QSOs}\label{sec: mock_sdss}
In Sec.~\ref{sec: lensmock}, we present mock catalogs of lensed QSOs and SNe for a 10-yr long LSST observation run. 
Here, we compare our mock catalogs with known lensed QSOs. For this purpose, we employ the statistical sample of lensed QSOs from the SDSS Quasar Lens Search~\citep[SQLS;][]{2006AJ....132..999O}, covering about the area of $8000~\mathrm{deg}^2$, provided by \citet{2012AJ....143..119I}, as well as the sample from Gaia \citep{2018A&A...616A...1G}, covering about the area of about $22000~\mathrm{deg}^2$.

\subsection{Comparing with the SQLS lens sample}
For a fair comparison with the SQLS statistical lensed QSO sample presented in \citet{2012AJ....143..119I}, we generate a subsample of our mock catalogs to include objects with image separations in the range of $1~\mathrm{arcsec} <\Delta \theta< 20~\mathrm{arcsec}$, an $i$-band magnitude difference less than 1.25 mag for doubles, and an effective point-spread function (PSF) magnitudes of $i_{\mathrm{eff}} < 19.1$ for QSOs with $0.6 <z< 2.2$.
Here the effective PSF magnitude is calculated by the effective magnification $\mu_{\mathrm{eff}}$, which is defined as~\citep{2012AJ....143..120O}
\eq{
\mu_{\mathrm{eff}}=\bar{\mu} \mu_{\mathrm{tot}}+(1-\bar{\mu}) \mu_{\mathrm{max}},
}
where $\mu_{\mathrm{tot}}$ is the total magnification, $\mu_{\mathrm{max}}$ is the magnification of the brightest image, and 
\eq{
\bar{\mu}=\frac{1}{2}[1+\tanh (1.76-1.78 \Delta\theta)],
}
with the maximum image separation $\Delta\theta$ in units of arcsec.

\begin{figure}
    \centering
    \includegraphics[width=\linewidth]{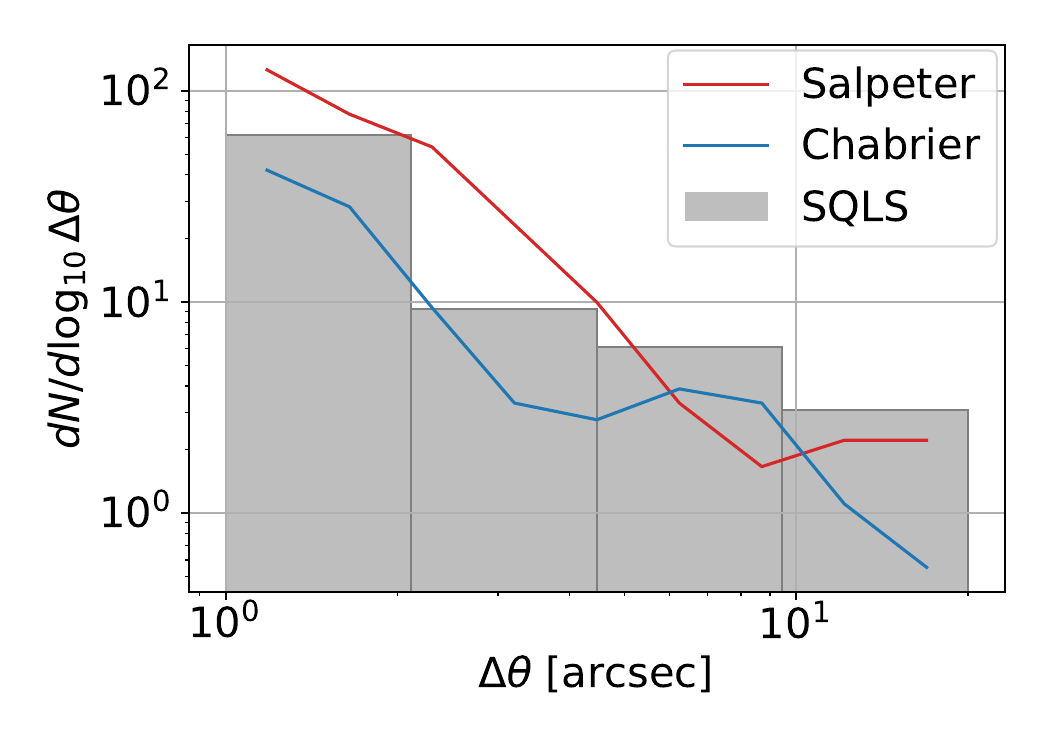}
    \caption{The number distribution of lensed QSOs as a function of the maximum image separations between multiple images from subsamples of our mock catalogs to construct SQLS-like mock catalogs. The red line shows the result for our fiducial Salpeter IMF mock catalogs, while the blue line shows the result for the Chabrier IMF mock catalog.  The grey histogram displays the number distribution of lensed QSOs in the SQLS statistical sample \citep{2012AJ....143..119I}.}
    \label{fig: dtheta_sdss}
\end{figure}

While the SQLS statistical sample contains 4 quad lenses out of a total of 26 lenses, our fiducial Salpeter IMF mock catalogs predict {$43\pm 6$} lensed QSOs, including {$9\pm 2$} quad lenses, where this interval corresponds to a $68\%$ confidence level from our mock catalogs of five LSST realizations. The numbers in our mock catalogs coincide with those in observations at the {$3\sigma$} level. In Fig.~\ref{fig: dtheta_sdss}, we show the distribution of the maximum image separation in our fiducial SQLS-like mock catalogs compared to the SQLS sample. It should be noted that the SQLS sample contains SDSS~J1004+4112 \citep{2003Natur.426..810I,2004ApJ...605...78O,2005PASJ...57L...7I} with the maximum image separation of about $15$~arcsec; OM10+ cannot predict such a large-separation lens, as can be seen from Fig.~\ref{fig: dtheta_dist_qso}. In contrast, our fiducial model indeed predicts such a large-separation lens in the SDSS. We also find that our fiducial model reproduces well the number of lensed QSOs for the whole image separation range.

In Fig.~\ref{fig: dtheta_sdss}, we also plot the result from the Chabrier IMF mock catalogs for comparison. We find that the Chabrier IMF cannot reproduce the observation.  Specifically, the Chabrier IMF mock catalogs predict a total of {$14 \pm 2$} lensed QSOs, of which {$4 \pm 1$} are expected to be quad lenses. This indicates that the abundance of lensed QSOs favors the Salpeter IMF over the Chabrier IMF. Interestingly, this result is in line with previous strong lensing studies favoring the Salpeter IMF rather than the Chabrier IMF~\citep[see e.g.,][]{2010ApJ...709.1195T,2014MNRAS.439.2494O, 2015ApJ...800...94S}. We note that there are several strong lensing studies that prefer Chabrier IMF rather than Salpeter IMF \citep[e.g.][]{2013MNRAS.434.1964S,2015MNRAS.449.3441S,2019A&A...630A..71S}, which implies the dependence of the stellar IMF on the stellar mass and redshift. Incorporating such mass and redshift dependence of the stellar IMF is left for future work.

\subsection{Comparing with the Gaia lens sample}
Following the analysis in \citet{2019MNRAS.483.4242L, 2023MNRAS.520.3305L}, we generate a subsample of our mock catalogs to include objects with image separations in the range of $1~\mathrm{arcsec} <\Delta \theta< 4~\mathrm{arcsec}$, the QSO redshift of $z< 4$, and $G$-band magnitudes of $G < 20.7$ for both images for double lenses and for at least three bright images for quad lenses. In our analysis we remove lensed QSOs without any information on the magnitude of the second brightest image for double lenses and the third brightest image for quad lenses.
 
To convert $i$-band magnitudes of lensed QSO images to Gaia $G$-band magnitudes, we use the SDSS spectroscopic QSO catalog cross-matched to Gaia \citep{2020ApJS..250....8L} to determine a $G-i$ relation, ignoring the redshift dependence and the scatter in the $G-i$ relation. Specifically we adopt the offset of $G-i = 0.15$. 

The Gaia lens sample is constructed based on \cta{lensed_qso_database_clemon} as well as the latest follow-up result presented in \cite{2023MNRAS.520.3305L} for 53.9\% of the full sky, which corresponds to the typical areas searched for lenses \citep{2023MNRAS.520.3305L}.
While the Gaia sample contains 215 lensed QSOs including 25 quad lenses, our fiducial Salpeter mock catalogs predict {$306\pm 25$} lensed QSOs, including {$51\pm 5$} quad lenses. This indicates that our mock catalogs are in agreement with the Gaia lens sample. 
It should be emphasized that in this Gaia-like selection, the quad fraction increases from $\sim 10\%$ for the LSST lens sample to {$\sim 17\%$}. As discussed in Fig.~\ref{fig: dN_mobs_cum}, this difference of the quad fraction simply reflects the difference of the survey depth between Gaia and LSST. 

\begin{figure}
    \centering
    \includegraphics[width=\linewidth]{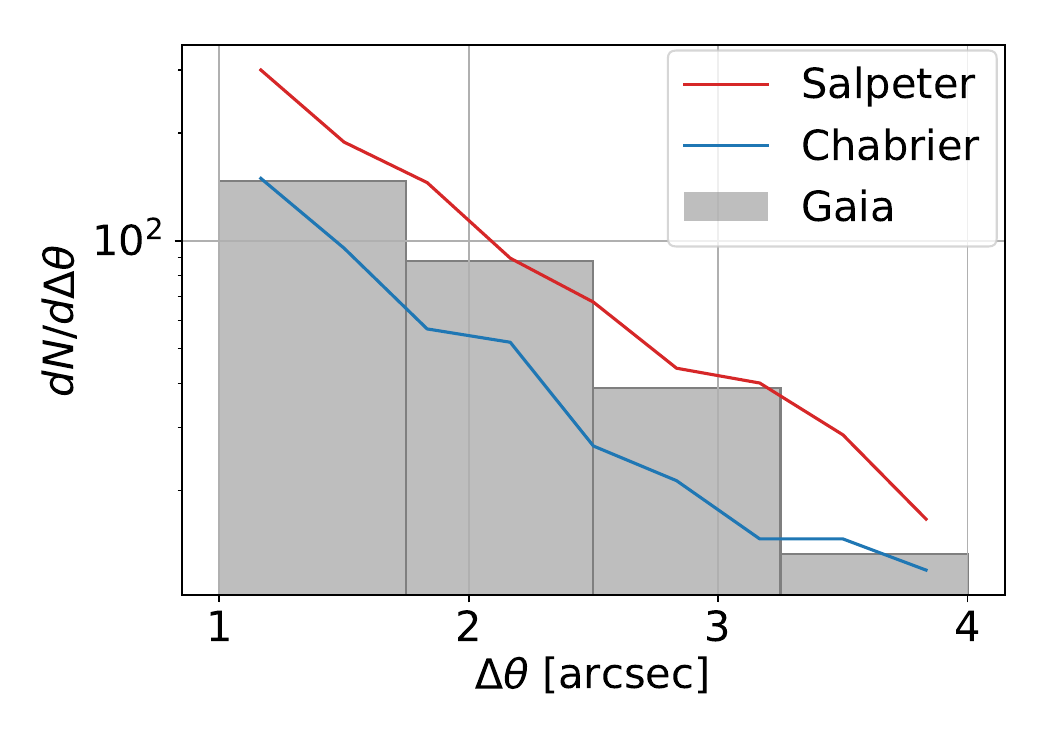}
    \caption{Similar to Fig.~\ref{fig: dtheta_sdss}, but for lensed QSOs observed in Gaia compiled from \cta{lensed_qso_database_clemon} and \cite{2023MNRAS.520.3305L}.}
    \label{fig: dtheta_gaia}
\end{figure}

In Fig.~\ref{fig: dtheta_gaia}, we show the number distribution as a function of the maximum image separation for our subsample of mock catalogs and the Gaia lens sample. We find that the image separation distribution in the whole range of $1~\mathrm{arcsec} < \Delta \theta < 4~\mathrm{arcsec} $ is reproduced well. In contrast, as in the case of SQLS, the Chabrier IMF mock catalogs predict a significantly smaller number of lensed QSOs in Gaia. Specifically, the Chabrier IMF mock catalogs predict a total of {$147\pm 6$} lensed QSOs, of which {$27 \pm 6$} are expected to be quad lenses.

\section{Conclusion}\label{sec: conclusion}
The Hubble tension has become a significant issue in cosmology. Time delays between multiple images of strong gravitational lenses offer one of the promising methods to determine $H_0$ independently of the distance ladder.
While $H_0$ has mainly been estimated from galaxy-scale lensed QSOs, cluster-scale lenses have also attracted attention in the context of $H_0$ measurements in recent years.
Both galaxy- and cluster-scale strong lens systems with measurable time delays are expected to be discovered in large numbers in future observations such as LSST.

We have adopted the halo model to generate mock catalogs of lensed QSOs and SNe covering galaxy-, group-, and cluster-scales. This represents a significant update of previous mock catalogs such as the one presented in OM10+, for which the SIE mass model is used and as a result only galaxy-scale lenses are considered. In our model, the mass distribution of individual lens objects is modeled as a combination of dark matter and stellar components, as described in Sec.~\ref{sec: lensmodel}. Another feature of our lens model is the inclusion of subhalos and satellite galaxies. In our model, galaxies properties change smoothly as a function of the halo mass from galaxy- to cluster-scale halos. In Sec.~\ref{sec: original_pop}, we discussed the validity of our model in populating galaxies within halos by comparing various observables, such as galaxy velocity dispersion functions and the stellar mass fundamental plane.

With this model, we have generated mock catalogs for lensed QSOs and SNe found in 
the baseline survey planned in LSST as described in Sec.~\ref{sec: lensmock}. We found that the LSST will discover about {4000} lensed QSOs and 200 SNe with resolved multiple images assuming the Salpeter IMF as shown in Table~\ref{tab: lensed_mocks}. We have presented the distributions of lens, source, and image properties from the mock catalogs. In addition to mock catalogs of multiple-image lens systems, we also created mock catalogs of highly magnified systems, including both multiple-image and single-image lenses, and discussed their statistical properties. We summarize below the highlights of our results and discuss our results in comparison with previous work such as OM10+.

\begin{itemize}
    \item In our fiducial calculation, we have assumed the Salpeter IMF, which predicts the similar numbers of lensed QSOs and SNs as compared with OM10+ as seen in Table~\ref{tab: lensed_mocks}. For comparison, we have also generated mock catalogs in the Chabrier IMF case. We have found that the predicted numbers of lensed QSOs and SNs decrease approximately by half by changing the IMF from Salpeter to the Chabrier, regardless of lens and source redshifts and image separations, which suggests the potential of strong lens statistics to probe the stellar IMF.
    \item We have provided the number distribution as a function of the maximum image separation tailored for the baseline survey planned by LSST in Fig.~\ref{fig: dtheta_dist_qso}. In addition to galaxy-scale lenses that are in agreement with the result of OM10+, our mock catalogs also include group- and cluster-scale lenses. Indeed our mock catalogs predict that LSST will find about 80 lensed QSOs with image separations larger than 10~arcsec.
    \item We have found that the probability of detecting highly magnified single-image systems, which is sometimes referred to as an intermediate lensed QSO, with high magnification, $\mu>10$, like J0025–0145 discovered in \citet{2023AJ....165..191Y} is around {$13\%$} among all the high-magnification gravitational lens events, which suggests that about {100} such systems for lensed QSOs will be present in LSST.
    \item  We have found that whilst highly magnified systems are typically produced by massive galaxies with halo masses of $M\approx 10^{12-13}~\Ms$, the non-negligible fraction of them correspond to member galaxies located in the outskirts of galaxy groups and clusters with halo masses of $M\approx 10^{14-15}~\Ms$. This is due to the additional magnification effect caused by the environmental convergence inside the massive groups and clusters. In contrast, lensing galaxies of multiply imaged QSOs and SNe typically reside in halos with masses of $M\approx 10^{12}~\Ms$.
    \item In Sec.~\ref{sec: mock_sdss}, we have discussed whether our mock catalogs can reproduce abundances and image separation distributions of known lensed QSO samples in the SDSS/SQLS and Gaia. We have confirmed that our mock catalog predictions of numbers, image separation distributions, and quad fractions are in good agreement with observations for our fiducial Salpeter IMF case. In contrast, for the Chabrier IMF case, predicted numbers are much smaller compared with observations. Our result supports the Salpeter-like IMF for massive galaxies.
\end{itemize}

We expect that this work will be useful for future astrophysical and cosmological applications of lensed QSO and SN systems that have been or will be searched and discovered.

We note that we have set a lower limit on the mass of halos to {$M_{\mathrm{hh}}\gtrsim 10^{10}~\Ms$} for main halos or {$M_{\mathrm{sh}}\gtrsim 10^{9}~\Ms$} for subhalos for numerical reasons. This means that the abundance of very small image separation lenses, such as sub-arcsecond lensed SNe discussed in \cite{2023NatAs...7.1098G}, can be underestimated. We leave the detailed comparison of the abundance of very small-separation lenses with observations to future work. 

Finally, our mock catalog generation code (\texttt{Strong Lensing Halo model-based mock catalogs~(SL-Hammocks)} used in this paper is published on the LSST DESC GitHub\footnote{\url{https://github.com/LSSTDESC/SL-Hammocks}}. In addition, catalogs of lensed QSOs and SNe presented in this paper are published on the LSST Strong Lensing Science GitHub\footnote{\url{https://github.com/LSST-strong-lensing/data_public}}. While in this paper we mainly focus on the LSST survey, this code can be applied to any ongoing and future time-domain optical imaging survey.

\acknowledgements
Author contributions: KTA mainly developed the SL-Hammocks code, conducted most of the analysis in the paper, and wrote the main body of the paper. MO conceived of and initiated the project and contributed to the development of the SL-Hammocks code, designing the validation tests, interpretations of results, and paper writing. SB provided a code to compute velocity dispersions of galaxies for validations, and contributed to the development and validations of the SL-Hammocks code. NK and PJM also contributed to the development and validations of the SL-Hammocks code. CL and AM provided suggestions on many aspects of the analysis and the write-ups.

{We thank Kai Liao for pointing out a bug in calculating lensing by satellite galaxies, which is now fixed.}
We thank anonymous referees for constructive comments.
This work was supported by JSPS KAKENHI Grant Numbers JP23K22531, JP20H05856, JP22K21349. The work of PJM was supported by the U.S. Department of Energy under contract number DE-AC02-76SF00515. This project has received funding from the European Union's Horizon Europe research and innovation programme under the Marie Sklodovska-Curie grant agreement No 101105725.

This paper has undergone internal review in the LSST Dark Energy
Science Collaboration. We would like to thank internal reviewers
Cameron Lemon and Anupreeta More for their role. The DESC acknowledges
ongoing support from the Institut National de Physique Nucl\'eaire et
de Physique des Particules in France; the  
Science \& Technology Facilities Council in the United Kingdom; and the
Department of Energy, the National Science Foundation, and the LSST 
Corporation in the United States.  DESC uses resources of the IN2P3 
Computing Center (CC-IN2P3--Lyon/Villeurbanne - France) funded by the 
Centre National de la Recherche Scientifique; the National Energy 
Research Scientific Computing Center, a DOE Office of Science User 
Facility supported by the Office of Science of the U.S.\ Department of
Energy under Contract No.\ DE-AC02-05CH11231; STFC DiRAC HPC Facilities, 
funded by UK BEIS National E-infrastructure capital grants; and the UK 
particle physics grid, supported by the GridPP Collaboration.  This 
work was performed in part under DOE Contract DE-AC02-76SF00515.

This work used the following packages; \texttt{NumPy} \citep{2020Natur.585..357H}, \texttt{SciPy} \citep{2020SciPy-NMeth}, \texttt{Matplotlib} \citep{2007CSE.....9...90H}, \texttt{COLOSSUS} \citep{2018ApJS..239...35D}, \texttt{glafic} \citep{2010PASJ...62.1017O,2021PASP..133g4504O}, \texttt{joblib} \citep{joblib}, \texttt{Astropy} \citep{2022ApJ...935..167A}.

\bibliography{article}

\end{document}